\documentclass[12pt]{article}
\pdfoutput=1

\usepackage[a4paper,text={16.8cm,22.4cm}]{geometry}
\usepackage{amsmath,amsfonts,braket,slashed,amssymb,tikz,bm,bbm,psfrag,graphicx,color,dsfont,euscript}
\usepackage{mathtools}
\usepackage{multicol}
\usepackage{ctable}
\usepackage[small,labelfont=bf]{caption}

\RequirePackage[sort&compress,square,comma,numbers]{natbib}
\allowdisplaybreaks
\newcommand{\A}{{\EuScript A}}
\newcommand{\X}{{\EuScript X}}
\newcommand{\G}{{\EuScript G}}
\newcommand{\Asl}{\rlap{\hspace{0.7mm}/}{\A}}
\newcommand{\Gsl}{\rlap{\hspace{0.2mm}/}{\G}}
\newcommand{\nsl}{\rlap{\hspace{0.25mm}/}{n}}
\newcommand{\nbsl}{\rlap{\hspace{0.25mm}/}{\bar n}}
\newcommand{\spac}{{\hspace{0.3mm}}}
\newcommand{\braces}[1]{[\hspace{-0.5mm}[#1]\hspace{-0.5mm}]}

\begin{document}

\begin{titlepage}

\begin{flushright}
\normalsize
MITP/20-047\\ 
September 14, 2020
% arXiv:2009.06779
% v1: September 14, 2020
% v2: November 18, 2020; updated with proof corrections on January 11, 2021
\end{flushright}

\vspace{1.0cm}
\begin{center}
\Large\bf\boldmath 
Factorization at Subleading Power and Endpoint Divergences in $h\to\gamma\gamma$ Decay:\\
II. Renormalization and Scale Evolution
\end{center}

\vspace{0.5cm}
\begin{center}
Ze Long Liu$^{a}$, Bianka Mecaj$^b$, Matthias Neubert$^{b,c,d}$ and Xing Wang$^b$\\
\vspace{0.7cm} 
{\sl ${}^a$Theoretical Division, Los Alamos National Laboratory, Los Alamos, NM 87545, U.S.A.\\[3mm]
${}^b$PRISMA$^+$ Cluster of Excellence \& Mainz Institute for Theoretical Physics\\
Johannes Gutenberg University, 55099 Mainz, Germany\\[3mm]
${}^c$Department of Physics \& LEPP, Cornell University, Ithaca, NY 14853, U.S.A.\\[3mm]
${}^d$Department of Physics, Universit\"at Z\"urich\\ 
Winterthurerstrasse 190, CH-8057 Z\"urich, Switzerland}
\end{center}

\vspace{0.8cm}
\begin{abstract}
Building on the recent derivation of a bare factorization theorem for the $b$-quark induced contribution to the $h\to\gamma\gamma$ decay amplitude based on soft-collinear effective theory, we derive the first renormalized factorization theorem for a process described at subleading power in scale ratios, where $\lambda=m_b/M_h\ll 1$ in our case. We prove two refactorization conditions for a matching coefficient and an operator matrix element in the endpoint region, where they exhibit singularities giving rise to divergent convolution integrals. The refactorization conditions ensure that the dependence of the decay amplitude on the rapidity regulator, which regularizes the endpoint singularities, cancels out to all orders of perturbation theory. We establish the renormalized form of the factorization formula, proving that extra contributions arising from the fact that ``endpoint regularization'' does not commute with renormalization can be absorbed, to all orders, by a redefinition of one of the matching coefficients. We derive the renormalization-group evolution equation satisfied by all quantities in the factorization formula and use them to predict the large logarithms of order $\alpha\spac\alpha_s^2\spac L^k$ in the three-loop decay amplitude, where $L=\ln(-M_h^2/m_b^2)$ and $k=6,5,4,3$. We find perfect agreement with existing numerical results for the amplitude and analytical results for the three-loop contributions involving a massless quark loop. On the other hand, we disagree with the results of previous attempts to predict the series of subleading logarithms $\sim\alpha\spac\alpha_s^n\spac L^{2n+1}$. 
\end{abstract}

\end{titlepage}

\tableofcontents
\newpage

\section{Introduction}
\renewcommand{\theequation}{1.\arabic{equation}}
\setcounter{equation}{0}

Soft-collinear effective theory (SCET) \cite{Bauer:2001yt,Bauer:2002nz,Beneke:2002ph} provides a convenient framework for addressing the problems of scale separation and factorization in high-energy physics using the powerful tools of effective field theory. Much recent work has focused on exploring the structure of factorization at subleading order in power counting -- a problem that turns out to be unexpectedly subtle and full of complexities. Specific applications discussed in the literature include the study of power corrections to event shapes \cite{Moult:2018jjd} and transverse-momentum distributions \cite{Ebert:2018gsn,Moult:2019vou}, the threshold factorization for the Drell-Yan process \cite{Beneke:2018gvs,Beneke:2019oqx}, and the factorization of power-suppressed contributions to Higgs-boson decays \cite{Liu:2019oav,Wang:2019mym}. One finds that such factorization theorems contain a sum over convolutions of Wilson coefficients with operator matrix elements, where the relevant SCET operators mix under renormalization. Several new complications arise, which do not occur at leading power. The most puzzling one is the appearance of endpoint-divergent convolution integrals over products of component functions each depending on a single scale \cite{Moult:2019mog,Beneke:2019kgv,Moult:2019uhz,Beneke:2019oqx,Moult:2019vou,Liu:2019oav,Wang:2019mym,Beneke:2020ibj}. In some sense, such endpoint divergences indicate a failure of dimensional regularization and the $\overline{\rm MS}$ subtraction scheme, because some of the $1/\epsilon^n$ pole terms are not removed by renormalizing the individual component functions, and hence naive scale separation is violated. Standard tools of renormalization theory are then insufficient to obtain well-defined, renormalized factorization theorems involving convergent convolutions over renormalized functions. Indeed, for none of the above-mentioned examples is it currently known how to formulate a theoretically consistent renormalized factorization theorem. This would, however, be needed in order to fully establish SCET as a versatile tool and apply it to several observables of phenomenological interest. 

In a recent paper \cite{Liu:2019oav}, two of us have started a detailed discussion of factorization at subleading power within the framework of SCET. As a concrete example, we have considered the decay amplitude for the radiative Higgs-boson decay $h\to\gamma\gamma$ mediated by the Higgs coupling to bottom quarks. In the limit $m_b\ll M_h$ one finds for the corresponding contribution to the decay amplitude
\begin{equation}\label{Mb}
   {\cal M}_b(h\to\gamma\gamma)
   = \frac{N_c\spac\alpha_b}{\pi}\,\frac{y_b}{\sqrt2}\,m_b\,g_\perp^{\mu\nu}\spac 
    \varepsilon_\mu^*(k_1)\,\varepsilon_\nu^*(k_2)\! 
    \left[ \frac{L^2}{2} - 2 + \frac{C_F\alpha_s}{4\pi}
    \left( - \frac{L^4}{12} - L^3 + \dots \right) + {\cal O}(\alpha_s^2) \right]\! ,
\end{equation}
where $L=\ln(M_h^2/m_b^2)-i\pi$ is the relevant large logarithm. Here $m_b$, $y_b$ and $\alpha_b=e_b^2\spac\alpha$ (with $e_b=-\frac13$) denote the mass, Yukawa coupling and electromagnetic coupling of the $b$ quark, and $\varepsilon_\mu(k_i)$ are the polarization vectors of the two photons. At leading double-logarithmic order, the large logarithms in this amplitude were resummed a long time ago using conventional tools of perturbative quantum field theory \cite{Kotsky:1997rq,Akhoury:2001mz} (see \cite{Liu:2017vkm,Liu:2018czl} for more recent related work), and it was found that
\begin{equation}
   {\cal M}_b^{\rm LL}(h\to\gamma\gamma)
   = \frac{N_c\spac\alpha_b}{\pi}\,\frac{y_b}{\sqrt2}\,m_b\,g_\perp^{\mu\nu}\spac 
    \varepsilon_\mu^*(k_1)\,\varepsilon_\nu^*(k_2)\, 
    \frac{L^2}{2} \sum_{n=0}^\infty \frac{2\Gamma(n+1)}{\Gamma(2n+3)} 
    \left( - \frac{C_F\alpha_s}{2\pi}\,L^2 \right)^n .
\end{equation}
However, we will see later that the sum of this series provides a poor numerical approximation to the decay amplitude. In order to go beyond the leading double-logarithmic approximation it is necessary to start from a consistent all-order factorization theorem, which properly separates the relevant scales in this process. In \cite{Liu:2019oav} we have factorized the decay amplitude and expressed it in terms of convolutions of bare matching coefficients with unrenormalized operator matrix elements. The convolution integrals contain endpoint divergences that require both dimensional and rapidity regulators. We have shown that, to all orders of perturbation theory, the endpoint divergences cancel out in the sum of all contributions to the factorization theorem, and they can be removed by suitably rearranging the factorization formula. 

Here we continue our study of the $b$-quark induced $h\to\gamma\gamma$ decay amplitude and derive a renormalized factorization theorem for this process. The establishment of such a factorization formula -- the first of its kind -- is the main accomplishment of the present work. In Section~\ref{sec:barefact} we recall the factorization theorem for the $b$-quark induced $h\to\gamma\gamma$ decay amplitude as derived in \cite{Liu:2019oav}. A crucial step in the derivation of this formula made use of two $D$-dimensional refactorization conditions, which were derived from the requirement that in the sum of all terms the dependence on the rapidity regulator must cancel to all orders of perturbation theory. In Section~\ref{sec:refact} we prove these refactorization conditions using techniques from SCET. The renormalized form of the factorization formula is derived in Section~\ref{sec:renfact}. We first discuss the renormalization of the various operators and matching coefficients and derive their renormalization-group (RG) evolution equations using standard tools of quantum field theory. There is one important subtlety related to the fact that the presence of hard cutoffs on the convolution integrals, which is a consequence of the regularization of endpoint divergences, does not commute with renormalization. In Section~\ref{sec:mismatch} we show that, to all orders of perturbation theory, the additional terms encountered when the cutoffs are moved from the bare to the renormalized convolutions amount to an extra contribution to one of the matching coefficients. The evolution equations satisfied by the renormalized matching coefficients and matrix elements are derived in Section~\ref{sec:RGEs}. Using these equations, we predict in Section~\ref{sec:largelogs} the large logarithms of order $\alpha_b\spac\alpha_s^2 L^k$ with $k=6,5,4,3$ in the three-loop decay amplitude, finding complete agreement with existing multi-loop results in the literature \cite{Harlander:2019ioe,Czakon:2020vql}. However, we do not confirm previous predictions for the series of subleading logarithms of order $\alpha_b\spac\alpha_s^n\spac L^{2n+1}$ \cite{Akhoury:2001mz,Anastasiou:2020vkr}, which were based on conventional resummation techniques. Section~\ref{sec:concl} contains our conclusions. Several technical details are relegated to a series of appendices. A short letter summarizing our main results has recently appeared in \cite{Liu:2020tzd}. There we have discussed the resummation of large logarithms at next-to-leading logarithmic order. A complete resummation of large logarithms in RG-improved perturbation theory is left for future work.

\section{Factorization formula in terms of bare objects}
\label{sec:barefact}
\renewcommand{\theequation}{2.\arabic{equation}}
\setcounter{equation}{0}
\begin{figure}[t]
\begin{center}
\includegraphics[width=0.448\textwidth]{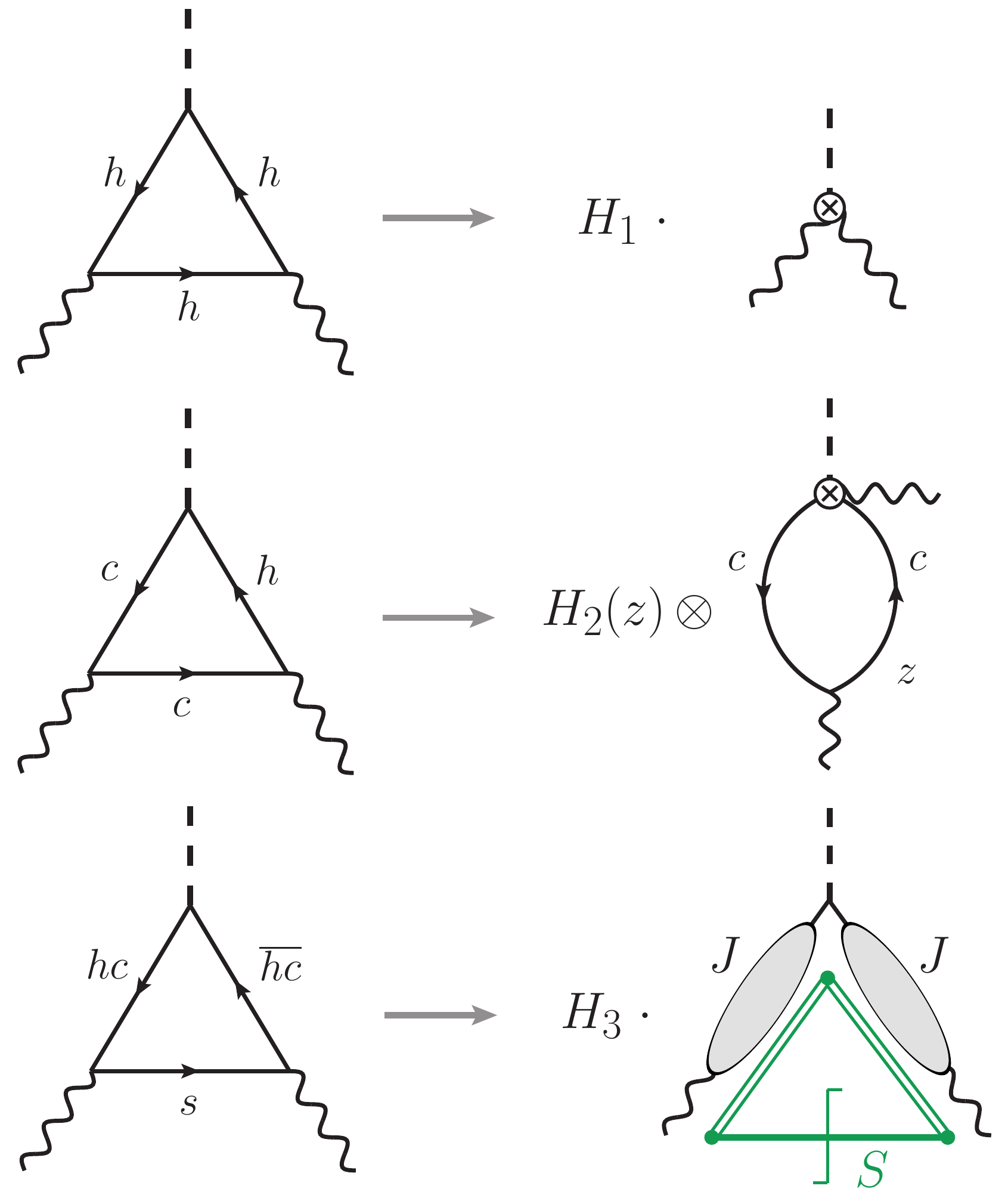} 
\caption{\label{fig:scet1} 
Leading regions of loop momenta ($h$: hard, $c$: $n_1$-collinear, $s$: soft, $hc$: $n_1$-hard-collinear, $\overline{hc}$: $n_2$-hard-collinear) contributing to the decay amplitude. The convolution symbol $\otimes$ in the second term means an integral over $z$.}
\end{center}
\vspace{-3mm}
\end{figure}

As shown in (\ref{Mb}), the $b$-quark induced contribution to the $h\to\gamma\gamma$ decay amplitude receives large logarithms of the form $\alpha_b\spac\alpha_s^{n-1} L^k$ with $k\le 2n$. In order to resum these logarithms in a systematic way it is necessary to factorize the amplitude into objects depending only on one of the three relevant scales, set by the Higgs-boson mass $M_h$, the mass $m_b$ of the bottom quark, and the intermediate scale $\sqrt{M_h\spac m_b}$. In \cite{Liu:2019oav}, two of us have derived a ``bare'' factorization theorem, which accomplishes this. It contains three terms consisting of bare (i.e.\ unrenormalized) SCET operators $O_i^{(0)}$ multiplied by bare Wilson coefficients $H_i^{(0)}$, which account for the hard matching corrections arising when the ``full theory'' (the Standard Model with the top quark integrated out) is matched onto SCET. In its simplest form, the factorization formula reads
\begin{equation}\label{barefact}
   {\cal M}_b(h\to\gamma\gamma)
   = H_1^{(0)} \langle\gamma\gamma|\,O_1^{(0)} |h\rangle 
    + 2 \int_0^1\!dz\,H_2^{(0)}(z)\,\langle\gamma\gamma|\,O_2^{(0)}(z)\,|h\rangle 
    + H_3^{(0)} \langle\gamma\gamma|\,O_3^{(0)} |h\rangle \,.
\end{equation}
The three terms corresponding to different regions of loop momenta giving rise to leading contributions to the decay amplitude ${\cal M}_b$, as illustrated in Figure~\ref{fig:scet1}. The operator 
\begin{equation}\label{O1def}
   O_1^{(0)} = \frac{m_{b,0}}{e_b^2}\,h\,\A_{n_1}^{\perp\mu}\,\A_{n_2,\mu}^\perp
\end{equation}
contains a Higgs field coupled to two collinear gauge fields describing photons moving along opposite light-like directions $n_1$ and $n_2\equiv\bar n_1$. Here and below, fields without an argument are located at the spacetime point $x=0$. The canonical choice of the reference vectors is $n_1^\mu=(1,0,0,1)$ and $n_2^\mu=(1,0,0,-1)$. This operator descents from full-theory graphs in which all internal momenta are hard, of order $M_h$. Next, the operator 
\begin{equation}\label{O2def}
   O_2^{(0)}(z) = h\,\Big[ \bar\X_{n_1}\gamma_\perp^\mu\,\frac{\nbsl_1}{2}\,
    \delta(z\spac\bar n_1\!\cdot k_1+i\bar n_1\!\cdot\partial)\,\X_{n_1} \Big]\,
    \A_{n_2,\mu}^\perp
\end{equation}
contains a Higgs field, an $n_2$-collinear photon field and two $n_1$-collinear $b$-quark fields, which share the momentum $k_1$ of the other photon. This operator is generated by full-theory graphs in which a loop momentum is collinear with the photon direction $n_1$ and carries virtuality of order $m_b$. The factor~2 in front of this contribution in (\ref{barefact}) arises because there is an analogous contribution with $n_1$ and $n_2$ interchanged. The symbols $\A_{n_1}^\mu$ and $\X_{n_1}$ in the above definitions denote effective photon and $b$-quark fields defined in SCET (the so-called ``gauge-invariant building blocks'' \cite{Bauer:2002nz,Hill:2002vw}), which differ from the ordinary quantum fields $A^\mu$ and $\psi$ in that they contain collinear Wilson lines in their definition and that they obey the constraints $\bar n_1\cdot\A_{n_1}=0$ and $\nsl_1\spac\X_{n_1}=0$. Note that the Feynman rule for the vector field $\A^\mu$ contains a factor of $e_b$, which is the reason why we have divided by $e_b^2$ in the definition of $O_1$. The symbol $\perp$ on 4-vectors indicates the components orthogonal to the light-cone basis vectors $n_1$ and $n_2$. Finally, the operator 
\begin{equation}\label{O3def}
   O_3^{(0)} = T\,\Big\{ h\,\bar\X_{n_1} \X_{n_2},
    i\!\int\!\!d^Dx\,{\cal L}_{q\,\xi_{n_1}}^{(1/2)}(x), 
    i\!\int\!\!d^Dy\,{\cal L}_{\xi_{n_2} q}^{(1/2)}(y) \Big\} 
    + \mbox{h.c.} 
\end{equation}
contains the time-ordered product of the scalar Higgs current with two subleading-power terms in the SCET Lagrangian \cite{Beneke:2002ph}, in which hard-collinear fields are coupled to a soft quark field. It arises from full-theory graphs containing a soft quark propagator between the two photons, with all momentum components of order $m_b$. In terms of gauge-invariant building blocks, the relevant subleading-power terms in the SCET Lagrangian read
\begin{equation}\label{Leffsubl}
\begin{aligned}
   {\cal L}_{q\,\xi_{n_1}}^{(1/2)}(x)
   &= \bar q_s(x_-) \big[ \Asl_{n_1}^\perp(x) + \Gsl_{n_1}^\perp(x) \big]\,\X_{n_1}(x) \,, \\
   {\cal L}_{\xi_{n_2} q}^{(1/2)}(y)
   &= \bar\X_{n_2}(y) \big[ \Asl_{n_2}^\perp(y) + \Gsl_{n_2}^\perp(y) \big]\,q_s(y_+) \,,
\end{aligned}
\end{equation}
where $\G_{n_1}^\mu$ is the building block for the hard-collinear gluon field. In interactions of hard-collinear fields with soft fields the soft field operators must be multipole expanded for consistency \cite{Beneke:2002ph,Beneke:2002ni}, and we denote $x_-^\mu=(\bar n_1\cdot x)\,\frac{n_1^\mu}{2}$ and $y_+^\mu=(\bar n_2\cdot y)\,\frac{n_2^\mu}{2}$. 

The $h\to\gamma\gamma$ matrix element of $O_3$ in (\ref{barefact}) can be factorized further into a convolution of two radiative jet functions with a soft function \cite{Liu:2019oav}, i.e.\ (for simplicity we use the default choices of the reference vectors $n_1$ and $n_2$, such that $\bar n_1\cdot k_1=\bar n_2\cdot k_2=M_h$)
\begin{equation}\label{O3ampl}
\begin{aligned}
   \langle\gamma\gamma|\,O_3^{(0)} |h\rangle 
   &= \frac{g_\perp^{\mu\nu}}{2} \int_0^\infty\!\frac{d\ell_+}{\ell_+}\!
    \int_0^\infty\!\frac{d\ell_-}{\ell_-} \\
   &\quad\times \Big[ J^{(0)}(M_h\ell_+)\,J^{(0)}(-M_h\ell_-)
    + J^{(0)}(-M_h\ell_+)\,J^{(0)}(M_h\ell_-) \Big]\,S^{(0)}(\ell_+\ell_-) \,.
\end{aligned}
\end{equation}
Contrary to the 4-vectors $x_-$ and $y_+$ introduced above, the integration variables $\ell_+$ and $\ell_-$ correspond to the light-cone components $n_1\cdot\ell$ and $n_2\cdot\ell$ of a soft 4-momentum $\ell^\mu$. Here and below we omit the photon polarization vectors when presenting expressions for operator matrix elements. The radiative jet function $J(p^2)$ has been studied first in the calculation of the decay amplitude for the rare exclusive decay $B^-\to\gamma\spac l^-\bar\nu_l$ in the context of QCD factorization \cite{Bosch:2003fc}. The soft function $S(w)$ is defined in terms of the discontinuity of a soft quark propagator dressed with soft Wilson lines oriented along the light-like directions $n_1$ and $n_2$. For a more detailed discussion about the derivation of the factorization theorem and the precise definition of the various SCET fields, operators, jet and soft functions the reader is referred to \cite{Liu:2019oav}, where we have shown that the three operators $O_i$ form a basis of ${\cal O}(\lambda^3)$ SCET operators contributing to the decay $h\to\gamma\gamma$, and that the sum of the three terms in (\ref{barefact}) correctly reproduces the decay amplitude at two-loop order. Here $\lambda\sim m_b/M_h$ is the expansion parameter of the effective theory.

Major complications arise from endpoint-divergent convolution integrals in the second and third term in (\ref{barefact}), which need to be properly identified and regularized. The integral over $z$ in the second term contains singularities at $z=0$ and $z=1$. Likewise, the integrals over $\ell_+$ and $\ell_-$ in (\ref{O3ampl}) contain singularities at $\ell_\pm=\infty$. Some of these endpoint divergences are regularized by the dimensional regulator $D=4-2\epsilon$, but others require an additional rapidity regulator \cite{Becher:2010tm,Chiu:2012ir,Li:2016axz}. In \cite{Liu:2019oav} we have regularized the rapidity divergences by means of an analytic regulator imposed on the convolution variables $z$ and $\ell_\pm$. The singular contributions in the rapidity regulator cancel in the sum of the second and third term of the factorization formula, but this requires that for $z\to 0$ (or 1) these two terms must have closely related structures to all orders of perturbation theory. Indeed, we have shown that this condition gives rise to the $D$-dimensional ``refactorization conditions''
\begin{equation}\label{refact}
\begin{aligned}
   \braces{\bar H_2^{(0)}(z)} 
   &= - H_3^{(0)} J^{(0)}(z M_h^2) \,, \\
   \braces{\langle\gamma\gamma|\,O_2^{(0)}(z)\,|h\rangle} 
   &= - \frac{g_\perp^{\mu\nu}}{2} \int_0^\infty\!\frac{d\ell_+}{\ell_+}\,
    J^{(0)}(-M_h\ell_+)\,S^{(0)}(zM_h\ell_+) \,, 
\end{aligned}
\end{equation}
which must hold to all orders of perturbation theory. The symbol $\braces{f(z)}$ means that one retains only the leading terms of a function $f(z)$ in the limit $z\to 0$ and neglects higher power corrections. We have rewritten the original function $H_2^{(0)}(z)$ as
\begin{equation}\label{eq09}
\vspace{-0.6mm}
   H_2^{(0)}(z) = \frac{\bar H_2^{(0)}(z)}{z(1-z)} \,,
\vspace{-0.5mm}
\end{equation}
where the new function $\bar H_2^{(0)}(z)$ contains logarithmic singularities only. With the help of the relations (\ref{refact}) one can rearrange the bare factorization formula (\ref{barefact}) in such a way that all endpoint and rapidity divergences are removed. The result is
\begin{equation}\label{fact4}
\begin{aligned}
   {\cal M}_b
   &= \left( H_1^{(0)} + \Delta H_1^{(0)} \right) \langle\gamma\gamma|\,O_1^{(0)} |h\rangle \\
   &\quad\mbox{}+ 2 \lim_{\delta\to 0}\,\int_\delta^{1-\delta}\!dz\,
    \bigg[ H_2^{(0)}(z)\,\langle\gamma\gamma|\,O_2^{(0)}(z)\,|h\rangle 
    - \frac{\braces{\bar H_2^{(0)}(z)}}{z}\,\braces{\langle\gamma\gamma|\,O_2^{(0)}(z)\,|h\rangle} \\
   &\hspace{3.1cm}\quad\mbox{}- \frac{\braces{\bar H_2^{(0)}(1-z)}}{1-z}\,
    \braces{\langle\gamma\gamma|\,O_2^{(0)}(1-z)\,|h\rangle} \bigg] \\
   &\quad\mbox{}+ g_\perp^{\mu\nu}\,\lim_{\sigma\to-1}\,H_3^{(0)}  
    \int_0^{M_h}\!\frac{d\ell_-}{\ell_-}\,\int_0^{\sigma M_h}\!\frac{d\ell_+}{\ell_+}\,
    J^{(0)}(M_h\ell_-)\,J^{(0)}(-M_h\ell_+)\,S^{(0)}(\ell_+\ell_-)\,\Big|_{\rm leading\;power} \,.
\end{aligned}
\end{equation}
Compared with \cite{Liu:2019oav} we have rewritten the second term in a different but equivalent way. The subtraction terms involving the $\braces{\dots}$ symbol remove the singularities at the endpoints $z=0$ and $z=1$, such that the limit $\delta\to 0$ is smooth. Note that both the matching coefficient $H_2^{(0)}(z)$ and the matrix element $\langle O_2^{(0)}(z)\rangle$ contain terms that are singular for $z=0,1$ and the two subtraction terms properly remove the singularities of the product of these two quantities. This generalizes a simple ``plus-type'' subtraction prescription for the bare operator proposed in \cite{Alte:2018nbn,Beneke:2019kgv}, which works only for cases where the relevant matching coefficient approaches a constant plus power-suppressed terms as $z\to 0$. 

\begin{figure}[t]
\begin{center}
\hspace{1.3cm}\includegraphics[width=0.47\textwidth]{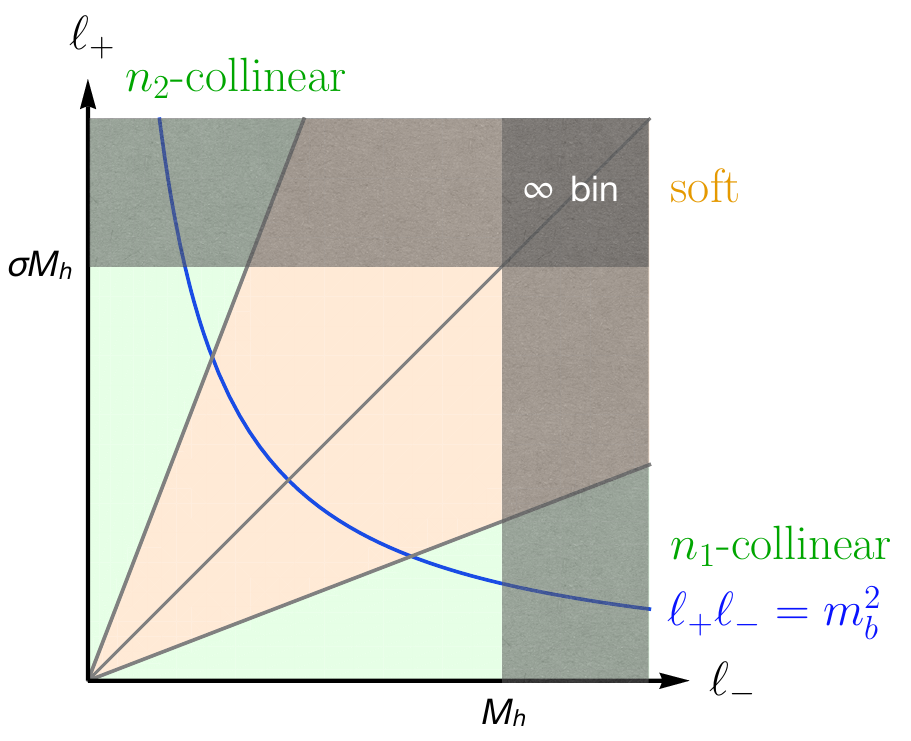} 
\vspace{2mm}
\caption{\label{fig:inftybin} 
Graphical illustration of the subtractions performed in the rearrangement of the factorization formula, which removes the endpoint divergences of the various contributions.}
\end{center}
\end{figure}

We stress that the form of (\ref{fact4}) is independent of the particular rapidity regularization scheme used to regularize the divergent convolution integrals in (\ref{barefact}) and (\ref{O3ampl}), as long as the same regularization scheme is applied consistently to all terms in the factorization theorem. The refactorization conditions ensure that the integrands of the second and third term in the factorization formula are identical in the singular regions (up to power-suppressed terms), and hence the endpoint divergences can be removed by a rearrangement of these terms, which leads to (\ref{fact4}).

Removing the endpoint divergences in the way described above comes at the price of introducing hard upper limits on the integrals over $\ell_+$ and $\ell_-$ in the last term of the factorization formula (\ref{fact4}), which originally have power counting $\ell_\pm={\cal O}(m_b)$.\footnote{It is an open question whether it is possible to formulate an alternative ``endpoint regularization scheme'' avoiding the hard cutoffs. If it exists, such a scheme would need to commute with the operation of renormalization in dimensional regularization, and hence in particular it would need to respect gauge invariance.} 
This gives rise to additional large rapidity logarithms in the matrix element of the operator $O_3$. They are a consequence of the so-called collinear anomaly, which results from the fact that a classical symmetry of the effective theory SCET$_{\rm II}$ under rescalings of the light-cone vectors $n_1$ and~$n_2$ is broken by quantum effects \cite{Becher:2010tm}. The presence of the upper limits also leads to some power-suppressed contributions of ${\cal O}(m_b^2/M_h^2)$ to the third term, which should be dropped for consistency. Moreover, to obtain the correct result for the decay amplitude the upper limit on the (positive) integration variable $\ell_+$ must be analytically continued from $M_h$ to $-M_h-i0$ after the integral has been evaluated, as indicated by the limit $\sigma\to-1$. In Figure~\ref{fig:inftybin} we show a graphical illustration of the rearrangement of the factorization formula that eliminates the endpoint divergences. The subtractions performed on the second term remove the shaded gray regions from the integrals in the third term. Note that in this process the hard region in which $|\ell_\pm|\ge M_h$ is subtracted twice. This over-subtraction needs to be corrected by adding back the ``infinity bin'' in the form of a contribution \begin{equation}
   \Delta H_1^{(0)}
   = - \lim_{\sigma\to-1}\,H_3^{(0)} \int_{M_h}^\infty\!\frac{d\ell_-}{\ell_-}\,
    \int_{\sigma M_h}^\infty\!\frac{d\ell_+}{\ell_+}\,
    J^{(0)}(M_h\ell_-)\,J^{(0)}(-M_h\ell_+)\,\frac{S_\infty^{(0)}(\ell_+\ell_-)}{m_{b,0}}
\end{equation}
to the matching coefficient of the operator $O_1^{(0)}$. Here $S_\infty^{(0)}(w)$ denotes the soft function in the limit where $w/m_{b,0}^2\to\infty$, as given in relation (\ref{Sinfty}) below.

In \cite{Liu:2019oav} we have presented explicit expressions for all quantities appearing in the factorization formula (\ref{fact4}) at next-to-leading order (NLO) in $\alpha_s$, corresponding to two-loop order for the decay amplitude. For completeness, the corresponding expressions are collected in Appendix~\ref{app:bareresults}. The main goal of the present work is to turn the bare factorization theorem into a formula involving renormalized matching coefficients and operator matrix elements. As we shall see this is a highly non-trivial task. The resulting formula provides the basis for a systematic resummation of the large logarithms $\ln(M_h^2/m_b^2)-i\pi$ to all orders of perturbation theory.

\section{SCET derivation of the refactorization conditions}
\label{sec:refact}
\renewcommand{\theequation}{3.\arabic{equation}}
\setcounter{equation}{0}

In this section we derive the refactorization conditions (\ref{refact}) using methods from SCET. This discussion is more technical than that in the remaining sections and draws heavily on SCET jargon as well as notations introduced and results derived in \cite{Liu:2019oav}. Our arguments should be relatively easy to follow for SCET practitioners. Readers not interested in a technical proof of the relations (\ref{refact}) can skip this discussion and proceed directly with Section~\ref{sec:renfact}.

\subsection[Refactorization condition for $\braces{\bar H_2^{(0)}(z)}$]{\boldmath Refactorization condition for $\braces{\bar H_2^{(0)}(z)}$}

The bare matching coefficient $H_2^{(0)}(z)$ has been calculated in \cite{Liu:2019oav} by computing the on-shell $h\to b(\bar z\spac k_1)\,\bar b(z\spac k_1)\,\gamma(k_2)$ amplitude for the decay of a Higgs boson into a pair of $n_1$-collinear bottom quarks and an $n_2$-collinear photon. Here and below we sometime use the symbol $\bar z\equiv 1-z$ for brevity. To simplify the matching calculation one sets the $b$-quark mass to zero and assigns momenta $\bar z k_1$ and $z k_1$ to the outgoing quark and anti-quark, respectively, where $k_1^2=0$. Then the only relevant momentum invariant is hard, $2k_1\cdot k_2=M_h^2$. In the absence of any non-zero low-energy scale the matrix element of the bare operator $O_2^{(0)}$ is given by its tree-level expression. Hence, one finds that 
\begin{equation}\label{H2match}
\begin{aligned}
   {\cal M}^{(0)}(h\to b\spac\bar b\spac\gamma)
   &= \int_0^1\!dz'\,H_2^{(0)}(z')\,
    \langle b(\bar z k_1)\,\bar b(z k_1)\,\gamma(k_2)|\,O_2^{(0)}(z')\,|h\rangle \\
   &= \frac{e_b}{M_h}\,H_2^{(0)}(z)\,\bar u(\bar z k_1)\,\rlap/\varepsilon_\perp^*(k_2)\,
    \frac{\nbsl_1}{2}\,v(z k_1) \,,
\end{aligned}
\end{equation}
where in the last step we have used that $\bar n_1\cdot k_1=M_h$. By calculating the same amplitude in the full theory and comparing the answer with this expression we have derived the result for the bare matching coefficient $H_2^{(0)}(z)$ given in (\ref{Hires}) of Appendix~\ref{app:bareresults}. 

In SCET, the momenta of $n_1$-collinear particles in the basis $(n_1\cdot p,n_2\cdot p,p_\perp)$ have the generic scaling $(\lambda^2,1,\lambda)$ in units of the hard scale $M_h$. In the case at hand the large ${\cal O}(1)$ components of the $b$-quark momenta are equal to $z$ and $\bar z$. Now consider the limit where $z\sim\lambda\ll 1$. Then the scaling of the outgoing anti-quark momentum becomes soft, $(\lambda^2,\lambda,\lambda)$. The process is now characterized by two different scales: the hard scale $M_h^2$ and the hard-collinear scale $z M_h^2\sim M_h\spac m_b$. In an intermediate effective theory called SCET$_{\rm I}$, the leading-order contribution to the decay amplitude in this limit can be written in the form
\begin{equation}
   \braces{{\cal M}^{(0)}(h\to b\spac\bar b\spac\gamma)}
   = H_3^{(0)}\,\langle b(k_1)\,\bar b(z k_1)\,\gamma(k_2)|\,
    T\,\Big\{ h\,\bar\X_{n_1} \X_{n_2}(0),
     i\!\int\!d^Dy\,{\cal L}_{\xi_{n_2} q}^{(1/2)}(y) \Big\}\,|h\rangle \,.
\end{equation}
The first operator in the time-ordered product describes the decay of the Higgs boson into an $n_1$-hard-collinear quark and an $n_2$-hard-collinear anti-quark. Hard matching corrections to this vertex are accounted for by the coefficient $H_3^{(0)}$. The second operator is an insertion of the subleading-power SCET Lagrangian given in (\ref{Leffsubl}), which couples a soft quark to an $n_2$-hard-collinear quark. There are no hard matching corrections to this Lagrangian \cite{Beneke:2002ph}. 

\begin{figure}[t]
\begin{center}
\includegraphics[width=0.57\textwidth]{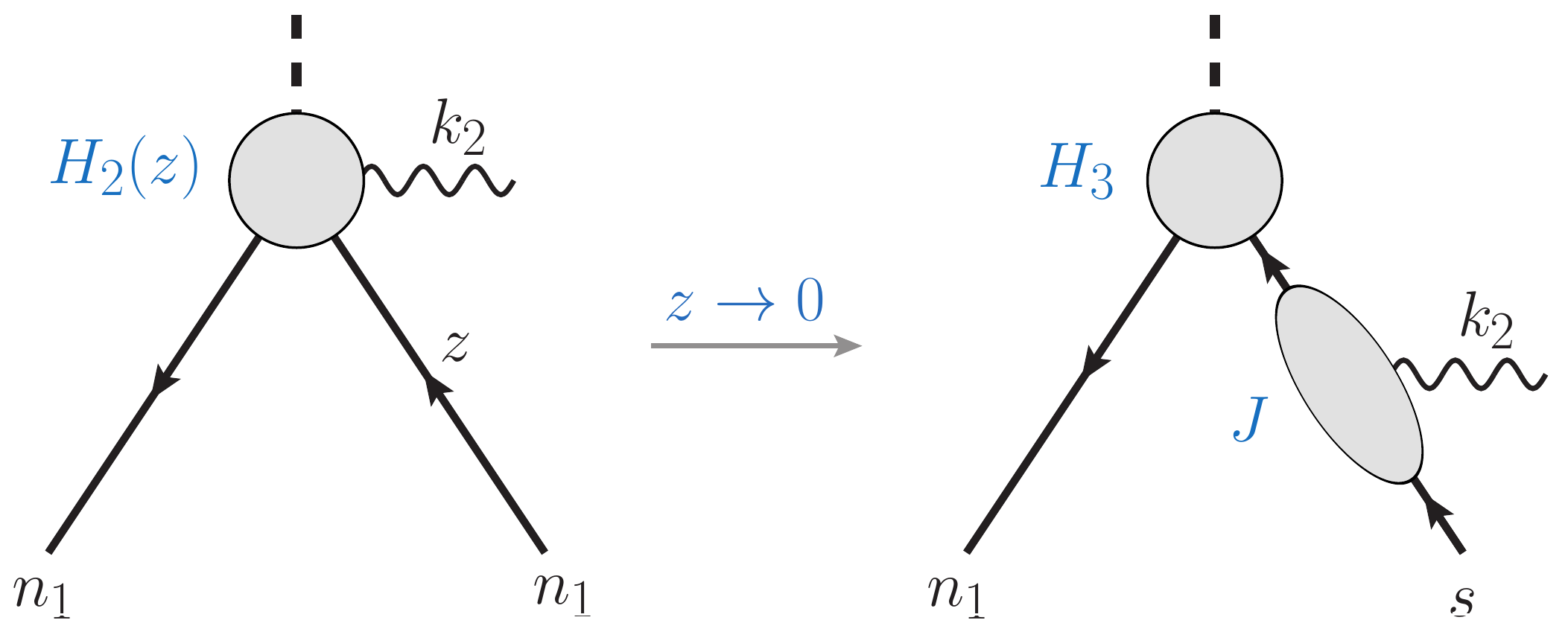} 
\vspace{2mm}
\caption{\label{fig:refact1} 
Graphical illustration of the refactorization condition for $\braces{\bar H_2^{(0)}(z)}$.}
\end{center}
\end{figure}

We now decouple soft gluons from the hard-collinear fields by performing the usual field redefinitions \cite{Bauer:2001yt}, but we do not change the names of the fields for simplicity. This leads to
\begin{equation}\label{M0eq11}
\begin{aligned}
   \braces{{\cal M}^{(0)}(h\to b\spac\bar b\spac\gamma)}
   &= H_3^{(0)}\,\langle b(k_1)\,\bar b(z k_1)\,\gamma(k_2)|\,
    T\,\Big\{ \bar\X_{n_1}(0)\spac S_{n_1}^\dagger(0)\,S_{n_2}(0)\,\X_{n_2}(0), \\
   &\hspace{7mm} i\!\int\!d^Dy\,\bar\X_{n_2}(y) 
    \left( \Asl_{n_2}^\perp(y) + \Gsl_{n_2}^\perp(y) \right) 
    S_{n_2}^\dagger(y_+)\,q_s(y_+) \Big\}\,|0\rangle \,.
\end{aligned}
\end{equation}
In this matrix element the different types of fields ($n_1$-hard-collinear, $n_2$-hard-collinear and soft) no longer interact with one another. We now match this result onto the low-energy effective theory called SCET$_{\rm II}$ by integrating out the hard-collinear fields. In this step we use the definition of the bare jet function $J^{(0)}(p^2)$ given in Appendix~\ref{app:jetsoft} to obtain (using that $n_2=\bar n_1$ and $n_1=\bar n_2$)
\begin{equation}\label{resujet1}
\begin{aligned}
   \braces{{\cal M}^{(0)}(h\to b\spac\bar b\spac\gamma)}
   &= e_b\,H_3^{(0)}\,
    \bar u_{n_1}(k_1)\,\frac{\nbsl_1}{2}\,\rlap/\varepsilon_\perp^*(k_2)\,v_s(z k_1)\,
    \frac{\bar n_2\cdot(k_2+z k_{1-})}{(k_2+z k_{1-})^2+i0}\,J^{(0)}\big( (k_2+z k_{1-})^2 \big) \\
   &= - \frac{e_b}{M_h}\,\frac{H_3^{(0)}}{z+i0}\,J^{(0)}(z M_h^2)\,
    \bar u(\bar z k_1)\,\rlap/\varepsilon_\perp^*(k_2)\,\frac{\nbsl_1}{2}\,v(z k_1) \,,
\end{aligned}
\end{equation}
where $k_{1-}=\bar n_1\cdot k_1\,\frac{n_1^\mu}{2}$, and hence $(k_2+z k_{1-})^2=z\spac\bar n_1\cdot k_1\,\bar n_2\cdot k_2=z M_h^2$. A graphical illustration of this result is shown in Figure~\ref{fig:refact1}. Note the important fact that, since we perform the calculation on-shell and for massless quarks, there is no soft or $n_1$-collinear scale in the problem, and hence the soft and $n_1$-collinear matrix elements are equal to their tree-level expressions. In particular, the soft Wilson lines do not give rise to any non-trivial contributions, and the soft matrix element simply provides a factor $e^{i zk_1\cdot y_+}\,v(z k_1)$. Matching this result with (\ref{H2match}) we obtain 
\begin{equation}
   \braces{H_2^{(0)}(z)} = \frac{\braces{\bar H_2^{(0)}(z)}}{z}
   = - \frac{H_3^{(0)}}{z}\,J(z M_h^2) \,,
\end{equation}
where we have used (\ref{eq09}) in the first step. This establishes the first relation in (\ref{refact}).

\subsection[Refactorization condition for $\braces{\langle O_2^{(0)}(z)\rangle}$]{\boldmath Refactorization condition for $\braces{\langle O_2^{(0)}(z)\rangle}$}

\begin{figure}[t]
\begin{center}
\includegraphics[width=0.48\textwidth]{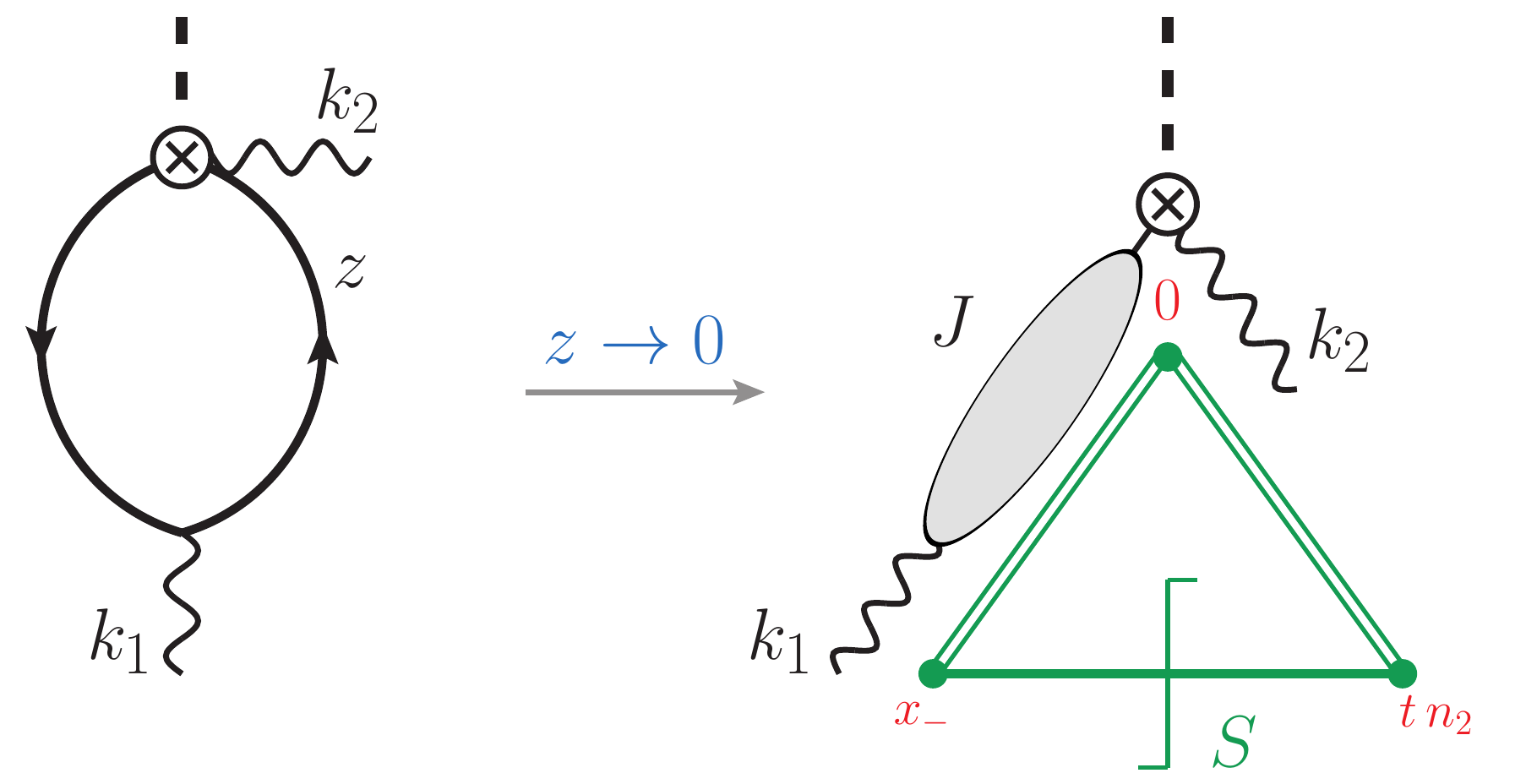} 
\vspace{2mm}
\caption{\label{fig:refact2} 
Graphical illustration of the refactorization condition for $\braces{\langle O_2^{(0)}(z)\rangle}$.}
\end{center}
\end{figure}

The derivation of the second refactorization condition is slightly more involved. In this case the soft contributions are non-zero, because the matrix element of the operator $O_2^{(0)}(z)$ depends on the $b$-quark mass. In fact, for $z\ll 1$ the two relevant scales are the soft scale $m_b^2$ and the hard-collinear scale $m_b^2/z\sim M_h\space m_b$. Our starting point is the position-space representation of the operator $O_2^{(0)}$ introduced in \cite{Liu:2019oav}, which reads
\begin{equation}
   O_2^{(0)}(t) = h(0)\,\bar\X_{n_1}(0)\,\gamma_\perp^\mu\,\frac{\nbsl_1}{2}\,\X_{n_1}(t\bar n_1)\,
    \A_{n_2,\mu}^\perp(0) \,.
\end{equation}
We now replace the $n_1$-collinear field $\X_{n_1}$ by a soft quark field $q_s$ and perform the soft decoupling transformation. This leads to (using that $\bar n_1=n_2$)
\begin{equation}\label{eq3.6}
   O_2^{(0)}(t)\to h(0)\,\bar\X_{n_1}(0)\,S_{n_1}^\dagger(0)\,S_{n_2}(0)\,
    \gamma_\perp^\mu\,\frac{\nsl_2}{2}\,S_{n_2}^\dagger(t n_2)\,q_s(t n_2)\,
    \A_{n_2,\mu}^\perp(0) \,.
\end{equation} 
The structure of the soft Wilson lines follows from the fact that the operator on the right-hand side derives from the amplitude in (\ref{M0eq11}) after integrating out the $n_2$-hard-collinear fields. We now need to evaluate the on-shell $h\to\gamma\gamma$ matrix element of this operator. To this end, we need an insertion of the subleading-power SCET Lagrangian, which turns the soft quark field back into an $n_1$-hard-collinear quark field. We thus obtain
\begin{equation}
\begin{aligned}
   \braces{\langle\gamma\gamma|\,O_2^{(0)}(t)\,|h\rangle}
   &= e_b\spac\langle\gamma(k_1)|\,T\,\Big\{ 
    \bar\X_{n_1}(0)\,S_{n_1}^\dagger(0)\,S_{n_2}(0)\,
    \rlap/\varepsilon_\perp^*(k_2)\,\frac{\nsl_2}{2}\,S_{n_2}^\dagger(t n_2)\,q_s(t n_2), \\
   &\hspace{2.9cm} i\!\int\!d^Dx\,\bar q_s(x_-)\,S_{n_1}(x_-)\,
    \big( \Asl_{n_1}^\perp(x) + \Gsl_{n_1}^\perp(x) \big)\,\X_{n_1}(x) \Big\}\,|0\rangle \,.
\end{aligned}
\end{equation}
This relation should be understood as a matching equation relating the matrix elements of the operators on the two sides of the equation. In the next step we use the definitions of the jet function and of the soft-quark soft function collected in Appendix~\ref{app:jetsoft}. Taking into account a minus sign from an odd number of interchanges of fermion fields, we find 
\begin{equation}\label{eq19}
\begin{aligned}
   \braces{\langle\gamma\gamma|\,O_2^{(0)}(t)\,|h\rangle}
   &= i\pi\,\text{tr}\bigg[ \rlap/\varepsilon_\perp^*(k_2)\,\frac{\nsl_2}{2}\,
    \rlap/\varepsilon_\perp^*(k_1)\,\frac{\nsl_1}{2} \bigg] 
    \int\!d^Dx \int\!\frac{d^D\ell}{(2\pi)^D} \int\!\frac{d^Dp}{(2\pi)^D}\,
    e^{-i\ell\cdot(t n_2-x_-)}\,{\cal S}_1^{(0)}(\ell) \\
   &\quad\times \frac{\bar n_1\cdot p}{p^2+i0}\,J^{(0)}(p^2)\,e^{-i(p-k_1)\cdot x} \\
   &= - \frac{i}{4\pi}\,\varepsilon_\perp^*(k_1)\cdot\varepsilon_\perp^*(k_2)
    \int_{-\infty}^\infty\!d\ell_-\,e^{-it\ell_-} \int_{-\infty}^\infty\!d\ell_+\,
    \frac{J^{(0)}(M_h\ell_+)}{\ell_+ +i0}\,{\cal S}^{(0)}(\ell_+\ell_-) \,,
\end{aligned}
\end{equation}
where $\ell_+=n_1\cdot\ell$ and $\ell_-=n_2\cdot\ell$, and the function ${\cal S}^{(0)}(\ell_+\ell_-)$ is defined as
\begin{equation}
   {\cal S}^{(0)}(\ell_+\ell_-) 
   = \int\frac{d^{D-2}\ell_\perp}{(2\pi)^{D-2}}\,\spac {\cal S}_1^{(0)}(\ell) \,. 
\end{equation}
Figure~\ref{fig:refact2} shows a graphical representation of the result (\ref{eq19}).

\begin{figure}[t]
\begin{center}
\includegraphics[width=0.55\textwidth]{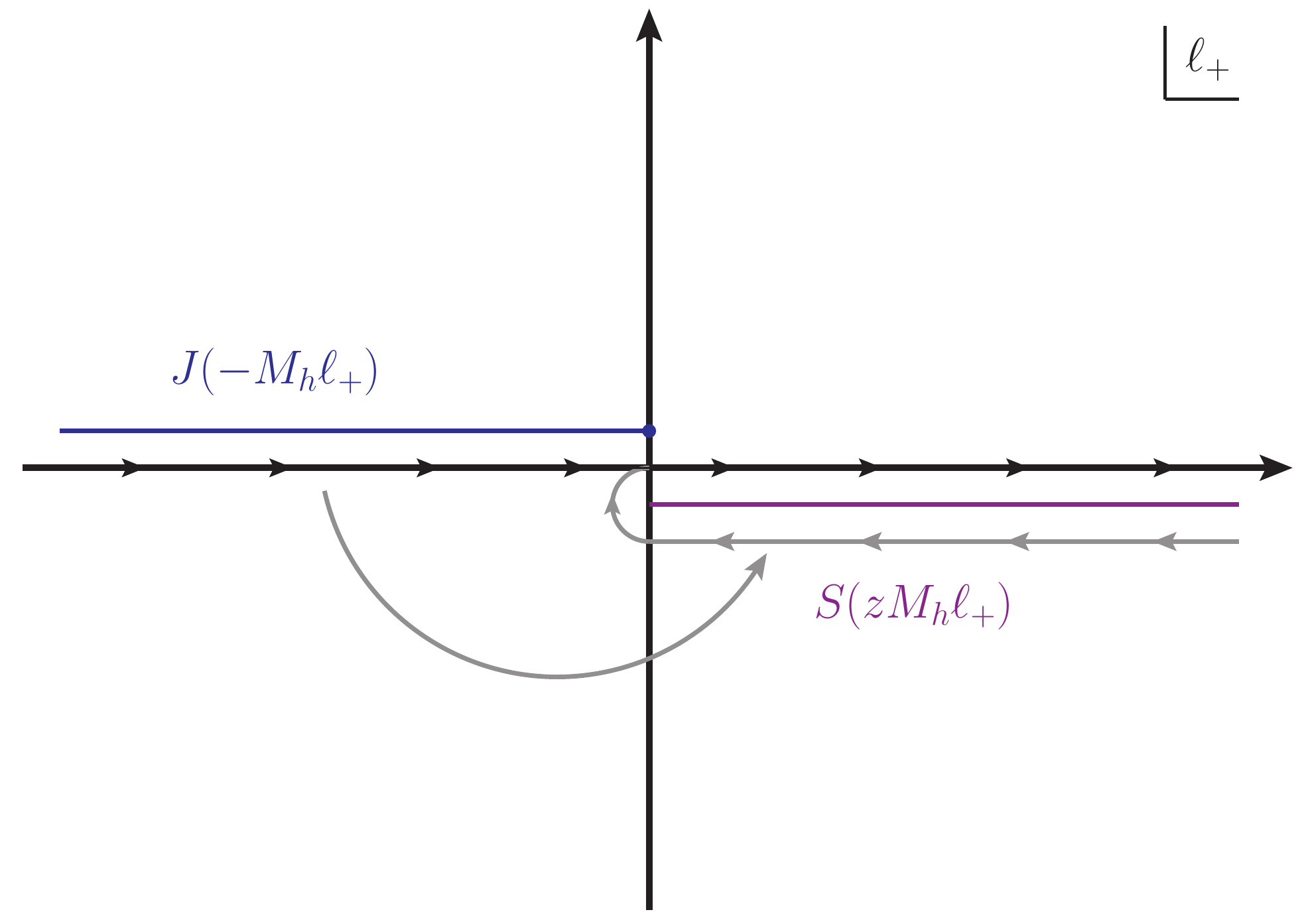} 
\vspace{2mm}
\caption{\label{fig:complex} 
Singularities of the integral (\ref{eq19}) in the complex $\ell_+$ plane.}
\end{center}
\end{figure}

We now switch back to momentum space and define
\begin{equation}
   \braces{\langle\gamma\gamma|\,O_2^{(0)}(z)\,|h\rangle}
   = \int\!\frac{dt}{2\pi}\,e^{-iz M_h t}\,
    \braces{\langle\gamma\gamma|\,O_2^{(0)}(t)\,|h\rangle} \,.
\end{equation}
We then find 
\begin{equation}
   \braces{\langle\gamma\gamma|\,O_2^{(0)}(z)\,|h\rangle}
   = \frac{i}{4\pi}\,\varepsilon_\perp^*(k_1)\cdot\varepsilon_\perp^*(k_2)
    \int_{-\infty}^\infty\!d\ell_+\,\frac{J^{(0)}(-M_h\ell_+)}{\ell_+ -i0}\,
    {\cal S}^{(0)}(zM_h\ell_+) \,,
\end{equation}
where we have relabeled the integration variable $\ell_+\to-\ell_+$ for later convenience. The singularities of the integrand in the complex $\ell_+$ plane are shown in Figure~\ref{fig:complex}. There is a pole at $\ell_+=i0$ from the denominator, a cut infinitesimally above the real axis for negative values of $\ell_+$ from the jet function, and a cut infinitesimally below the real axis for positive values of $\ell_+$ from the soft function. The integral is thus non-zero only if $\ell_+\ge 0$ and we can deform the contour such that it wraps around the cut in the lower half-plane. This leads to
\begin{equation}
   \braces{\langle\gamma\gamma|\,O_2^{(0)}(z)\,|h\rangle}
   = - \frac12\,\varepsilon_\perp^*(k_1)\cdot\varepsilon_\perp^*(k_2)
    \int_0^\infty\!\frac{d\ell_+}{\ell_+}\,J^{(0)}(-M_h\ell_+)\,S^{(0)}(z M_h\ell_+) \,,
\end{equation}
where
\begin{equation}
   S^{(0)}(\ell_+\ell_-) = \frac{1}{2\pi i}\,
    \Big[ {\cal S}^{(0)}(\ell_+\ell_- +i0) - {\cal S}^{(0)}(\ell_+\ell_- -i0) \Big]
\end{equation}
is the discontinuity of the soft function ${\cal S}^{(0)}(\ell_+\ell_-)$. This proves the second refactorization condition in (\ref{refact}).

\section{Renormalized factorization formula}
\label{sec:renfact}
\renewcommand{\theequation}{4.\arabic{equation}}
\setcounter{equation}{0}

The main goal of this work is to establish the renormalized factorization formula 
\begin{equation}\label{renfact}
\begin{aligned}
   {\cal M}_b
   &= H_1(\mu)\,\langle O_1(\mu)\rangle \\
   &\quad\mbox{}+ 2 \int_0^1\!dz\,\bigg[ H_2(z,\mu)\,\langle O_2(z,\mu)\rangle 
    - \frac{\braces{\bar H_2(z,\mu)}}{z}\,\braces{\langle O_2(z,\mu)\rangle} 
    - \frac{\braces{\bar H_2(\bar z,\mu)}}{\bar z}\,\braces{\langle O_2(\bar z,\mu)\rangle} \bigg] \\
   &\quad\mbox{}+ g_\perp^{\mu\nu}\,\lim_{\sigma\to-1}\,H_3(\mu) 
    \int_0^{M_h}\!\frac{d\ell_-}{\ell_-}\,\int_0^{\sigma M_h}\!\frac{d\ell_+}{\ell_+}\,
    J(M_h\ell_-,\mu)\,J(-M_h\ell_+,\mu)\,S(\ell_+\ell_-,\mu)\,\Big|_{\rm leading\;power} \,,
\end{aligned}
\end{equation}
which is structurally equivalent to the bare formula (\ref{fact4}). We have omitted the external states in the matrix elements for brevity. It is not at all evident that such a formula exists, because, as we will show, the presence of cutoffs on some of the integrals does not commute with the operation of renormalization. We will show that this complication gives rise to some additional contributions, which to all orders of perturbation theory can be absorbed into the definition of the matching coefficient $H_1(\mu)$.

\subsection{Parameter renormalization}

In a first step, we relate the bare parameters entering the decay amplitude ${\cal M}_b$ to the corresponding renormalized parameters. These are the mass of the $b$ quark (which enters in the matrix elements of the operators $O_i$), its Yukawa coupling (which enters in the expressions for the matching coefficients $H_i$), as well as the gauge couplings of QCD and QED. We write the relevant renormalization conditions in the $\overline{\rm MS}$ subtraction scheme as
\begin{equation}
\begin{aligned}
   y_{b,0} &= \mu^\epsilon\spac Z_y\,y_b(\mu) \,, ~~\quad 
    & m_{b,0} &= Z_m\,m_b(\mu) \,, \\
   \alpha_0 &= \mu^{2\epsilon}\spac Z_\alpha\,\alpha \,,
    & \alpha_{s,0} &= \mu^{2\epsilon}\spac Z_{\alpha_s}\spac\alpha_s(\mu) \,.
\end{aligned}
\end{equation}
The factor of $\mu^\epsilon$ from the renormalization of the Yukawa coupling multiplies the entire decay amplitude. It can be ignored, since after parameter renormalization the amplitude is finite and the limit $\epsilon\to 0$ is smooth. In our analysis we consider QCD radiative corrections only. To first order in $\alpha_s\equiv\alpha_s(\mu)$, we then have
\begin{equation}\label{Zfactors}
   Z_y = Z_m = 1 - 3 C_F\,\frac{\alpha_s}{4\pi\epsilon} + {\cal O}(\alpha_s^2) \,, \qquad
   Z_{\alpha_s} = 1 - \beta_0\,\frac{\alpha_s}{4\pi\epsilon} + {\cal O}(\alpha_s^2) \,,
\end{equation}
and $Z_\alpha=1$. Here $\beta_0=\frac{11}{3}\,C_A-\frac43\,T_F\spac n_f$ is the first coefficient of the QCD $\beta$-function, with $n_f=5$ being the number of active quark flavors. 

In our analysis we will sometimes use the $b$-quark pole mass $m_b$ instead of the running mass $m_b(\mu)$. At NNLO the relation between the two quantities is given by \cite{Tarrach:1980up}
\begin{equation}\label{polemass}
\begin{aligned}
   m_b(\mu) 
   &= m_b\,\bigg\{ 1 + \frac{C_F\alpha_s}{4\pi} \left( 3 L_m - 4 \right) \\
   &\hspace{1.3cm}\mbox{}+ C_F \left( \frac{\alpha_s}{4\pi} \right)^2\! \left[ 
    \left( \frac92\,C_F - \frac32\,\beta_0 \right) L_m^2
    + \left( - \frac{21}{2}\,C_F + \frac{185}{6}\,C_A - \frac{26}{3}\,T_F\spac n_f \right) L_m 
    + \dots \right] \\
   &\hspace{1.3cm}\mbox{}+ {\cal O}(\alpha_s^3) \bigg\} \,,
\end{aligned}
\end{equation}
where $L_m=\ln(m_b^2/\mu^2)$, and for the purposes of this work we do not need the scale-independent two-loop contribution denoted by the dots. This relation, as well as the relations in (\ref{Zfactors}), are known to very high orders of perturbation theory.

\subsection{Operator renormalization}
\label{subsec:4.2}

The matrix elements of the bare operators $O_{1,2}^{(0)}$ as well as the bare jet and soft functions $J^{(0)}$ and $S^{(0)}$ contain ultraviolet (UV) divergences not eliminated by the renormalization of the bare parameters. These divergences must be removed by renormalizing the operators themselves, allowing for the possibility of operator mixing. In recent work we have studied the renormalization properties of the jet function \cite{Liu:2020ydl} and the soft function \cite{Liu:2020eqe}. We now discuss the renormalization of the remaining operators at first order in $\alpha_s$. We find that the operators $O_1$ and $O_2$ mix under renormalization, in such a way that 
\begin{equation}\label{O2ren}
\begin{aligned}
   O_1(\mu) &= Z_{11}\,O_1^{(0)} \,, \\
   O_2(z,\mu) &= \int_0^1\!dz'\,Z_{22}(z,z')\,O_2^{(0)}(z') + Z_{21}(z)\,O_1^{(0)} \,. 
\end{aligned}
\end{equation}
From the definition of the operator $O_1$ in (\ref{O1def}) it follows that
\begin{equation}\label{Z11Zm}
   Z_{11} = Z_m^{-1} 
   = 1 + \frac{3}{\epsilon}\,\frac{C_F\alpha_s}{4\pi} + {\cal O}(\alpha_s^2) \,.
\end{equation}
The relation $Z_{11}=Z_m^{-1}$ holds to all orders of QCD perturbation theory, since the quantum fields in the definition of $O_1$ do not carry color. This factor is known to very high orders of perturbation theory.

\begin{figure}[t]
\begin{center}
\includegraphics[width=0.65\textwidth]{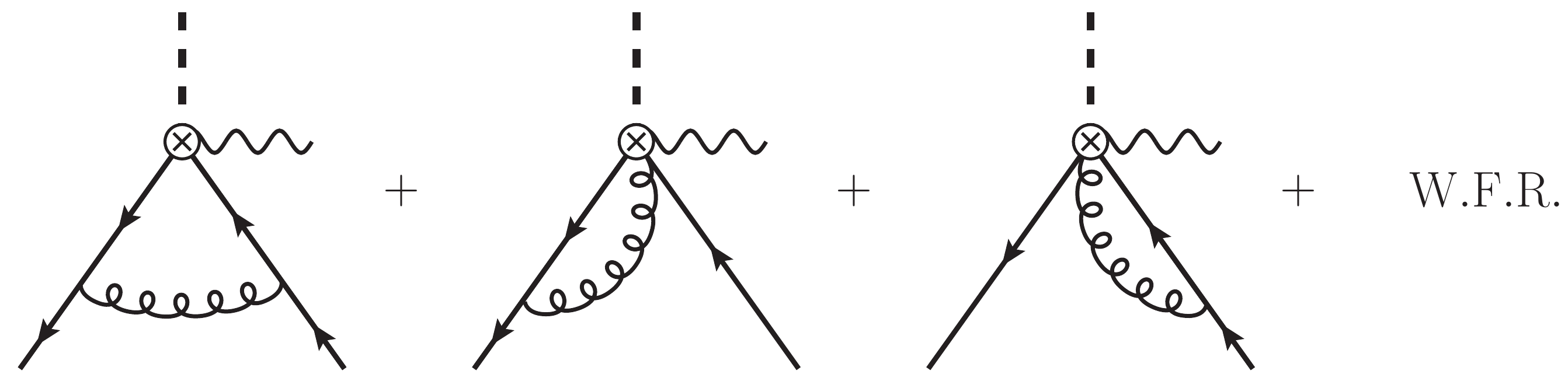} 
\vspace{2mm}
\caption{\label{fig:Z22graphs} 
One-loop diagrams contributing to the calculation of the diagonal renormalization factor $Z_{22}$. In the second and third graph the gluon is emitted from the Wilson lines contained in the definition of the collinear quark fields. These graphs must be supplemented by wave-function renormalization.}
\end{center}
\end{figure}

The diagonal renormalization factor $Z_{22}$ can be derived by studying the UV divergences of the matrix element of the operator $O_2(z)$ defined in (\ref{O2def}) between an initial-state Higgs boson and a final state consisting of a photon with momentum $k_2$ and a pair of $n_1$-collinear bottom quarks sharing the total momentum $k_1$. The relevant one-loop diagrams are shown in Figure~\ref{fig:Z22graphs}. From a straightforward calculation, we obtain at NLO in QCD
\begin{equation}\label{Z22}
\begin{aligned}
   Z_{22}(z,z') &= \left( 1 + \frac{C_F\alpha_s}{4\pi\epsilon} \right) \delta(z-z') \\
   &\quad\mbox{}- \frac{C_F\alpha_s}{2\pi\epsilon}\,\frac{1}{z'(1-z')} \left[ 
    z(1-z')\,\frac{\theta(z'-z)}{z'-z} + z'(1-z)\,\frac{\theta(z-z')}{z-z'} \right]_+
    + {\cal O}(\alpha_s^2) \,,
\end{aligned}
\end{equation}
where $z,z'\in[0,1]$, and the plus distribution is defined in the usual way. The above result is the well-known Brodsky-Lepage kernel \cite{Lepage:1979zb,Chernyak:1983ej}. This is not surprising, because the colored fields in the operator $O_2$ in (\ref{O2def}) have the same structure as the quark fields entering the definition of the leading-twist light-cone distribution amplitude of a transversely polarized vector meson. Before Fourier transformation, they form the non-local structure \cite{Liu:2019oav}
\begin{equation}
   \bar\X_{n_1}(0)\,\gamma_\perp^\mu\,\frac{\nbsl_1}{2}\,\X_{n_1}(t\bar n_1)
   = \bar\psi(0)\,\gamma_\perp^\mu\,\frac{\nbsl_1}{2}\,[0,t\bar n_1]\,\psi(t\bar n_1) \,,
\end{equation}
where $[0,t\bar n_1]$ denotes a finite-length Wilson line along the light-like direction $\bar n_1$, and $\psi(x)$ is the conventional quark field in QCD. The two-loop expression for $Z_{22}$ can in principle be derived from results available in the literature \cite{Dittes:1983dy,Mikhailov:1984ii,Brodsky:1984xk,Mueller:1993hg,Mueller:1994cn}, but it is not needed for our purposes in this work. It will be useful to rewrite the above result in the equivalent form
\begin{equation}\label{thisexpression}
\begin{aligned}
   Z_{22}(z,z') &= \left[ 1 - \frac{C_F\alpha_s}{4\pi\epsilon}\, 
    \Big( 2\ln z + 2\ln(1-z) + 3 \Big) \right] \delta(z-z') \\
   &\quad\mbox{}- \frac{C_F\alpha_s}{2\pi\epsilon}\,z(1-z) \left[ 
    \frac{1}{z'(1-z)}\,\frac{\theta(z'-z)}{z'-z} 
    + \frac{1}{z(1-z')}\,\frac{\theta(z-z')}{z-z'} \right]_+ + {\cal O}(\alpha_s^2) \,.
\end{aligned}
\end{equation}
The operator $O_2$ is not only renormalized multiplicatively, but it mixes with $O_1$ under renormalization. The factor $Z_{21}(z)$ can be derived from the condition that the $h\to\gamma\gamma$ matrix element of the renormalized operator $O_2(\mu)$ in (\ref{O2ren}) be free of UV divergences. Starting from the expression for the matrix element of the bare operator $O_2^{(0)}(z)$ presented in \cite{Liu:2019oav} and shown explicitly in (\ref{O2mel}), applying the factor $Z_{22}$ to this expression and renormalizing the bare parameters $\alpha_{b,0}$, $\alpha_{s,0}$ and $m_{b,0}$, we find that some $1/\epsilon$ poles remain, which must be removed by the counterterm $Z_{21}\,\langle O_1^{(0)}\rangle$. In this way we obtain
\begin{equation}\label{thatexpression}
\begin{aligned}
   Z_{21}(z) 
   &= \frac{N_c\spac\alpha_b}{2\pi}\,\bigg\{ - \frac{1}{\epsilon}
    + \frac{C_F\alpha_s}{4\pi}\,\bigg[ \frac{\ln z+\ln(1-z)}{\epsilon^2} \\
   &\hspace{1.9cm}\mbox{}- \frac{1}{\epsilon} \left( \frac{\ln^2 z+\ln^2(1-z)}{2} 
    - 2\ln z\ln(1-z) + \frac{11}{2} - \frac{\pi^2}{3} \right) \!\bigg] 
    + {\cal O}(\alpha_s^2) \bigg\} \,.
\end{aligned}
\end{equation}

Given these results, the renormalized $h\to\gamma\gamma$ matrix element $\braces{\langle O_2(z,\mu)\rangle}$ can be obtained in two equivalent ways: either by taking the limit $z\to 0$ in the expression for $\langle O_2(z,\mu)\rangle$, or by starting from the bare matrix elements $\braces{\langle O_2^{(0)}(z)\rangle}$ and $\langle O_1^{(0)}(z)\rangle$ and applying appropriate limiting expressions for the renormalization factors $Z_{22}$ and $Z_{21}$. We find
\begin{equation}\label{eq4.5}
   \braces{\langle O_2(z,\mu)\rangle } 
   = \int_0^\infty\!dz'\,\braces{Z_{22}(z,z')}\,\braces{\langle O_2^{(0)}(z')\rangle}
    + \braces{Z_{21}(z)}\,\langle O_1^{(0)}\rangle \,. 
\end{equation}
The renormalization factor $\braces{Z_{22}(z,z')}$ can be obtained from (\ref{thisexpression}) by taking the limit $z,z'\to 0$ (with $z\sim z'$), leading to
\begin{equation}\label{bracesZ22}
   \braces{Z_{22}(z,z')} = \left[ 1 - \frac{C_F\alpha_s}{4\pi\epsilon}\, 
    \Big( 2\ln z + 3 \Big) \right] \delta(z-z') - \frac{C_F\alpha_s}{2\pi\epsilon}\,z 
    \left[ \frac{\theta(z'-z)}{z'(z'-z)} + \frac{\theta(z-z')}{z(z-z')} \right]_+ 
    + {\cal O}(\alpha_s^2) \,.
\end{equation}
Likewise, taking the limit $z\to 0$ in (\ref{thatexpression}), we derive 
\begin{equation}\label{bracesZ21}
   \braces{Z_{21}(z)} 
   = \frac{N_c\spac\alpha_b}{2\pi}\,\bigg\{ - \frac{1}{\epsilon} + \frac{C_F\alpha_s}{4\pi}\, 
    \bigg[ \frac{\ln z}{\epsilon^2} - \frac{1}{\epsilon} \left( \frac{\ln^2 z}{2} 
    + \frac{11}{2} - \frac{\pi^2}{3} \right) \!\bigg] + {\cal O}(\alpha_s^2) \bigg\} \,.
\end{equation}
Notice the important fact that, contrary to (\ref{O2ren}), the convolution integral now runs over the interval $z'\in[0,\infty)$. The reason is that $z$ and $z'$ are treated as variables which take infinitesimally small values.

According to (\ref{O3def}) the operator $O_3$ is defined in terms of a time-ordered product of the scalar current $J_S=h\,\bar\X_{n_1}\X_{n_2}$ with two subleading-power Lagrangian insertions. In this product only the current gets renormalized. The corresponding renormalization factor is known to three-loop order and given in eq.~(A.2) of \cite{Becher:2009qa}. At first order in $\alpha_s$ it reads
\begin{equation}\label{Z33}
   Z_{33} = Z_{J_S}
   = 1 + \frac{C_F\alpha_s}{4\pi} \left[ - \frac{2}{\epsilon^2}
    + \frac{2}{\epsilon} \left( L_h - \frac32 \right) \right] + {\cal O}(\alpha_s^2) \,,
\end{equation}
where $L_h=\ln(-M_h^2/\mu^2)$. Here and below we use the implicit definition $-M_h^2\equiv-M_h^2-i0$ to fix the sign of the imaginary part of the logarithm. When the $h\to\gamma\gamma$ matrix element of $O_3$ is expressed in terms of a convolution over jet and soft functions, as shown in (\ref{O3ampl}), relation (\ref{Z33}) implies a relation between $Z_{33}$ and the renormalization factors $Z_J$ and $Z_S$ of the jet function and the soft function, which has been given in \cite{Liu:2020eqe}. It can be written as
\begin{equation}\label{simplerelaZ}
   Z_S(w,w') 
   = \frac{w}{w'}\,Z_{33} \int_0^\infty\!\frac{dx}{x}\,
    Z_J^{-1}\Big(\frac{M_h w'}{x\ell_+},\frac{M_h w}{\ell_+}\Big)\,
    Z_J^{-1}(-x M_h\ell_+,-M_h\ell_+) \,,
\end{equation}
which despite appearance is independent of the choice of $\ell_+$. This result implies a simple relation between the anomalous dimensions of the matching coefficient $H_3$ and of the jet and soft functions, which was conjectured in \cite{Liu:2020eqe} and will be given in relation (\ref{simplerela}) in Section~\ref{subsec:5.1}.

In general, it is well known that time-ordered products such as in (\ref{O3def}) can mix under renormalization with local operators of the same order in power counting (see e.g.\ \cite{Beneke:2018rbh} for a discussion in the context of SCET). In our case, however, the presence of the UV cutoffs on the integrals over $\ell_+$ and $\ell_-$ in (\ref{fact4}) and (\ref{renfact}) prevents such a mixing. The renormalized matrix element of $O_3$, which depends on several different physical scales, is expressed in (\ref{renfact}) as a double convolution over renormalized jet and soft functions. The calculation of the renormalized jet function $J(p^2)$ at two-loop order and the study of its RG evolution equation have been discussed in \cite{Liu:2020ydl}, while the renormalization of the soft function $S(w)$ at one-loop order and the derivation of its two-loop RG equation have been studied in \cite{Liu:2020eqe}. When the renormalized expressions are used, we find that (omitting the limit $\sigma\to -1$ for brevity)
\begin{equation}\label{eq22}
\begin{aligned}
   &\phantom{=} g_\perp^{\mu\nu}\,H_3(\mu) \int_0^{M_h}\!\frac{d\ell_-}{\ell_-}\,
    \int_0^{\sigma M_h}\!\frac{d\ell_+}{\ell_+}\,
    J(M_h\ell_-,\mu)\,J(-M_h\ell_+,\mu)\,S(\ell_+\ell_-,\mu)\,\Big|_{\rm leading\;power} \\
   &= g_\perp^{\mu\nu}\,H_3^{(0)}\! \int_0^{M_h}\!\frac{d\ell_-}{\ell_-}\,
    \int_0^{\sigma M_h}\!\frac{d\ell_+}{\ell_+}\,
    J^{(0)}(M_h\ell_-)\,J^{(0)}(-M_h\ell_+)\,S^{(0)}(\ell_+\ell_-)\,\Big|_{\rm leading\;power} 
    + \delta H_1^{(0)} \langle O_1^{(0)}\rangle \,,
\end{aligned}
\end{equation}
where
\begin{equation}\label{Z31eff}
   \delta H_1^{(0)} = \frac{N_c\spac\alpha_b}{\pi}\,\frac{y_{b,0}}{\sqrt2}\,
    \frac{C_F\alpha_s}{4\pi}\,8\zeta_3 \left( - \frac{1}{\epsilon} + L_h \right) 
    + {\cal O}(\alpha_s^2) \,. 
\end{equation}
It is tempting to interpret this extra contribution as a mixing of the operator $O_3$ with $O_1$, but in reality its origin lies in the fact that imposing the upper cutoffs on the convolution integrals over $\ell_+$ and $\ell_-$ does not commute with renormalization. It is thus more appropriate to treat the extra term as a contribution to the bare matching coefficient $H_1^{(0)}$. We will come back to this issue in Section~\ref{sec:mismatch}.

\subsection{Renormalized matrix elements}
\label{subsec:4.3}

With the renormalization factors fixed, we now proceed to derive the $h\to\gamma\gamma$ matrix elements of the renormalized operators in the factorization formula (\ref{renfact}). For the case of $O_1$ we trivially obtain 
\begin{equation}\label{O1res}
   \langle O_1(\mu) \rangle = m_b(\mu)\,g_\perp^{\mu\nu} \,.
\end{equation}
For the matrix element of $O_2$ we find
\begin{equation}\label{O2res}
\begin{aligned}
   \langle O_2(z,\mu) \rangle
   &= \frac{N_c\spac\alpha_b}{2\pi}\,m_b(\mu)\,g_\perp^{\mu\nu}\,
    \bigg\{ - L_m + \frac{C_F\alpha_s}{4\pi}\,\bigg[ 
    L_m^2\,\Big( \ln z + \ln(1-z) + 3 \Big) \\
   &\hspace{3.7cm}\mbox{}- L_m \left( \ln^2 z + \ln^2(1-z) - 4\ln z\ln(1-z) 
    + 11 - \frac{2\pi^2}{3} \right) \\
   &\hspace{3.7cm}\mbox{}+ F(z) + F(1-z) \bigg] + {\cal O}(\alpha_s^2) \bigg\} \,,
\end{aligned}
\end{equation}
where
\begin{equation}
\begin{aligned}
   F(z) &= \frac{\ln^3 z}{6} + z\ln^2 z - \ln^2 z\ln(1-z) - \ln z\ln(1-z) - \frac{1+3z}{2} \ln z \\
   &\quad\mbox{}- \left( 4\ln z + 2z \right) \mbox{Li}_2(z) + 6\,\mbox{Li}_3(z) 
    + \frac{11}{2} - 4\zeta_3 \,.
\end{aligned}
\end{equation}
Note that we use the running $b$-quark mass $m_b(\mu)$ in the prefactor but the pole mass $m_b$ in the argument of the logarithm $L_m=\ln(m_b^2/\mu^2)$. Besides being convenient this is not unnatural, because the linear factor of $m_b(\mu)$ in each matrix element plays the role of a running coupling, whereas the quantity $m_b$ appearing in the arguments of the logarithm $L_m$ is due to phase-space effects. If desired, one can always switch back from one choice to the other using relation (\ref{polemass}). In the limit $z\to 0$ the above expression simplifies to
\begin{equation}
\begin{aligned}
   \braces{\langle O_2(z,\mu) \rangle}
   &= \frac{N_c\spac\alpha_b}{2\pi}\,m_b(\mu)\,g_\perp^{\mu\nu}\,
    \bigg\{ - L_m + \frac{C_F\alpha_s}{4\pi}\,\bigg[ 
    L_m^2\,\Big( \ln z + 3 \Big) 
    - L_m \left( \ln^2 z + 11 - \frac{2\pi^2}{3} \right) \\
   &\hspace{3.73cm}\mbox{}+ \frac{\ln^3 z}{6} - \frac{\ln z}{2} + 11 - \frac{\pi^2}{3} - 2\zeta_3 
    \bigg] + {\cal O}(\alpha_s^2) \bigg\} \,.
\end{aligned}
\end{equation}
Note that the same result is obtained using relation in (\ref{eq4.5}) along with the renormalization factors given in (\ref{bracesZ22}) and (\ref{bracesZ21}). 

To obtain the renormalized matrix element of $O_3$, we start from the expressions for the renormalized jet and soft functions. They are \cite{Bosch:2003fc,Liu:2020ydl} 
\begin{equation}
   J(p^2,\mu) 
   = 1 + \frac{C_F\alpha_s}{4\pi} \left[ \ln^2\!\bigg(\frac{-p^2-i0}{\mu^2}\bigg)
    - 1 - \frac{\pi^2}{6} \right] + {\cal O}(\alpha_s^2) \,,
\end{equation}
and \cite{Liu:2020eqe}
\begin{equation}\label{Swres}
   S(w,\mu) = - \frac{N_c\spac\alpha_b}{\pi}\,m_b(\mu)\,
    \Big[ S_a(w,\mu)\,\theta(w-m_b^2) + S_b(w,\mu)\,\theta(m_b^2-w) \Big] \,,
\end{equation}
with
\begin{equation}
\begin{aligned}
   S_a(w,\mu) &= 1 + \frac{C_F\alpha_s}{4\pi} \bigg[ - L_w^2 - 6 L_w + 12 - \frac{\pi^2}{2} 
    + 2\,\mbox{Li}_2\Big(\frac{1}{\hat w}\Big) \\
   &\hspace{2.53cm}\mbox{}- 4\ln\!\Big(1-\frac{1}{\hat w}\Big)\,
    \bigg( L_m + 1 + \ln\!\Big(1-\frac{1}{\hat w}\Big) + \frac32 \ln\hat w \bigg) \bigg] 
    + {\cal O}(\alpha_s^2) \,, \\
   S_b(w,\mu) &= \frac{C_F\alpha_s}{\pi}\,\ln(1-\hat w)\,\big[ L_m + \ln(1-\hat w) \big] 
    + {\cal O}(\alpha_s^2) \,.
\end{aligned}
\end{equation}
Here $\hat w=w/m_b^2$ and $L_w=\ln(w/\mu^2)$. The result for the function $S_a(w,\mu)$ takes this relatively simple form only if one uses the pole mass in the argument of the $\theta(w-m_b^2)$ distribution in (\ref{Swres}). When these expressions are used in the double convolution integral shown in the third term of (\ref{renfact}), one obtains
\begin{equation}\label{O3ren}
\begin{aligned}
   \langle O_3(\mu) \rangle
   &= - \frac{N_c\spac\alpha_b}{\pi}\,m_b(\mu)\,g_\perp^{\mu\nu}
    \bigg\{ \frac{L^2}{2} + \frac{C_F\alpha_s}{4\pi}\,\bigg[ \frac{5}{12}\,L^4 - L^3
    + \left( 5 - \frac{5\pi^2}{12} \right) L^2 
    + \left( \frac{2\pi^2}{3} + 8\zeta_3 \right) L \\
   &\hspace{3.95cm}\mbox{}- 4\zeta_3 - \frac{\pi^4}{9} + \frac12\,L_m^2\,L^2
    + L_m \left( L^3 - 3 L^2 - 8\zeta_3 \right) \bigg] + {\cal O}(\alpha_s^2) \bigg\} \,.
\end{aligned}
\end{equation}
Contrary to the matrix elements of $O_1$ and $O_2$ this expression contains the large rapidity logarithm $L=\ln(-M_h^2/m_b^2)$, which is a consequence of the collinear anomaly. The fact that the integrals over $\ell_+$ and $\ell_-$ in (\ref{renfact}) run from the soft region (with $\ell_+\ell_-\sim m_b^2$) up to values of ${\cal O}(M_h)$ generates up to two powers of $L$ for each loop order in addition to the logarithms $L_m$ associated with the soft scale $m_b$. In other examples where the collinear anomaly appears, the rapidity logarithms take on a simpler form and (typically) exponentiate \cite{Becher:2010tm}. In the present case their structure is more complicated, because the rapidity logarithms arise from a double integral over a rather complicated integrand. In order to resum these logarithms it is necessary to factorize the matrix element into a convolution over jet and soft functions, each of which depends on a different scale, and then solve the RG evolution equations of these various functions.

\subsection{Renormalized matching coefficients}
\label{subsec:4.4}

Ignoring the cutoffs on the convolutions integrals in (\ref{renfact}) for a moment, one would conclude that the UV divergences of the bare matching coefficients $H_i^{(0)}$ are removed by applying the inverse matrix of renormalization factors 
\begin{equation}\label{Zijinv}
   \bm{Z}^{-1} = \left( \begin{array}{cccc}
    Z_{11}^{-1} & 0 & 0 & 0 \\
    Z_{21}^{-1} & Z_{22}^{-1} & 0 & 0 \\
    \braces{Z_{21}^{-1}} & 0 & \braces{Z_{22}^{-1}} & 0 \\
    0 & 0 & 0 & Z_{33}^{-1}
   \end{array} \right)
\end{equation}
from the right to the row vector of bare matching coefficients $(H_1^{(0)},2H_2^{(0)},-2\braces{H_2^{(0)}},H_3^{(0)})$, where
\begin{equation}\label{Zinvelements}
\begin{aligned}
   Z_{21}^{-1}(z) 
   &= - \int_0^1\!dz'\,Z_{22}^{-1}(z,z')\,Z_{21}(z')\,Z_{11}^{-1} \,, \\[-1mm]
   \braces{Z_{21}^{-1}(z)} 
   &= - \int_0^\infty\!dz'\,\braces{Z_{22}^{-1}(z,z')}\,\braces{Z_{21}(z')}\,Z_{11}^{-1} \,.
\end{aligned}
\end{equation}
Specifically, one would then derive
\begin{equation}\label{Hiren}
\begin{aligned}
   H_2(z,\mu) 
   &= \int_0^1\!dz'\,H_2^{(0)}(z')\,Z_{22}^{-1}(z',z) \,, \\
   \frac{\braces{\bar H_2(z,\mu)}}{z} 
   &= \int_0^\infty\!dz'\,\frac{\braces{\bar H_2^{(0)}(z')}}{z'}\,\braces{Z_{22}^{-1}(z',z)} \,, \\
   H_3(\mu) &= H_3^{(0)} Z_{33}^{-1} \,,
\end{aligned}
\end{equation}
as well as
\begin{equation}\label{H1naive}
\begin{aligned}
   H_1(\mu)
   &= \left( H_1^{(0)} + \Delta H_1^{(0)} - \delta H_1^{(0)} \right) Z_{11}^{-1} 
    + 2 \int_0^1\!dz\,H_2^{(0)}(z)\,Z_{21}^{-1}(z) \\
   &\quad\mbox{}- 2 \int_0^\infty\!dz\,\frac{\braces{\bar H_2^{(0)}(z)}}{z}\,\braces{Z_{21}^{-1}(z)} 
    - 2 \int_0^\infty\!d\bar z\,\frac{\braces{\bar H_2^{(0)}(\bar z)}}{\bar z}\,
    \braces{Z_{21}^{-1}(\bar z)} \,. \qquad \text{(naively)} \\
\end{aligned}
\end{equation} 
At ${\cal O}(\alpha_s)$ most elements of the inverse matrix $\bm{Z}^{-1}$ can be obtained from the corresponding renormalization factors $Z_{ij}$ by simply changing the sign in front of $\alpha_s$. The only exception are the entries $Z_{21}^{-1}$ and $\braces{Z_{21}^{-1}}$, for which we find
\begin{equation}
\begin{aligned}
   Z_{21}^{-1}(z) 
   &= \frac{N_c\spac\alpha_b}{2\pi}\,\bigg\{ \frac{1}{\epsilon} + \frac{C_F\alpha_s}{4\pi}\, 
    \bigg[ \frac{\ln z+\ln(1-z)}{\epsilon^2} \\
   &\hspace{1.8cm}\mbox{}+ \frac{1}{\epsilon} \left( \frac{\ln^2 z+\ln^2(1-z)}{2} 
    - 2\ln z\ln(1-z) + \frac{11}{2} - \frac{\pi^2}{3} \right) \bigg] 
    + {\cal O}(\alpha_s^2) \bigg\} \,, \\
   \braces{Z_{21}^{-1}(z)} 
   &= \frac{N_c\spac\alpha_b}{2\pi}\,\bigg\{ \frac{1}{\epsilon} + \frac{C_F\alpha_s}{4\pi}\, 
    \bigg[ \frac{\ln z}{\epsilon^2} + \frac{1}{\epsilon} \left( \frac{\ln^2 z}{2} 
    + \frac{11}{2} - \frac{\pi^2}{3} \right) \bigg] + {\cal O}(\alpha_s^2) \bigg\} \,.
\end{aligned}
\end{equation}

The three relations in (\ref{Hiren}) indeed provide the correct renormalization conditions for the corresponding matching coefficients. Using the expressions for the bare matching coefficients derived in \cite{Liu:2019oav} and collected in (\ref{Hires}) and (\ref{H2bar}), we find
\begin{equation}\label{H2ren}
\begin{aligned}   
   H_2(z,\mu) 
   &= \frac{y_b(\mu)}{\sqrt 2} \frac{1}{z(1-z)} \bigg\{ 1 \!+\! \frac{C_F\alpha_s}{4\pi}
    \Big[ 2 L_h \hspace{-0.3mm}\big(\hspace{-0.3mm} \ln z \spac\!+\! \ln(1-z) \big) 
    \!+\! \ln^2 z \!+\! \ln^2(1-z) \!-\! 3 \Big] \!+\! {\cal O}(\alpha_s^2) \!\bigg\} , \\
   \braces{\bar H_2(z,\mu)} 
   &= \frac{y_b(\mu)}{\sqrt 2} \left[ 1 + \frac{C_F\alpha_s}{4\pi}
    \left( 2 L_h \ln z + \ln^2 z - 3 \right) + {\cal O}(\alpha_s^2) \right] , \\
   H_3(\mu) 
   &= \frac{y_b(\mu)}{\sqrt 2} \left[ - 1 + \frac{C_F\alpha_s}{4\pi}
    \left( L_h^2 + 2 - \frac{\pi^2}{6} \right) + {\cal O}(\alpha_s^2) \right] ,
\end{aligned}
\end{equation}
where $L_h=\ln(-M_h^2/\mu^2)$, and $y_b(\mu)$ denotes the running $b$-quark Yukawa coupling. The result for $\braces{\bar H_2(z,\mu)}$ can be obtained in two ways: either by using the second relation in (\ref{Hiren}) or by taking the limit $z\to 0$ in the expression for $H_2(z,\mu)$. Both methods lead to the same result.

The expression for $H_1$ shown in (\ref{H1naive}) is problematic, because the integrals over $z$ and $\bar z$ extending up to infinity are divergent and indeed undefined. To see this, note that at lowest order in perturbation theory $\braces{Z_{21}^{-1}(z)}$ is a constant, while $\braces{\bar H_2^{(0)}(z)}=1$. In order to obtain a well-behaved expression we need to restrict the integration to the interval $z\in[0,1]$, like in the first term. We thus define
\begin{equation}\label{H1mu}
\begin{aligned}
   H_1(\mu) &= \left( H_1^{(0)} + \Delta H_1^{(0)} - \delta H_1^{(0)}  
    - \delta'\hspace{-0.5mm} H_1^{(0)} \right) Z_{11}^{-1} \\
   &\quad\mbox{}+ 2 \lim_{\delta\to 0}\,\int_\delta^{1-\delta}\!dz\,\bigg[ 
    H_2^{(0)}(z)\,Z_{21}^{-1}(z) - \frac{\braces{\bar H_2^{(0)}(z)}}{z}\,\braces{Z_{21}^{-1}(z)} 
    - \frac{\braces{\bar H_2^{(0)}(\bar z)}}{\bar z}\,\braces{Z_{21}^{-1}(\bar z)} \bigg] \,,
\end{aligned}
\end{equation}
where the sum of the three terms in the second line is now well defined and free of endpoint singularities, such that the limit $\delta\to 0$ is smooth. The quantity $\delta'\hspace{-0.5mm} H_1^{(0)}$ accounts for the mismatch of integration limits, which one encounters when equating the factorization formula (\ref{renfact}) expressed in terms of renormalized quantities with formula (\ref{barefact}) expressed in terms of bare quantities (recall that both correctly reproduce the decay amplitude). After a straightforward calculation we find that
\begin{equation}\label{deltaH1def}
\begin{aligned}
   \delta'\hspace{-0.5mm} H_1^{(0)} 
   &= 4 \int_0^1\!dz\,\frac{\braces{\bar H_2^{(0)}(z)}}{z}\,
    \frac{\braces{\langle O_2^{(0)}(z)\rangle}}{\langle O_1^{(0)}\rangle} \\  
   &\quad\mbox{}- 4 \int_0^\infty\!dz\,\frac{\braces{\bar H_2^{(0)}(z)}}{z}
    \int_0^\infty\!dz'\,\frac{\braces{\langle O_2^{(0)}(z')\rangle}}{\langle O_1^{(0)}\rangle}
    \int_0^1\!dz''\,\braces{Z_{22}^{-1}(z,z'')}\,\braces{Z_{22}(z'',z')} \\
   &\quad\mbox{}+ 4 \left[ \int_0^1\!dz \int_0^\infty\!dz' - \int_0^\infty\!dz \int_0^1\!dz' \right]
    \frac{\braces{\bar H_2^{(0)}(z)}}{z}\,\braces{Z_{22}^{-1}(z,z')}\,
    \braces{Z_{21}(z')} \,.
\end{aligned}
\end{equation}
Starting from the expressions for the bare quantities derived in \cite{Liu:2019oav} and given in Appendix~\ref{app:bareresults}, and using our results for the various renormalization factors, we find that $\delta'\hspace{-0.5mm} H_1^{(0)}={\cal O}(\alpha_s^2)$. It is then straightforward to obtain from (\ref{H1mu})
\begin{equation}
   H_1(\mu) = \frac{N_c\spac\alpha_b}{\pi}\,\frac{y_b(\mu)}{\sqrt 2}\,\bigg\{ 
    - 2 + \frac{C_F\alpha_s}{4\pi} \left[ - \frac{\pi^2}{3}\,L_h^2 
    + \left( 12 + 8\zeta_3 \right) L_h - 36 - \frac{2\pi^2}{3} - \frac{11\pi^4}{45} \right]
    + {\cal O}(\alpha_s^2) \bigg\} \,.
\end{equation}

\subsection{Higher-order analysis of cutoff effects}
\label{sec:mismatch}

In our discussion so far we have glanced over an important subtlety related to the cutoffs on the various terms in the bare and renormalized factorization theorems (\ref{fact4}) and (\ref{renfact}). In (\ref{H1mu}) the renormalized matching coefficient $H_1(\mu)$ is expressed in terms of bare quantities and renormalization factors. However, it is far from obvious that the sum of the terms on the right-hand side is indeed only a function of the hard scale $-M_h^2$ and independent of the soft scale set by the $b$-quark mass. Indeed, the definition of $\delta'\hspace{-0.5mm} H_1^{(0)}$ in (\ref{deltaH1def}) involves the matrix element of the bare operator $\braces{\langle O_2^{(0)}\rangle}$, which {\em does\/} depend on the $b$-quark mass, see (\ref{O2mel}). Likewise, the expression for $\delta H_1^{(0)}$ following from (\ref{eq22}) contains the renormalized and bare soft functions, both of which depend on the $b$-quark mass, see (\ref{Swres}) and (\ref{Ssplit}). However, we will now show that the sum
\begin{equation}
   \delta H_1^{(0),\,{\rm tot}}
   \equiv \delta H_1^{(0)} + \delta'\hspace{-0.5mm} H_1^{(0)} \,,
\end{equation}
which enters in (\ref{H1mu}), is independent of the $b$-quark mass to all orders of perturbation theory. This combined quantity is thus truly a hard subtraction term.

From the definition (\ref{eq22}) it follows that
\begin{equation}
\begin{aligned}
   \delta H_1^{(0)}\spac m_{b,0}
   &= H_3(\mu)\!\int_0^{M_h}\!\frac{d\ell_-}{\ell_-}\!\int_0^{\sigma M_h}\!\frac{d\ell_+}{\ell_+}\,
    J(M_h\ell_-,\mu)\,J(-M_h\ell_+,\mu)\,S(\ell_+\ell_-,\mu)\spac\Big|_{\rm leading\;power} \\
   &\mbox{}- H_3^{(0)} \int_0^{M_h}\!\frac{d\ell_-}{\ell_-}\,
    \int_0^{\sigma M_h}\!\frac{d\ell_+}{\ell_+}\,
    J^{(0)}(M_h\ell_-)\,J^{(0)}(-M_h\ell_+)\,S^{(0)}(\ell_+\ell_-)\,\Big|_{\rm leading\;power} \,,
\end{aligned}
\end{equation}
where the factor $m_{b,0}$ on the left-hand side stems from the matrix element $\langle O_1^{(0)}\rangle$. Throughout this section we do not write out the limit $\sigma\to-1$ explicitly, but it is understood in all expressions where $\sigma$ occurs. Rewriting the renormalized jet and soft functions in the first line in terms of the corresponding bare functions, using relations derived in \cite{Liu:2020ydl,Liu:2020eqe} and expressing the renormalized matching coefficient $H_3(\mu)$ in terms of the bare one using the last relation in (\ref{Hiren}), the right-hand side of this equation can be put in the form
\begin{equation}\label{eq41}
\begin{aligned}
   &\phantom{=} H_3^{(0)} \int_0^\infty\!\frac{d\rho_-}{\rho_-} 
    \int_0^\infty\!\frac{d\rho_+}{\rho_+}\,
    S^{(0)}(\rho_+\rho_-) \int_0^\infty\!d\omega_- \int_0^\infty\!d\omega_+\,
    J^{(0)}(M_h\omega_-)\,J^{(0)}(-M_h\omega_+) \\
   &\quad\times \int_0^{M_h}\!d\ell_-\,Z_J^{-1}(M_h\rho_-,M_h\ell_-)\,Z_J(M_h\ell_-,M_h\omega_-) \\
   &\quad\times \int_0^{\sigma M_h}\!d\ell_+\,Z_J^{-1}(-M_h\rho_+,-M_h\ell_+)\,
    Z_J(-M_h\ell_+,-M_h\omega_+)\,\Big|_{\rm leading\;power} \\
   &\quad\mbox{}- H_3^{(0)} \int_0^{M_h}\!\frac{d\rho_-}{\rho_-}
    \int_0^{\sigma M_h}\!\frac{d\rho_+}{\rho_+}\,
    S^{(0)}(\rho_+\rho_-)\,J^{(0)}(M_h\rho_-)\,J^{(0)}(-M_h\rho_+)\,\Big|_{\rm leading\;power} \,.
\end{aligned}
\end{equation}
The quantity $Z_J$ denotes the renormalization factor of the jet function defined as 
\begin{equation}
\begin{aligned}
   J(\pm M_h\ell,\mu) &= \int_0^\infty\!d\ell'\,Z_J(\pm M_h\ell,\pm M_h\ell')\,
    J^{(0)}(\pm M_h\ell') \,.
\end{aligned}   
\end{equation}
At one-loop order one finds \cite{Bosch:2003fc,Liu:2020ydl}
\begin{equation}\label{ZJexpr}
   Z_J(\pm M_h\ell,\pm M_h\ell') 
   = \left[ 1 + \frac{C_F\alpha_s}{4\pi} \left( - \frac{2}{\epsilon^2} 
    + \frac{2}{\epsilon} \ln\frac{\mp M_h\ell}{\mu^2} \right) \right] \delta(\ell-\ell') 
    + \frac{C_F\alpha_s}{2\pi\epsilon}\,\ell\,\Gamma(\ell,\ell')
    + {\cal O}(\alpha_s^2) \,,
\end{equation}
where
\begin{equation}
   \Gamma(\ell,\ell') 
   = \left[ \frac{\theta(\ell-\ell')}{\ell(\ell-\ell')} 
    + \frac{\theta(\ell'-\ell)}{\ell'(\ell'-\ell)} \right]_+
\end{equation}
is the symmetric Lange-Neubert kernel \cite{Lange:2003ff}. Note that the quantity $Z_J$ satisfies the symmetry relation
\begin{equation}
   Z_J(p^2,p^{\prime\spac 2}) 
   = \frac{p^2}{p^{\prime\spac 2}}\,Z_J(p^{\prime\spac 2},p^2) \,.
\end{equation}
As shown in (\ref{simplerelaZ}), the renormalization factor of the soft function can be expressed in terms of the same object \cite{Liu:2020eqe}. If it was not for the cutoffs on the integrals, the quantity in (\ref{eq41}) would evaluate to zero, because the integrals over the products of $Z_J$ factors in the second and third lines would yield $\delta(\rho_\pm-\omega_\pm)$. 

Only the terms involving the matrix element $\braces{\langle O_2^{(0)}\rangle}$ in the definition (\ref{deltaH1def}) depend on the $b$-quark mass. We can use the refactorization conditions (\ref{refact}) to eliminate $\braces{\bar H_2^{(0)}}$ and $\braces{\langle O_2^{(0)}\rangle}$ from these expressions and rewrite them in terms of $H_3^{(0)}$ and the bare jet and soft functions. Moreover, the first refactorization condition implies a connection between the renormalization factors $Z_{33}$, $\braces{Z_{22}}$ and $Z_J$, which yields the relation
\begin{equation}\label{Z22rela}
   Z_{33}\,\braces{Z_{22}^{-1}(z,z')}
   = M_h\,Z_J(z M_h^2,z' M_h^2) = M_h\,\frac{z}{z'}\,Z_J(z' M_h^2,z M_h^2) \,.
\end{equation}
This allows us to express the product of renormalization factors in the second line of (\ref{deltaH1def}) in terms of $Z_J$ and $Z_J^{-1}$. We find
\begin{equation}\label{deltaprime}
\begin{aligned}
   \delta'\hspace{-0.5mm} H_1^{(0)}\spac m_{b,0} \Big|_{\,\braces{\langle O_2^{(0)}\rangle}} 
   &= -2 H_3^{(0)} \int_0^\infty\!\frac{d\rho_-}{\rho_-} \int_0^\infty\!\frac{d\rho_+}{\rho_+}\,
    S^{(0)}(\rho_+\rho_-)\,J^{(0)}(-M_h\rho_+) \int_0^\infty\!d\omega_-\,J^{(0)}(M_h\omega_-) \\
   &\quad\times \int_0^{M_h}\!d\ell_-\,Z_J^{-1}(M_h\rho_-,M_h\ell_-)\,Z_J(M_h\ell_-,M_h\omega_-) \\
   &\quad\mbox{}+2 H_3^{(0)} \int_0^{M_h}\!\frac{d\rho_-}{\rho_-}
    \int_0^\infty\!\frac{d\rho_+}{\rho_+}\,
    S^{(0)}(\rho_+\rho_-)\,J^{(0)}(M_h\rho_-)\,J^{(0)}(-M_h\rho_+) \,.
\end{aligned}
\end{equation}
The two terms in this result are structurally similar to those appearing in (\ref{eq41}). Indeed, it is possible to rearrange these terms in such a way that
\begin{equation}\label{difficult}
\begin{aligned}
   \delta H_1^{(0),\,{\rm tot}}
   &= H_3^{(0)} \int_0^\infty\!\frac{d\rho_-}{\rho_-} \int_0^\infty\!\frac{d\rho_+}{\rho_+}\,
    \frac{S^{(0)}(\rho_+\rho_-)}{m_{b,0}} \int_0^\infty\!d\omega_- \int_0^\infty\!d\omega_+\,
    J^{(0)}(M_h\omega_-)\,J^{(0)}(-M_h\omega_+) \\
   &\quad\times \int_{M_h}^\infty\!d\ell_-\,Z_J^{-1}(M_h\rho_-,M_h\ell_-)\,
    Z_J(M_h\ell_-,M_h\omega_-) \\
   &\quad\times \int_{\sigma M_h}^\infty\!d\ell_+\,Z_J^{-1}(-M_h\rho_+,-M_h\ell_+)\,
    Z_J(-M_h\ell_+,-M_h\omega_+)\,\Big|_{\rm leading\;power} \\
   &\quad\mbox{}- H_3^{(0)} \int_{M_h}^\infty\!\frac{d\rho_-}{\rho_-}
    \int_{\sigma M_h}^\infty\!\frac{d\rho_+}{\rho_+}\,
    \frac{S^{(0)}(\rho_+\rho_-)}{m_{b,0}}\,J^{(0)}(M_h\rho_-)\,J^{(0)}(-M_h\rho_+)\,
    \Big|_{\rm leading\;power} \\
   &\quad\mbox{}+ 4 \left[ \int_0^1\!dz \int_0^\infty\!dz' 
    - \int_0^\infty\!dz \int_0^1\!dz' \right]
    \frac{\braces{\bar H_2^{(0)}(z)}}{z}\,\braces{Z_{22}^{-1}(z,z')}\,
    \braces{Z_{21}(z')} \,.
\end{aligned}
\end{equation}
Details on the derivation of this result are given in Appendix~\ref{app:mismatch}. It is important to remember that we only need the leading-power terms in this expression. In the second integral (fourth line) the variables $\rho_\pm$ are both in the hard region, and hence the arguments of the soft and jet functions are all of order $M_h^2$. For the first integral, the variables $\ell_\pm$ are in the hard region, and this forces the variables $\rho_\pm$ and $\omega_\pm$ to be in the hard region as well (see Appendix~\ref{app:mismatch} for more details). In fact, we can recast relation (\ref{difficult}) in the alternative form 
\begin{equation}
\begin{aligned}
   \delta H_1^{(0),\,{\rm tot}}
   &= H_3(\mu) \int_{M_h}^\infty\!\frac{d\rho_-}{\rho_-}
    \int_{\sigma M_h}^\infty\!\frac{d\rho_+}{\rho_+}\,
    \frac{S^{(\epsilon)}(\rho_+\rho_-,\mu)}{m_{b,0}}\,J^{(\epsilon)}(M_h\rho_-,\mu)\,
    J^{(\epsilon)}(-M_h\rho_+,\mu)\,\Big|_{\rm leading\;power} \\
   &\quad\mbox{}- H_3^{(0)} \int_{M_h}^\infty\!\frac{d\rho_-}{\rho_-}
    \int_{\sigma M_h}^\infty\!\frac{d\rho_+}{\rho_+}\,
    \frac{S^{(0)}(\rho_+\rho_-)}{m_{b,0}}\,J^{(0)}(M_h\rho_-)\,J^{(0)}(-M_h\rho_+)\,
    \Big|_{\rm leading\;power} \\
   &\quad\mbox{}+ 4 \left[ \int_0^1\!dz \int_0^\infty\!dz' 
    - \int_0^\infty\!dz \int_0^1\!dz' \right]
    \frac{\braces{\bar H_2^{(0)}(z)}}{z}\,\braces{Z_{22}^{-1}(z,z')}\,
    \braces{Z_{21}(z')} \,,
\end{aligned}
\end{equation}
where in the first term the dimensional regulator $\epsilon=(4-D)/2$ must be kept in place after renormalization in order to regularize the divergences for $\rho_\pm\to\infty$, as indicated by the superscript ``($\epsilon$)''. This form makes it explicit that the quantity $\delta H_1^{(0),\,{\rm tot}}$ only depends on the hard scale $-M_h^2$. In order to obtain the leading-power terms we can therefore simply set $m_{b,0}=0$ in the ratio $S^{(0)}(\rho_+\rho_-)/m_{b,0}$ in (\ref{difficult}), in which case we obtain from (\ref{Ssplit}) and (\ref{SaSb})
\begin{equation}\label{Sinfty}
   \frac{S_\infty^{(0)}(w)}{m_{b,0}} 
   \equiv \frac{S^{(0)}(w)}{m_{b,0}} \bigg|_{m_{b,0}\to 0}
   = - \frac{N_c\spac\alpha_{b,0}}{\pi}\,\theta(w) 
   \left[ \frac{e^{\epsilon\gamma_E}}{\Gamma(1-\epsilon)}\,w^{-\epsilon} 
    + \frac{C_F\alpha_{s,0}}{4\pi}\,C_1(\epsilon)\,w^{-2\epsilon} 
    + {\cal O}(\alpha_s^2) \right] ,
\end{equation}
with $C_1(\epsilon)$ given in (\ref{C1C2eps}). In order to calculate the quantity $\delta H_1^{(0),\,{\rm tot}}$ at leading order in $\alpha_s$ we use the lowest-order expressions for the bare jet and soft functions. We also note that the integral in the last line of (\ref{difficult}) vanishes at first order in $\alpha_s$. After a straightforward calculation we find 
\begin{equation}\label{eq55}
   \delta H_1^{(0),\,{\rm tot}}
   = \frac{y_{b,0}}{\sqrt2}\,\frac{N_c\spac\alpha_{b,0}}{\pi}\,
    \frac{C_F\alpha_{s,0}}{4\pi} \left( - M_h^2 -i0 \right)^{-\epsilon} 
    \frac{e^{\epsilon\gamma_E}}{\Gamma(1-\epsilon)}\,\frac{4}{\epsilon^3}\,
    \Big[ H(\epsilon) + H(-\epsilon) \Big] + {\cal O}(\alpha_s^2) \,,
\end{equation}
where $H(\epsilon)=\psi(1+\epsilon)+\gamma_E$ is the harmonic-number function. This generalizes relation (\ref{Z31eff}) to higher orders in $\epsilon$.

In terms of this quantity, the correct all-order definition of the renormalized matching coefficient $H_1(\mu)$ is obtained as
\begin{equation}
\begin{aligned}
   H_1(\mu) &= \left( H_1^{(0)} + \Delta H_1^{(0)} - \delta H_1^{(0),\,{\rm tot}} \right)
    Z_{11}^{-1} \\
   &\quad\mbox{}+ 2 \lim_{\delta\to 0}\,\int_\delta^{1-\delta}\!dz\,\bigg[ 
    H_2^{(0)}(z)\,Z_{21}^{-1}(z) - \frac{\braces{\bar H_2^{(0)}(z)}}{z}\,\braces{Z_{21}^{-1}(z)} 
    - \frac{\braces{\bar H_2^{(0)}(\bar z)}}{\bar z}\,\braces{Z_{21}^{-1}(\bar z)} \bigg] \,.
\end{aligned}
\end{equation}

\subsection{Contributions to the decay amplitude}

As a cross check we evaluate the three terms $T_i$ shown in the three lines of the renormalized factorization theorem (\ref{renfact}) using the results for the matrix elements and matching coefficients obtained above. This yields
\begin{equation}
\begin{aligned}
   T_1 &= {\cal M}_0 \left\{ - 2 + \frac{C_F\alpha_s}{4\pi}\,\bigg[ 
    - \frac{\pi^2}{3}\,L_h^2 + \left( 12 + 8\zeta_3 \right) L_h - 36 - \frac{2\pi^2}{3}
    - \frac{11\pi^4}{45} \bigg] \right\} , \\
   T_2 &= {\cal M}_0\,\frac{C_F\alpha_s}{4\pi}\,\bigg[\spac\frac{2\pi^2}{3}\,L_h L_m 
    - \frac{\pi^2}{3}\,L_m^2 + \frac{2\pi^2}{3} + 8\zeta_3 + \frac{7\pi^4}{45} \bigg] \,, \\
   T_3 &= {\cal M}_0\,\bigg\{ \frac{L^2}{2} + \frac{C_F\alpha_s}{4\pi}\,\bigg[\spac  
    \frac{5L^4}{12} + \left( L_m - 1 \right) L^3 
    + \left( 4 - \frac{\pi^2}{3} + \frac{L_m^2}{2} - \frac{L_h^2}{2} - 3 L_m \right) L^2 \\
   &\hspace{3.95cm}\mbox{}+ \left( \frac{2\pi^2}{3} + 8\zeta_3 \right) L - 8\zeta_3 L_m 
    - 4\zeta_3 - \frac{\pi^4}{9} \bigg] \bigg\} \,, 
\end{aligned}
\end{equation}
where we have defined
\begin{equation}
   {\cal M}_0 = \frac{N_c\spac\alpha_b}{\pi}\,\frac{y_b(\mu)}{\sqrt2}\,m_b(\mu)\,
    \varepsilon_\perp^*(k_1)\cdot\varepsilon_\perp^*(k_2) \,.
\end{equation}
Adding up the three contributions we correctly reproduce the QCD amplitude 
\begin{equation}
\begin{aligned}
   {\cal M}_b
   &= {\cal M}_0\,\bigg\{\! \left( \frac{L^2}{2} - 2 \right) 
    + \frac{C_F\alpha_s(\mu)}{4\pi}\,\bigg[ - \frac{L^4}{12} - L^3 
    + \left( 4 - \frac{2\pi^2}{3} \right) L^2 \\
   &\hspace{16mm}\mbox{}+ \left( 12 + \frac{2\pi^2}{3} + 16\zeta_3 \right) L 
    - 36 + 4\zeta_3 - \frac{\pi^4}{5} - \left( 3L^2 - 12 \right) \ln\frac{m_b^2}{\mu^2} \bigg] 
    \bigg\} 
\end{aligned}
\end{equation}
up to higher-order corrections in $\alpha_s$ and $m_b^2/M_h^2$.

\section{RG evolution equations}
\label{sec:RGEs}
\renewcommand{\theequation}{5.\arabic{equation}}
\setcounter{equation}{0}

In the previous section we have accomplished the main goal of this work: the establishment of the renormalized factorization formula (\ref{renfact}). We have derived explicit expressions for the renormalized matrix elements and matching coefficients to first order in $\alpha_s$ (corresponding to the two-loop order for the decay amplitude), and we have shown that to all orders of perturbation theory the effects of the upper cutoffs on the convolution integrals over $\ell_+$ and $\ell_-$ (and on the $z$ integral over the functions inside the braces $\braces{\dots}$) is not in conflict with factorization. We will now derive the RG evolution equations for the various objects in (\ref{renfact}), which set the basis for the systematic resummation of the large logarithms in the decay amplitude. 

In terms of the various renormalization factors $Z_{ij}$ derived in Section~\ref{subsec:4.2} we obtain the corresponding anomalous dimensions $\gamma_{ij}$ in the usual way, i.e.\
\begin{equation}\label{gammaij}
   \gamma_{ij} = \left( 2\alpha_b\,\frac{\partial}{\partial\alpha_b} 
    + 2\alpha_s\,\frac{\partial}{\partial\alpha_s} \right) Z_{ij}^{(1)} \,,
\end{equation}
where $Z_{ij}^{(1)}$ denotes the coefficient of the single $1/\epsilon$ pole in $Z_{ij}$. In this way, we obtain the diagonal elements
\begin{equation}\label{gammaiires}
\begin{aligned}
   \gamma_{11} &= \frac{3C_F\alpha_s}{2\pi} + {\cal O}(\alpha_s^2) \,, \\
   \gamma_{22}(z,z') &= - \frac{C_F\alpha_s}{\pi}\,\bigg\{\!
    \left[ \ln z + \ln(1-z) + \frac32 \right] \delta(z-z') \\
   &\hspace{2.2cm}\mbox{}+ z(1-z) \left[ \frac{1}{z'(1-z)}\,\frac{\theta(z'-z)}{z'-z} 
    + \frac{1}{z(1-z')}\,\frac{\theta(z-z')}{z-z'} \right]_+ \!\bigg\} 
    + {\cal O}(\alpha_s^2) \,, \\
   \braces{\gamma_{22}(z,z')} &= - \frac{C_F\alpha_s}{\pi}\,\bigg\{\!
    \left( \ln z + \frac32 \right) \delta(z-z') 
    + z \left[ \frac{\theta(z'-z)}{z'(z'-z)} + \frac{\theta(z-z')}{z(z-z')} \right]_+ 
    \!\bigg\} + {\cal O}(\alpha_s^2) \,, \\
   \gamma_{33} &= \frac{C_F\alpha_s}{\pi} \left( L_h - \frac32 \right) + {\cal O}(\alpha_s^2) \,,
\end{aligned}   
\end{equation}
as well as the off-diagonal elements
\begin{equation}
\begin{aligned}
   \gamma_{21}(z) 
   &= - \frac{N_c\spac\alpha_b}{\pi} \left\{ 1 + \frac{C_F\alpha_s}{4\pi}
    \left[ \ln^2 z + \ln^2(1-z) - 4\ln z\ln(1-z) + 11 - \frac{2\pi^2}{3} \right] 
    + {\cal O}(\alpha_s^2) \right\} , \\
   \braces{\gamma_{21}(z)} 
   &= - \frac{N_c\spac\alpha_b}{\pi} \left\{ 1 + \frac{C_F\alpha_s}{4\pi}
    \left( \ln^2 z + 11 - \frac{2\pi^2}{3} \right) + {\cal O}(\alpha_s^2) \right\} .
\end{aligned}   
\end{equation}
Note that both $Z_{ij}$ and $\gamma_{ij}$ are scale-dependent quantities, but we suppress this dependence for the sake of simplicity of the notation. 

The diagonal elements of the anomalous-dimension matrix are also known to higher orders in $\alpha_s$. Relation (\ref{Z11Zm}) implies that $\gamma_{11}=-\gamma_m$ is determined in terms of the anomalous dimension of the quark mass, defined as
\begin{equation}
   \frac{d}{d\ln\mu}\,m_b(\mu) = \gamma_m\,m_b(\mu) \,.
\end{equation}
The quantity $\gamma_m$ is known to five-loop order \cite{Luthe:2016xec}. The two-loop expression for $\gamma_{22}$ (and with it $\braces{\gamma_{22}}$) can in principle be derived from \cite{Dittes:1983dy,Mikhailov:1984ii,Brodsky:1984xk,Mueller:1993hg,Mueller:1994cn}. The anomalous dimension $\gamma_{33}$, which according to (\ref{Z33}) is equal to the anomalous dimension of two-jet current operators in SCET, can to all orders be written in the form \cite{Becher:2009qa}
\begin{equation}\label{gamma33}
   \gamma_{33} = \Gamma_{\rm cusp}(\alpha_s)\,\ln\frac{-M_h^2}{\mu^2} + 2\gamma_q(\alpha_s) \,,
\end{equation}
where $\Gamma_{\rm cusp}$ is the light-like cusp anomalous dimension in the fundamental representation of SU$(N_c)$ \cite{Korchemskaya:1992je}, and $\gamma_q$ is the anomalous dimension of the quark field in light-cone gauge. The cusp anomalous dimension has recently been calculated to four-loop order \cite{Henn:2019swt}, while $\gamma_q$ is known to three loops. It can be determined from the three-loop expression for the divergent part of the on-shell quark form factor in QCD 
\cite{Moch:2005id,Becher:2006mr}.

\subsection{RG equations for the operator matrix elements}
\label{subsec:5.1}
 
From the renormalization conditions (\ref{O2ren}) and (\ref{eq4.5}) it follows that the matrix elements of the renormalized operators satisfy the RG evolution equations
\begin{equation}\label{OiRGEs}
\begin{aligned}
   \frac{d}{d\ln\mu}\,\langle O_1(\mu)\rangle 
   &= - \gamma_{11}\,\langle O_1(\mu)\rangle \,, \\
   \frac{d}{d\ln\mu}\,\langle O_2(z,\mu)\rangle 
   &= - \int_0^1\!dz'\,\gamma_{22}(z,z')\,\langle O_2(z',\mu)\rangle 
    - \gamma_{21}(z)\,\langle O_1(\mu)\rangle \,, \\
   \frac{d}{d\ln\mu}\,\braces{\langle O_2(z,\mu)\rangle} 
   &= - \int_0^\infty\!dz'\,\braces{\gamma_{22}(z,z')}\,\braces{\langle O_2(z',\mu)\rangle} 
    - \braces{\gamma_{21}(z)}\,\langle O_1(\mu)\rangle \,.
\end{aligned}
\end{equation}
We have checked that these equations are satisfied to ${\cal O}(\alpha_s)$.

As mentioned earlier, in order to resum all large logarithms contained in the matrix element of the operator $O_3$ one must factorize the matrix element in the form
\begin{equation}
   \langle O_3(\mu)\rangle
   = g_\perp^{\mu\nu}\,\lim_{\sigma\to-1}\, 
    \int_0^{M_h}\!\frac{d\ell_-}{\ell_-}\,\int_0^{\sigma M_h}\!\frac{d\ell_+}{\ell_+}\,
    J(M_h\ell_-,\mu)\,J(-M_h\ell_+,\mu)\,S(\ell_+\ell_-,\mu)\,\Big|_{\rm leading\;power}
\end{equation}
and solve the RG equations for the jet and soft functions separately. We have derived the corresponding evolution equations at two-loop order in two recent papers. For the jet function one finds \cite{Bosch:2003fc,Liu:2020ydl}
\begin{equation}\label{Jevol}
   \frac{d}{d\ln\mu}\,J(p^2,\mu) 
   = - \int_0^\infty\!dx\,\gamma_J(p^2,x p^2)\,J(x p^2,\mu) \,,
\end{equation}
which in this form holds for both space-like and time-like values of $p^2$. The anomalous dimension is given by
\begin{equation}\label{gammaJ}
\begin{aligned}
   \gamma_J(p^2,x p^2) 
   &= \left[ \Gamma_{\rm cusp}(\alpha_s)\,\ln\frac{-p^2}{\mu^2}
    - \gamma'(\alpha_s) \right] \delta(1-x) + \Gamma_{\rm cusp}(\alpha_s)\,\Gamma(1,x) \\
   &\quad\mbox{}+ C_F \left( \frac{\alpha_s}{2\pi} \right)^2 \frac{\theta(1-x)}{1-x}\,h(x) 
    + {\cal O}(\alpha_s^3) \,,
\end{aligned}
\end{equation}
where 
\begin{equation}\label{hdef}
   h(x) = \ln x \left[ \beta_0 
    + 2C_F \left( \ln x - \frac{1+x}{x}\,\ln(1-x) - \frac32 \right) \right] .
\end{equation}
The local terms (with $x=1$) can to all orders be expressed in terms of the cusp anomalous dimension and an anomalous dimension $\gamma'(\alpha_s)$, which was recently obtained at two-loop order \cite{Liu:2020ydl}. Since the plus distribution contained in $\Gamma(1,x)$ is linked with the logarithmic term, it is also multiplied by $\Gamma_{\rm cusp}$. However, starting at two-loop order additional non-local terms arise, whose explicit form was obtained in \cite{Liu:2020ydl} by using the RG invariance of the $B^-\to\gamma\spac l^-\bar\nu$ decay rate along with the calculation of the two-loop anomalous dimension of the $B$-meson light-cone distribution amplitude (LCDA) performed in \cite{Braun:2019wyx}.

The RG equation for the soft function is tightly linked to that of the jet function \cite{Liu:2020eqe}. One finds\footnote{The quantity $\gamma_S(w,w/x)$ in this relation is connected with the original definition of the anomalous dimension $\gamma_S(w,w';\mu)$ in \cite{Liu:2020eqe} by $\gamma_S(w,w';\mu)=(w/w^{\prime\spac 2})\,\gamma_S(w,w/x)$, where $w'=w/x$.}  
\begin{equation}\label{Sevol}
   \frac{d}{d\ln\mu}\,S(w,\mu) = - \int_0^\infty\!dx\,\gamma_S(w,w/x)\,S(w/x,\mu) \,,
\end{equation}
where 
\begin{equation}\label{gammaS}
\begin{aligned}
   \gamma_S(w,w/x) 
   &= - \left[ \Gamma_{\rm cusp}(\alpha_s)\,\ln\frac{w}{\mu^2} - \gamma_s(\alpha_s) \right]
    \delta(1-x) - 2\Gamma_{\rm cusp}(\alpha_s)\,\Gamma(1,x) \\
   &\quad\mbox{}- 2 C_F \left( \frac{\alpha_s}{2\pi} \right)^2 
    \frac{\theta(1-x)}{1-x}\,h(x) + {\cal O}(\alpha_s^3) \,,
\end{aligned}
\end{equation} 
with
\begin{equation}\label{gams}
   \gamma_s(\alpha_s) = 2\gamma_q(\alpha_s) + 2\gamma'(\alpha_s) \,.
\end{equation}
Via this relation the quantity $\gamma_s$ is known to two-loop order. As defined above, the anomalous dimensions of the matching coefficient $H_3$ and of the jet and soft functions obey the simple relation
\begin{equation}\label{simplerela}
   \gamma_{33} = \gamma_J\bigg(\frac{M_h w}{\ell_+},x\,\frac{M_h w}{\ell_+}\bigg) 
    + \gamma_J(-M_h\ell_+,-x M_h\ell_+) + \gamma_S(w,w/x) \,,
\end{equation}
which is a consequence of relation (\ref{simplerelaZ}). In this form it is easy to see that the variable $\ell_+$ drops out from the result. 

\subsection{RG equations for the matching coefficients}
 
The renormalized matching coefficients obey the evolution equations
\begin{equation}\label{HiRGEs}
\begin{aligned}
   \frac{d}{d\ln\mu}\,H_1(\mu) 
   &= D_{\rm cut}(\mu) + \gamma_{11}\,H_1(\mu) \\[-1mm]
   &\quad\mbox{}+ 2\int_0^1\!dz\,\bigg[ 
    H_2(z,\mu)\,\gamma_{21}(z) - \frac{\braces{\bar H_2(z,\mu)}}{z}\,\braces{\gamma_{21}(z)} 
    - \frac{\braces{\bar H_2(\bar z,\mu)}}{\bar z}\,\braces{\gamma_{21}(\bar z)} \bigg] \,, \\
   \frac{d}{d\ln\mu}\,H_2(z,\mu) 
   &= \int_0^1\!dz'\,H_2(z',\mu)\,\gamma_{22}(z',z) \,, \\
   \frac{d}{d\ln\mu}\,\braces{\bar H_2(z,\mu)}
   &= \int_0^\infty\!dz'\,\braces{\bar H_2(z',\mu)}\,\frac{z}{z'}\,\braces{\gamma_{22}(z',z)} \,, \\
   \frac{d}{d\ln\mu}\,H_3(\mu) &= \gamma_{33}\,H_3(\mu) \,,
\end{aligned}
\end{equation}
where the quantity
\begin{equation}\label{gam31}
   D_{\rm cut}(\mu) = - \frac{N_c\spac\alpha_b}{\pi}\,\frac{y_b(\mu)}{\sqrt2}\,
    \frac{C_F\alpha_s}{4\pi}\,16\zeta_3 + {\cal O}(\alpha_s^2)
\end{equation}
in the first equation results from non-trivial effects of the cutoffs on the scale evolution. Once again, we have checked that all of these equations are satisfied to ${\cal O}(\alpha_s)$. 

The evolution equation for the matching coefficient $H_1(\mu)$ calls for a more careful discussion. We have seen in Section~\ref{sec:mismatch} that the definition of this coefficient in higher orders is quite subtle and requires a careful treatment of the effects of the cutoffs on the various convolution integrals. In this section we will derive the evolution equation for $H_1$ beyond one-loop order using the RG invariance of the decay amplitude on the left-hand side of (\ref{renfact}). Explicitly, we evaluate
\begin{equation}\label{RGinvariance}
   \frac{d}{d\ln\mu}\,{\cal M}_b = 0
   = \left[ \left( \frac{d}{d\ln\mu} - \gamma_{11} \right) H_1(\mu) \right] \langle O_1(\mu)\rangle
    + \frac{d}{d\ln\mu}\,T_2(\mu) + \frac{d}{d\ln\mu}\,T_3(\mu) \,,
\end{equation}
where $T_2$ and $T_3$ denote the second and third lines in the factorization formula (\ref{renfact}). The scale dependence of the third term has been studied in our recent work \cite{Liu:2020eqe}, where we have shown that
\begin{equation}\label{eq78}
\begin{aligned}
   \frac{d}{d\ln\mu}\,T_3(\mu) 
   &= g_\perp^{\mu\nu}\,\lim_{\sigma\to -1}\,H_3(\mu) \int_0^\infty\!dx\,K(x) \\
   &\quad\times \Bigg[ \int_{M_h}^{M_h/x}\!\frac{d\ell_-}{\ell_-}
    \int_0^{\sigma M_h}\!\frac{d\ell_+}{\ell_+}\,
    J(xM_h\ell_-,\mu)\,J(-M_h\ell_+,\mu)\,S(\ell_+\ell_-,\mu) \\
   &\hspace{1.0cm}\mbox{}+ \int_0^{M_h}\!\frac{d\ell_-}{\ell_-} 
    \int_{\sigma M_h}^{\sigma M_h/x}\!\frac{d\ell_+}{\ell_+}\,
    J(M_h\ell_-,\mu)\,J(-xM_h\ell_+,\mu)\,S(\ell_+\ell_-,\mu) \Bigg]_{\rm leading\;power} \,, 
\end{aligned}
\end{equation}
where the kernel $K(x)$ is given by
\begin{equation}\label{Kdef}
   K(x) = \Gamma_{\rm cusp}(\alpha_s)\,\Gamma(1,x)
    + C_F \left( \frac{\alpha_s}{2\pi} \right)^2 \frac{\theta(1-x)}{1-x}\,h(x)
    + {\cal O}(\alpha_s^3) \,.
\end{equation}
This quantity appears in the non-local terms of the anomalous dimensions $\gamma_J$ and $\gamma_S$ in (\ref{gammaJ}) and (\ref{gammaS}). The two terms on the right-hand side of relation (\ref{eq78}) are, in fact, identical, as can be seen by redefining the integration variables according to $\ell_\pm\to\sigma\ell_\mp$. Note the important fact that the result is {\em not\/} given by a hard function. Indeed, at leading order in perturbation theory the last integral evaluates to
\begin{equation}
   \int_0^{\sigma M_h}\!\frac{d\ell_+}{\ell_+}\,J(-M_h\ell_+,\mu)\,S(\ell_+\ell_-,\mu) 
   = - \frac{N_c\spac\alpha_b}{\pi}\,m_b\,\ln\frac{\sigma M_h\ell_-}{m_b^2} + {\cal O}(\alpha_s) \,,
\end{equation}
which is sensitive to the low scale $m_b$.

Using the evolution equations (\ref{OiRGEs}) and (\ref{HiRGEs}) it is straightforward to show that the scale dependence of the second term is given by
\begin{equation}
\begin{aligned}
   \frac{d}{d\ln\mu}\,T_2(\mu)
   &= -2 \int_0^1\!dz \left[ H_2(z,\mu)\,\gamma_{21}(z)
    - \frac{\braces{\bar H_2(z,\mu)}}{z}\,\braces{\gamma_{21}(z)}
    - \frac{\braces{\bar H_2(\bar z,\mu)}}{\bar z}\,\braces{\gamma_{21}(\bar z)} \right]
    \langle O_1(\mu)\rangle \\
   &\quad\mbox{}+ 4 \left[ \int_0^1\!dz \int_0^\infty\!dz' - \int_0^\infty\!dz \int_0^1\!dz' \right]
    \frac{\braces{\bar H_2(z,\mu)}}{z}\,\braces{\gamma_{22}(z,z')}\,
    \braces{\langle O_2(z',\mu)\rangle} \,.
\end{aligned}
\end{equation}
The terms shown in the first line arise from the ``normal'' operator mixing of $O_2$ into $O_1$. They account for the second line in the evolution equation for $H_1(\mu)$ shown in (\ref{HiRGEs}). The quantity in the second line is again a ``left-over contribution'' related to the presence of the cutoffs in the factorization formula. Note that the local terms in $\braces{\gamma_{22}(z,z')}$ proportional to $\delta(z-z')$ in (\ref{gammaiires}) give no contribution here. From (\ref{Z22rela}) it follows that
\begin{equation}
   \braces{\gamma_{22}(z,z')} 
   = - \frac{z}{z^{\prime\,2}}\,K\Big(\frac{z}{z'}\Big) + \text{local terms.}
\end{equation}
Substituting $x=z/z'$ and using the renormalized version of the first refactorization condition in (\ref{refact}),
\begin{equation}\label{refact1renorm}
   \braces{\bar H_2(z,\mu)} = - H_3(\mu)\,J(z M_h^2,\mu) \,, 
\end{equation}
we obtain for the left-over terms
\begin{equation}\label{T2evol}
   \frac{d}{d\ln\mu}\,T_2(\mu) \bigg|_\text{left-over}
   = 4 H_3(\mu) \int_0^\infty\!dx\,K(x) \int_{M_h}^{M_h/x}\!\frac{d\ell_-}{\ell_-}\,
    J(x M_h\ell_-,\mu)\,\braces{\langle O_2(\ell_-/M_h,\mu)\rangle} \,. 
\end{equation}
The refactorization theorem for the bare operator $O_2^{(0)}$ in (\ref{refact}),
\begin{equation}
   \braces{\langle O_2^{(0)}(z)\rangle}
   = - \frac{g_\perp^{\mu\nu}}{2} \int_0^\infty\!\frac{d\ell_+}{\ell_+}\,
    J^{(0)}(-M_h\ell_+)\,S^{(0)}(z M_h\ell_+) \,,
\end{equation}
suggests that this result closely resembles the structure of (\ref{eq78}). We would thus like to establish a similar relation that holds after renormalization. However, the integral on the right-hand side contains endpoint divergences, which are regularized by the dimensional regulator $\epsilon$. In other words, the matrix element of the bare operator $\braces{\langle O_2^{(0)}(z)\rangle}$ contains some $1/\epsilon$ poles not contained in the bare jet and soft functions. Their presence makes the derivation of the renormalized relation non-trivial. After a careful analysis we find that
\begin{equation}\label{refac2new}
   \braces{\langle O_2(z,\mu)\rangle}
   = - \frac{g_\perp^{\mu\nu}}{2} \int_0^{\sigma M_h}\!\frac{d\ell_+}{\ell_+}\,
    J(-M_h\ell_+,\mu)\,S(z M_h\ell_+,\mu)\,\bigg|_{\rm leading\;power}
    + \Delta_{21}(z,\mu)\,\langle O_1(\mu)\rangle \,,
\end{equation}
where all quantities are renormalized and free of UV divergences. Our explicit result for the quantity $\Delta_{21}(z)$ reads
\begin{equation}
\begin{aligned}
   \Delta_{21}(z,\mu)
   &= \braces{Z_{21}(z)}\,Z_{11}^{-1}\!
    - \frac{z M_h}{2}\,Z_{33}\,Z_{11}^{-1} \int_{\sigma M_h}^\infty\!d\ell_+ 
    \int_0^\infty\!d\omega_+ \int_0^\infty\!\frac{d\rho_+}{\rho_+} 
    \int_0^\infty\!\frac{d\rho_-}{\rho_-}\,Z_J^{-1}(M_h\rho_-,z M_h^2) \\
   &\quad\times Z_J^{-1}(-M_h\rho_+,-M_h\ell_+)\,
    Z_J(-M_h\ell_+,-M_h\omega_+)\,J^{(0)}(-M_h\omega_+)\,
    \frac{S_\infty^{(0)}(\rho_+\rho_-)}{m_{b,0}} \,.
\end{aligned}
\end{equation}
All integration variables are in the hard region, and hence the bare soft function can be replaced by its asymptotic form $S_\infty^{(0)}$, see (\ref{Sinfty}). The two terms on the right-hand side of this relation are UV divergent, but by construction their sum is finite when expressed in terms of renormalized parameters. After a somewhat cumbersome calculation, we find
\begin{equation}\label{Delta21}
\begin{aligned}
   \Delta_{21}(z,\mu)
   &= \frac{N_c\spac\alpha_b}{\pi}\,\Bigg\{ - \frac12 \left( L_h + \ln z \right) \\
   &\quad\mbox{}+ \frac{C_F\alpha_s}{4\pi}\,\bigg[\
    \left( \frac12 \ln z + \frac32 \right) L_h^2
    + \left( \frac12 \ln^2 z + 3\ln z - \frac{11}{2} + \frac{\pi^2}{3} \right) L_h \\
   &\hspace{2.2cm}\mbox{}+ \frac{1}{12} \ln^3 z + \frac32 \ln^2 z 
    + \left( \frac{\pi^2}{3} - \frac{23}{4} \right) \ln z
    + \frac{11}{2} - \frac{\pi^2}{2} - 5\zeta_3 \bigg]
    + {\cal O}(\alpha_s^2) \Bigg\} \,.
\end{aligned}
\end{equation}
Like the renormalized matrix element on the left-hand side of (\ref{refac2new}), the quantity $\Delta_{21}$ contains single-logarithmic terms of order $\alpha_b\spac\alpha_s^n L_h^{n+1}$ in higher orders of perturbation theory.  

Combining the results (\ref{eq78}) and (\ref{T2evol}), and using relation (\ref{refac2new}), all terms involving the soft function cancel out, and we obtain the exact formula
\begin{equation}\label{eq89}
\begin{aligned}
   \frac{d}{d\ln\mu}\,\Big[\spac T_3(\mu) + T_2(\mu) \Big] 
   &= 4 H_3(\mu) \int_0^\infty\!dx\,K(x) \int_{1}^{1/x}\!\frac{dz}{z}\,
    J(xz M_h^2,\mu)\,\Delta_{21}(z,\mu)\,\langle O_1(\mu)\rangle \\
   &\hspace{-2.8cm}\mbox{}- 2 \int_0^1\!dz \left[ H_2(z,\mu)\,\gamma_{21}(z)
    - \frac{\braces{\bar H_2(z,\mu)}}{z}\,\braces{\gamma_{21}(z)}
    - \frac{\braces{\bar H_2(\bar z,\mu)}}{\bar z}\,\braces{\gamma_{21}(\bar z)} \right]
    \langle O_1(\mu)\rangle \,.
\end{aligned}
\end{equation}
Using this result along with (\ref{RGinvariance}), and comparing the answer with the evolution equation for $H_1(\mu)$ shown in (\ref{HiRGEs}), we find 
\begin{equation}\label{gam31resu}
\begin{aligned}
   D_{\rm cut}(\mu) 
   &= -4 H_3(\mu) \int_0^\infty\!dx\,K(x) \int_{1}^{1/x}\!\frac{dz}{z}\,
    J(xz M_h^2,\mu)\,\Delta_{21}(z,\mu) \\
   &= 4 \int_0^\infty\!dx\,K(x) \int_{1}^{1/x}\!\frac{dz}{z}\,
    \braces{\bar H_2(xz,\mu)}\,\Delta_{21}(z,\mu) \,.
\end{aligned}
\end{equation}
In the second step we have used the renormalized refactorization condition (\ref{refact1renorm}). It is now explicit that this quantity depends only on the hard scale $L_h$. Performing the integrals over $x$ and $z$, and using the explicit expressions for the quantities $K(x)$ and $\Delta_{21}$ given above, we can compute the first two expansion coefficients of $D_{\rm cut}(\mu)$ in powers of $\alpha_s$. We find
\begin{equation}
   D_{\rm cut}(\mu) = - \frac{N_c\spac\alpha_b}{\pi}\,\frac{y_b(\mu)}{\sqrt2}
    \left[ \frac{C_F\alpha_s}{4\pi}\,16\zeta_3 
    + \left( \frac{\alpha_s}{4\pi} \right)^2 d_{{\rm cut},\spac 2} + {\cal O}(\alpha_s^3) \right] ,
\end{equation}
where
\begin{equation}
\begin{aligned}
   d_{{\rm cut},\spac 2}
   &= C_F^2 \left[ -48\zeta_3\,L_h^2 + \left( - 48\zeta_3 + \frac{8\pi^4}{15} \right) L_h
    + 136\zeta_3 + \frac{4\pi^4}{5} - 32\zeta_5 + \frac{16\pi^2}{3}\,\zeta_3 \right] \\
   &\quad\mbox{}+ C_F C_A \left( - \frac{176}{3}\,\zeta_3\,L_h + \frac{1072}{9}\,\zeta_3 
    - \frac{44\pi^4}{45} - \frac{16\pi^2}{3}\,\zeta_3 \right) \\
   &\quad\mbox{}+ C_F T_F\spac n_f \left( \frac{64}{3}\,\zeta_3\,L_h - \frac{320}{9}\,\zeta_3 
    + \frac{16\pi^4}{45} \right) .
\end{aligned}
\end{equation} 
Interestingly, this result is expressed entirely in terms of $\zeta_n$ values. The leading-order term agrees with (\ref{gam31}). In the two-loop term $n_f=5$ is the number of light quark flavors in the relevant scale interval between $m_b$ and $M_h$. 

The quantity $D_{\rm cut}(\mu)$ exhibits single-logarithmic behavior in higher orders. To see this, note that $\Delta_{21}\ni\alpha_b\spac\alpha_s^n L_h^{n+1}$, $\braces{\bar H_2}\ni\alpha_s^n L_h^n$ and $K={\cal O}(\alpha_s)$, which implies
\begin{equation}
   D_{\rm cut}(\mu) \ni \alpha_b\left(\alpha_s L_h\right)^n .
\end{equation}
Note also that instead of calculating $D_{\rm cut}$ directly one can recast the second relation in (\ref{gam31resu}) in the form
\begin{equation}\label{Dcutrewrite}
   D_{\rm cut}(\mu) = \int_0^\infty\!\frac{dz}{z}\,\braces{\bar H_2(z,\mu)}\,\gamma_{\rm cut}(z) \,,
\end{equation}
where
\begin{equation}
\begin{aligned}
   \gamma_{\rm cut}(z) 
   &= \frac{N_c\spac\alpha_b}{\pi}\,\frac{C_F\alpha_s}{4\pi}\,\bigg\{ 
    8\theta(1-z)\,\Big[ L_h \ln(1-z) - \mbox{Li}_2(z) \Big] \\
   &\hspace{2.9cm}\mbox{}- 8\theta(z-1) \left[ L_h \ln\Big(1-\frac{1}{z}\Big) 
    + \mbox{Li}_2\Big(\frac{1}{z}\Big) \right] \!\bigg\} 
    + {\cal O}(L_h^2\spac\alpha_s^2) \,.
\end{aligned}
\end{equation}
The two-loop coefficient is straightforward to calculate and contains polylogarithms of fourth order. An advantage of the form (\ref{Dcutrewrite}) is that it brings the evolution equation for $H_1$ in (\ref{HiRGEs}) to a more canonical form. However, one finds that the new anomalous dimension $\gamma_{\rm cut}\ni\alpha_b\left(\alpha_s L_h\right)^n$ contains higher powers of the logarithm $L_h$ in higher orders of perturbation theory, and it is therefore not of the Sudakov type. 

This logarithmic behavior in higher orders has an important implication for the solution of the RG evolution equations. As long as it is not known how to resum the logarithmic terms in $\gamma_{\rm cut}$ or $D_{\rm cut}$, it is impossible to systematically integrate the evolution equation for $H_1(\mu)$ from a high matching scale down to a scale $\mu\ll M_h$. We are thus forced to choose a value for the factorization scale $\mu$ that is of order the Higgs-boson mass. The challenge is then to solve the evolution equations for the operator matrix elements in (\ref{OiRGEs}), (\ref{Jevol}) and (\ref{Sevol}) in order to evolve these matrix elements up to the same scale $\mu\sim M_h$. 

Let us mention an interesting and indeed fortunate coincidence in this context. In \cite{Liu:2020eqe} it has been found that the solution of the RG evolution equation for the soft function $S(w,\mu)$ requires that the factorization scale $\mu$ must be larger than the matching scale $\mu_s$, at which the initial condition for the soft function can be calculated in fixed-order perturbation theory. Because of the fact that the argument $w=\ell_+\ell_-$ of the soft function is integrated from the soft region ($w\sim m_b^2$) up into the hard region ($w\sim M_h^2$), it is necessary to perform a {\em dynamical scale setting\/} $\mu_s^2\sim\ell_+\ell_-$ under the integral when solving the RG equation \cite{Liu:2020tzd}. Hence, it is necessary, also from this point of view, that the factorization scale $\mu$ is chosen of order the hard scale $M_h$, so as to ensure that $\mu>\mu_s$ for all values of the integrand. As a final comment, we stress that for the method of dynamical scale setting to be consistent, it is crucially important that the renormalized soft function $S(w,\mu)$ in (\ref{Swres}) does not develop any large logarithms $\ln^n(w/m_b^2)$ in the limit where $w\gg m_b^2$. Otherwise, it would be necessary the refactorize the soft function in the limit where the variable $w$ approaches the hard region.

\section{Large logarithms in the three-loop decay amplitude}
\label{sec:largelogs}
\renewcommand{\theequation}{6.\arabic{equation}}
\setcounter{equation}{0}

The RG evolution equations established in the previous section, along with the explicit expressions for the relevant anomalous dimensions, provide the basis for a systematic resummation of the large logarithms $L=\ln(-M_h^2/m_b^2)$ in the $b$-quark induced $h\to\gamma\gamma$ decay amplitude. In the factorization theorem (\ref{renfact}) contribution from the last term is enhanced by at least two powers of logarithms with respect to the other terms, because the integrals over $\ell_+$ and $\ell_-$ provide a logarithmic enhancement. A careful analysis reveals that, for generic values of $\mu$, the first two terms in the factorization formula, $T_1$ and $T_2$, yield terms of ${\cal O}(\alpha_b\spac\alpha_s^n L^{n+1})$ to the decay amplitude, while the third term, $T_3$, yields terms of ${\cal O}(\alpha_b\spac\alpha_s^n L^{2n+2})$. In particular, starting at first order in $\alpha_s$ the series of leading and subleading logarithms receives contributions from this last term only. For this reason, it is often stated in the literature that the large double-logarithmic corrections arise from the region in which the quark propagator connecting the two photons carries a soft momentum (``soft quark contribution'' \cite{Kotsky:1997rq,Akhoury:2001mz,Liu:2017vkm}). 

With the results derived in this paper it is possible to perform a resummation of large logarithms at (almost) NLO in RG-improved perturbation theory, in which all terms enhanced by large logarithms are exponentiated, while contributions not in the exponent are suppressed by powers of $\alpha_s$. This requires the ${\cal O}(\alpha_s)$ expressions for all matching coefficients and matrix elements as well as for the jet and soft functions, each evaluated at its characteristic scale. The various anomalous dimensions in the evolution equations are needed at two-loop order in QCD, while the three-loop expressions are needed for the cusp anomalous dimension and the $\beta$-function. With the exception of $\gamma_{21}$, all these ingredients are known or, in the case of $\gamma_{22}$, can be derived from existing results. 

While conceptually straightforward, performing the resummation at NLO in RG-improved perturbation theory is technically challenging because of the complicated structure of the RG evolution equations for the soft and jet functions, the need to perform a dynamical scale setting for which the matching scales float with the integration variables $\ell_+$ and $\ell_-$ \cite{Liu:2020eqe}, and the analytic continuation $\sigma\to-1$ that needs to be performed in $T_3$ and requires extending the running coupling $\alpha_s(\mu)$ into the complex plane. We leave a detailed discussion of these technicalities for future work, mentioning however that for the case of $T_3$ the resummation has been studied at LO in RG-improved perturbation theory in \cite{Liu:2020tzd}. Instead, here we will use the RG equations derived in Section~\ref{sec:RGEs} to predict the terms in the three-loop $h\to\gamma\gamma$ decay amplitude that are enhanced by at least three powers of the large logarithm $L$. Solving the RG evolution equations (\ref{OiRGEs}), (\ref{Jevol}), (\ref{Sevol}) and (\ref{HiRGEs}) iteratively in perturbation theory, it is straightforward to derive the necessary higher-order logarithmic contributions to the various operator matrix elements and matching coefficients. 

\subsection{Higher-order logarithms in the matrix elements}

The renormalized matrix elements of the operators $O_i$, the jet function and the soft function have been derived in Section~\ref{subsec:4.3} at first non-trivial order in $\alpha_s$. The result for the matrix element of $O_1$ in (\ref{O1res}) is exact without any higher-order corrections. For the matrix element of $O_2$, we extend relation (\ref{O2res}) in the form
\begin{equation}
\begin{aligned}
   \langle O_2(z,\mu) \rangle
   &= \frac{N_c\spac\alpha_b}{2\pi}\,m_b(\mu)\,g_\perp^{\mu\nu} \bigg\{ 
    \!-\! L_m \!+ {\cal O}(\alpha_s) - C_F\! \left( \frac{\alpha_s}{4\pi} \right)^2\! 
    \Big[ L_m^3\,\Big( f_3(z) +\! f_3(1-z) \Big) \!+ {\cal O}(L_m^2) \Big] \bigg\} ,
\end{aligned}
\end{equation}
where the evolution equation (\ref{OiRGEs}) yields
\begin{equation}
   f_3(z) = C_F \left( \frac23\,\ln^2 z + 4 \ln z + 3 \right)
    + \frac{\beta_0}{3} \left( \ln z + \frac32 \right) .
\end{equation}
To derive the terms of ${\cal O}(\alpha_b\spac\alpha_s^2\spac L_m^2)$ would require knowledge of the two-loop coefficient of the anomalous dimension $\gamma_{22}$. The matrix element $\braces{\langle O_2(z,\mu) \rangle}$ can be readily derived by taking the limit $z\to 0$ in the above expression.

There is no need to derive the higher-order logarithmic terms for the jet function, because the complete two-loop expression for $J(p^2,\mu)$ has been obtained in \cite{Liu:2020ydl}. We thus turn directly to the case of the soft function. The iterative solution of the RG equation (\ref{Sevol}) involves some rather complicated integrals, which need to be simplified using various identities for polylogarithms. We present the results as higher-order contributions to the coefficient functions $S_a$ and $S_b$ defined in (\ref{Swres}). Recall that in this equation we use the running $b$-quark mass in the prefactor but the pole mass everywhere else. We parameterize our results in the form
\begin{equation}
\begin{aligned}
   S_a(w,\mu) 
   &= 1 + {\cal O}(\alpha_s) + C_F \left( \frac{\alpha_s}{4\pi} \right)^2 
    \bigg[\, \frac{C_F}{2}\spac L_w^4 + \bigg( 6 C_F + \frac{\beta_0}{3} \bigg)\spac L_w^3 
    + r_2\spac L_w^2 + r_1\spac L_w + {\cal O}(L_w^0) \\
   &\hspace{4.95cm}\mbox{}+ s_{3a}(\hat w)\spac L_m^3 + s_{2a}(\hat w)\spac L_m^2 
    + s_{1a}(\hat w)\spac L_m + {\cal O}(L_m^0) \bigg] \,, \\
   S_b(w,\mu) 
   &= {\cal O}(\alpha_s) + C_F \left( \frac{\alpha_s}{4\pi} \right)^2 
    \Big[ s_{3b}(\hat w)\spac L_m^3 + s_{2b}(\hat w)\spac L_m^2 
    + s_{1b}(\hat w)\spac L_m + {\cal O}(L_m^0) \Big] \,,
\end{aligned}
\end{equation}
where the functions $s_{ia}(\hat w)$ vanish in the limit $\hat w\to\infty$, while the functions $s_{ib}(\hat w)$ vanish for $\hat w\to 0$. For the coefficients $r_i$ we obtain
\begin{equation}
\begin{aligned}
   r_2 &= \left( 6 + \frac{\pi^2}{2} \right) C_F 
    + \left( \frac{32}{9} + \frac{\pi^2}{3} \right) C_A - \frac{16}{9}\,T_F\spac n_f \,, \\
   r_1 &= \left( - 75 + 3\pi^2 \right) C_F 
    + \left( - \frac{1297}{27} + \frac{11\pi^2}{9} - 2\zeta_3 \right) C_A
    + \left( \frac{428}{27} - \frac{4\pi^2}{9} \right) T_F\spac n_f \,,
\end{aligned}
\end{equation}
while the functions $s_{ia}(\hat w)$ are given by
\begin{align}
   s_{3a}(\hat w) 
   &= 4 C_F \ln\!\Big( 1 - \frac{1}{\hat w} \Big) , \notag\\
   s_{2a}(\hat w) 
   &= C_F \left[ 28 \ln\!\Big( 1 - \frac{1}{\hat w} \Big) 
    + 20 \ln^2\!\Big( 1 - \frac{1}{\hat w} \Big) 
    + 18 \ln\hat w\,\ln\!\Big( 1 - \frac{1}{\hat w} \Big) 
    + 10\,\mbox{Li}_2\Big( \frac{1}{\hat w} \Big) \right] \notag\\
   &\quad\mbox{}+ 2\beta_0 \ln\!\Big( 1 - \frac{1}{\hat w} \Big) \,, \notag\\
   s_{1a}(\hat w) 
   &= C_F\,\bigg[ \left( - 24 + \frac{22\pi^2}{3} \right) \ln\!\Big( 1 - \frac{1}{\hat w} \Big)
    + 48 \ln^2\!\Big( 1 - \frac{1}{\hat w} \Big) + 32 \ln^3\!\Big( 1 - \frac{1}{\hat w} \Big) \notag\\
   &\hspace{1.3cm}\mbox{}+ 60 \ln\hat w\,\ln\!\Big( 1 - \frac{1}{\hat w} \Big)
    + 72 \ln\hat w\,\ln^2\!\Big( 1 - \frac{1}{\hat w} \Big) 
    + 24\ln^2\hat w\,\ln\!\Big( 1 - \frac{1}{\hat w} \Big) \notag\\
   &\hspace{1.3cm}\mbox{}+ \left[ 12 + 12 \ln\hat w 
    - 8 \ln\!\Big( 1 - \frac{1}{\hat w} \Big) \right] \mbox{Li}_2\Big( \frac{1}{\hat w} \Big)
    - 48\,\mbox{Li}_3\Big( 1 - \frac{1}{\hat w} \Big) + 48\zeta_3 \bigg] \notag\\
   &\quad\mbox{}+ C_A \left( - \frac{16}{3} + \frac{4\pi^2}{3} \right) 
    \ln\!\Big( 1 - \frac{1}{\hat w} \Big) \notag\\
   &\quad\mbox{}+ \beta_0 \left[ - \frac83\,\ln\!\Big( 1 - \frac{1}{\hat w} \Big)
    + 4 \ln^2\!\Big( 1 - \frac{1}{\hat w} \Big) 
    + 6 \ln\hat w\,\ln\!\Big( 1 - \frac{1}{\hat w} \Big)
    - 2\,\mbox{Li}_2\Big( \frac{1}{\hat w} \Big) \right] . 
\end{align}
In the derivation of these results one needs the two-loop coefficients of the anomalous dimensions $\Gamma_{\rm cusp}$, $\gamma_s$ and $\gamma_m$, which are collected in Appendix~\ref{app:anomdims}. Similarly, for the functions $s_{ib}(\hat w)$ we find
\begin{equation}
\begin{aligned}
   s_{3b}(\hat w) 
   &= - 4 C_F \ln(1-\hat w) \,, \\[2mm]
   s_{2b}(\hat w) 
   &= - C_F\,\Big[ 24 \ln(1-\hat w) + 20 \ln(1-\hat w)^2 - 4 \ln\hat w \ln(1-\hat w)  
    + 4\,\mbox{Li}_2(\hat w) \Big] - 2\beta_0 \ln(1-\hat w) \,, \\
   s_{1b}(\hat w) 
   &= C_F\,\bigg[ \left( 48 - \frac{22\pi^2}{3} \right) \ln(1-\hat w)
    - 32 \ln^2(1-\hat w) - 32 \ln^3(1-\hat w) \\
   &\hspace{1.4cm}\mbox{}- 12 \ln\hat w\,\ln(1-\hat w) + 24\ln\hat w\,\ln^2(1-\hat w) 
    + 8 \ln^2\hat w\,\ln(1-\hat w) \\
   &\hspace{1.4cm}\mbox{}+ \Big(\! -4 + 8\ln\hat w + 8 \ln(1-\hat w) \Big)\,\mbox{Li}_2(\hat w)
    - 8\,\mbox{Li}_3(\hat w) + 48\,\mbox{Li}_3(1-\hat w) - 48\zeta_3 \bigg] \\
   &\quad\mbox{}+ C_A \left( \frac{16}{3} - \frac{4\pi^2}{3} \right) \ln(1-\hat w) \\
   &\quad\mbox{}+ \beta_0 \left[ \frac{20}{3}\,\ln(1-\hat w) - 4 \ln^2(1-\hat w)
    + 4 \ln\hat w\,\ln(1-\hat w) + 4\,\mbox{Li}_2(\hat w) \right] .
\end{aligned}
\end{equation}
To compute the remaining ${\cal O}(\alpha_s^2\spac L_{w,m}^0)$ terms in the soft function would require a complete two-loop calculation.

\subsection{Higher-order logarithms in the matching coefficients}

The renormalized matching coefficients have been discussed in Section~\ref{subsec:4.4}. Higher-order corrections to the coefficients $H_2$ and $H_3$ can be derived straightforwardly by perturbatively solving the corresponding evolution equations in (\ref{HiRGEs}). We extend the first relation in (\ref{H2ren}) in the form
\begin{equation}\label{H2ho}
\begin{aligned}
   H_2(z,\mu) 
   &= \frac{y_b(\mu)}{\sqrt 2}\,\frac{1}{z(1-z)}\,\bigg\{ 1 + {\cal O}(\alpha_s)
    + C_F \left( \frac{\alpha_s}{4\pi} \right)^2 \Big[ L_h^2\,\Big( f_2(z) + f_2(1-z) \Big)
    + {\cal O}(L_h) \Big] \bigg\} \,,
\end{aligned}
\end{equation}
where
\begin{equation}
   f_2(z) = 2 C_F \ln^2 z - \beta_0 \ln z \,.
\end{equation}
To derive the terms of ${\cal O}(\alpha_s^2\spac L_h)$ would require knowledge of the two-loop coefficient of the anomalous dimension $\gamma_{22}$ in (\ref{OiRGEs}). The matching coefficient $\braces{H_2(z,\mu)}$ can be readily derived by taking the limit $z\to 0$ in the above expression.

The second relation in (\ref{H2ren}) can be extended as
\begin{equation}
   H_3(\mu) = - \frac{y_b(\mu)}{\sqrt 2} \left[ 1 + {\cal O}(\alpha_s) 
    + C_F \left( \frac{\alpha_s}{4\pi} \right)^2 
    \bigg( \frac{C_F}{2}\spac L_h^4 + \frac{\beta_0}{3}\spac L_h^3 + c_2\spac L_h^2 + c_1\spac L_h 
    + {\cal O}(L_h^0) \bigg) \right] ,
\end{equation}
with
\begin{equation}
\begin{aligned}   
   c_2 &= \left( 2 - \frac{\pi^2}{6} \right) C_F
    + \left( - \frac{67}{9} + \frac{\pi^2}{3} \right) C_A + \frac{20}{9}\,T_F\spac n_f \,, \\
   c_1 &= \left( -2\pi^2 + 24\zeta_3 \right) C_F
    + \left( \frac{242}{27} + \frac{11\pi^2}{9} - 26\zeta_3 \right) C_A
    + \left( - \frac{112}{27} - \frac{4\pi^2}{9} \right) T_F\spac n_f \,.
\end{aligned}
\end{equation}
In deriving these coefficients we have used the two-loop expression for the anomalous dimension $\gamma_{33}$ in (\ref{gamma33}), which is given in Appendix~\ref{app:anomdims}. To determine the ${\cal O}(\alpha_s^2\spac L_h^0)$ contribution would require a complete two-loop calculation of $H_3$.

Given the higher-order logarithmic corrections to $H_2$ shown in (\ref{H2ho}), it is straightforward to integrate the first evolution equation in (\ref{HiRGEs}) perturbatively. In this way we obtain
\begin{equation}
   H_1(\mu) = \frac{N_c\spac\alpha_b}{\pi}\,\frac{y_b(\mu)}{\sqrt 2} \left[
    - 2 + {\cal O}(\alpha_s) + C_F \left( \frac{\alpha_s}{4\pi} \right)^2
    \Big( k_3 L_h^3 + {\cal O}(L_h^2) \Big) \right] ,
\end{equation}
where
\begin{equation}
   k_3 = \left( \frac{2\pi^2}{3} - \frac{16}{3}\,\zeta_3 \right) C_F
    + \frac{2\pi^2}{9}\,\beta_0 \,.
\end{equation}
The derivation of the ${\cal O}(\alpha_b\spac\alpha_s^2\spac L_h^2)$ term would require knowledge of the ${\cal O}(\alpha_s^2\spac L_h)$ contribution to $H_2(z,\mu)$, which in turn needs the two-loop coefficient of the anomalous dimension $\gamma_{22}$.

\subsection{Higher-order logarithmic contributions to the decay amplitude}

Given the above higher-order results for the matrix elements and matching coefficients, it is straightforward to derive the higher-order logarithmic corrections to the $b$-quark induced $h\to\gamma\gamma$ decay amplitude at ${\cal O}(\alpha_s^2 L^k)$ with $k\ge 3$. We find 
\begin{align}\label{finres}
   {\cal M}_b
   &= {\cal M}_0(\mu)\,\Bigg\{\! \left( \frac{L^2}{2} - 2 \right) 
    + \frac{C_F\alpha_s(\mu)}{4\pi}\,\bigg[ - \frac{L^4}{12} - L^3 
    + \left( 4 - \frac{2\pi^2}{3} \right) L^2 \notag\\
   &\hspace{2.9cm}\mbox{}+ \left( 12 + \frac{2\pi^2}{3} + 16\zeta_3 \right) L 
    - 36 + 4\zeta_3 - \frac{\pi^4}{5} - \left( 3L^2 - 12 \right) L_m \bigg] \notag\\[2cm]
   &\hspace{2.1cm}\mbox{}+ C_F \left( \frac{\alpha_s(\mu)}{4\pi} \right)^2
    \Bigg[ \frac{C_F}{90}\,L^6 + \left( \frac{C_F}{10} + \frac{\beta_0}{20} \right) L^5
    + d_4\spac L^4 + d_3\spac L^3 \\
   &\hspace{2.9cm}\mbox{}+ L_m \left[ \left( \frac{C_F}{2} + \frac{\beta_0}{12} \right) L^4 
    + \big( 6 C_F + \beta_0 \big)\spac L^3 + d_{1,2}\spac L^2 \right] \notag\\
   &\hspace{2.9cm}\mbox{}+ L_m^2\spac L^2 \left( 9 C_F + \frac32\,\beta_0 \right) 
    + \dots \Bigg] + {\cal O}\bigg(\frac{m_b^2}{M_h^2}\bigg) \Bigg\} \,, \notag
\end{align}
where the dots refer to three-loop terms containing less than three powers of logarithms ($L$ or $L_m$). The higher-order expansion coefficients are
\begin{equation}
\begin{aligned}
   d_4 &= \left( \frac56 + \frac{\pi^2}{18} \right) C_F 
    + \left( \frac{8}{27} + \frac{\pi^2}{36} \right) C_A
    - \frac{4}{27}\,T_F\spac n_f \,, \\
   d_3 &= \left( - \frac{17}{2} + \frac{7\pi^2}{9} + \frac{20}{3}\,\zeta_3 \right) C_F
    + \left( - \frac{199}{18} + \frac{44\pi^2}{27} - 4\zeta_3 \right) C_A
    + \left( \frac{22}{9} - \frac{16\pi^2}{27} \right) T_F\spac n_f \,, \\
   d_{1,2} &= \left( - \frac{51}{2} + 4\pi^2 \right) C_F
    + \left( - \frac{185}{6} + \frac{22\pi^2}{9} \right) C_A
    + \left( \frac{26}{3} - \frac{8\pi^2}{9} \right) T_F\spac n_f \,.
\end{aligned}
\end{equation}
The amplitude is scale-independent to the order we are working, meaning that the $\mu$ dependence of the running couplings $y_b(\mu)$, $m_b(\mu)$ contained in ${\cal M}_0(\mu)$ and of $\alpha_s(\mu)$ is compensated by the terms containing $L_m=\ln(m_b^2/\mu^2)$. 

\begin{figure}[t]
\begin{center}
\includegraphics[width=0.65\textwidth]{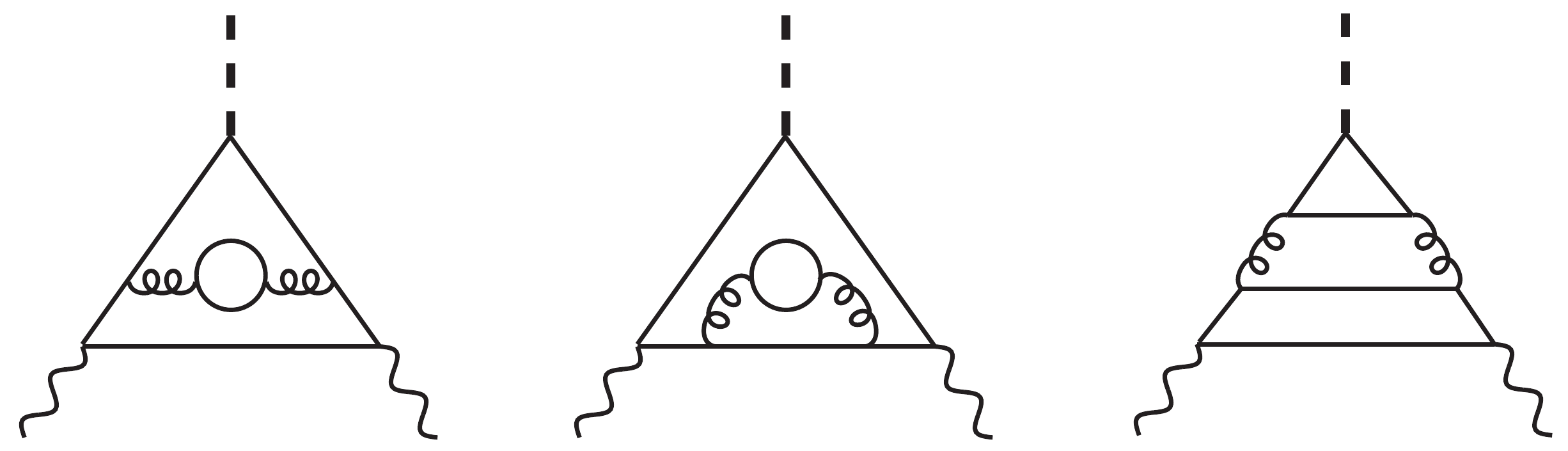} 
\vspace{2mm}
\caption{\label{fig:3loop} 
Examples of three-loop Feynman diagrams containing a second quark loop in addition to the $b$-quark loop connected to the Higgs boson.}
\end{center}
\end{figure}

As a cross check of our results, we now compare expression (\ref{finres}) for the decay amplitude with the results of previous calculations. To this end we need to perform transformations to different renormalization schemes. First, we express the running parameters $m_b(\mu)$ and $y_b(\mu)=\sqrt2\,m_b(\mu)/v$ in the prefactor ${\cal M}_0(\mu)$ in terms of the $b$-quark pole mass, using relation (\ref{polemass}). We then eliminate the remaining scale dependence by making the choice $\mu^2=\hat\mu_h^2\equiv-M_h^2-i0$ in the running coupling $\alpha_s(\mu)$. In this ``on-shell scheme'' (OS), we find that the amplitude takes the form
\begin{equation}\label{amplOS}
\begin{aligned}
   {\cal M}_b
   &= \frac{N_c\spac\alpha_b}{\pi}\,\frac{m_b^2}{v}\,
    \varepsilon_\perp^*(k_1)\cdot\varepsilon_\perp^*(k_2) \\
   &\times \Bigg\{ \frac{L^2}{2} - 2  
    + \frac{C_F\spac\alpha_s(\hat\mu_h)}{4\pi} \left[ - \frac{L^4}{12} - L^3 
    - \frac{2\pi^2}{3}\,L^2 + \left( 12 + \frac{2\pi^2}{3} + 16\zeta_3 \right) L 
    - 20 + 4\zeta_3 - \frac{\pi^4}{5} \right] \\
   &\hspace{0.8cm}\mbox{}+ C_F \left(\! \frac{\alpha_s(\hat\mu_h)}{4\pi} \!\right)^2
    \left[ \frac{C_F}{90}\,L^6 + \left( \frac{C_F}{10} - \frac{\beta_0}{30} \right) L^5
    + d_4^{\spac\rm OS}\spac L^4 + d_3^{\spac\rm OS}\spac L^3 + \dots \right] \!\Bigg\} \,,
\end{aligned}
\end{equation}
where
\begin{equation}
\begin{aligned}
   d_4^{\spac\rm OS} &= \left( \frac32 + \frac{\pi^2}{18} \right) C_F 
    + \left( - \frac{91}{27} + \frac{\pi^2}{36} \right) C_A
    + \frac{32}{27}\,T_F\spac n_f \,, \\
   d_3^{\spac\rm OS} &= \left( - \frac12 + \frac{7\pi^2}{9} + \frac{20}{3}\,\zeta_3 \right) C_F
    + \left( - \frac{199}{18} - \frac{22\pi^2}{27} - 4\zeta_3 \right) C_A
    + \left( \frac{22}{9} + \frac{8\pi^2}{27} \right) T_F\spac n_f \,,
\end{aligned}
\end{equation}
and the dots refer to terms containing less than three powers of logarithms. The contributions to the decay amplitude of ${\cal O}(\alpha_b\spac\alpha_s^2\spac n_f)$ have been calculated in closed analytic form in \cite{Harlander:2019ioe}, and we find full agreement with the results obtained by these authors. Moreover, recently the entire three-loop $gg\to h$ amplitude has been calculated in numerical form  \cite{Czakon:2020vql}. The authors of this paper have kindly repeated their calculation for the case of $h\to\gamma\gamma$ and found the following numerical result for the three-loop coefficient inside the rectangular bracket in (\ref{amplOS}) (to much higher accuracy than indicated here):
\begin{equation}\label{Marco}
\begin{aligned}
   C_F \left[ \frac{C_F}{90}\,L^6 + \dots \right]
   &= 0.01975\spac L^6 - 0.31111\spac L^5 - 8.74342\spac L^4 - 68.6182\spac L^3
    + \dots \\[-1mm]
%    - 117.541\spac L^2 + 1236.52\spac L - 794.055 \\
   &\quad\mbox{}+ \big( 0.02963\spac L^5 + 0.79012\spac L^4 + 3.57918\spac L^3 
    + \dots \big)\,n_l \\[1.5mm]
%    + 7.45126\spac L^2 - 115.737\spac L + 154.325)\,n_l \\
   &\quad\mbox{}+ \big( - 0.04444\spac L^5 - 0.09877\spac L^4 - 2.26947\spac L^3
    + \dots \big)\,n_b \,,
%    + 22.7436\spac L^2 - 106.254\spac L + 212.072 \big)\,n_b \,,
\end{aligned}
\end{equation}
where we only show the coefficients of the logarithmic terms of order $L^3$ and higher. The terms in the second line refer to three-loop diagrams containing a quark loop with $n_l$ massless flavors in addition to the $b$-quark loop connecting to the Higgs boson; see Figure~\ref{fig:3loop} for some representative examples. The terms shown in the third line refer to the same diagrams, but now with a $b$-quark rather than a massless quark propagating in the second fermion loop (where $n_b=1$). In \cite{Czakon:2020vql} the authors defined the running coupling in the $\overline{\rm MS}$ scheme with massive-quark decoupling. Its dependence on the renormalization scale is given by the $\beta$-function for $n_l=4$ massless quarks, i.e.\ $\alpha_s^{(n_l)}(\mu)$. Since in our case the massive $b$-quark is much lighter than the Higgs boson, it is more appropriate to work in a scheme in which one uses the running coupling defined with $n_f=n_l+1$ active quark flavors, as we have done in (\ref{amplOS}). The relevant conversion relation is \cite{Ovrut:1980uv}
\begin{equation}
   \alpha_s^{(n_l)}(\hat\mu_h) = \alpha_s^{(n_f)}(\hat\mu_h)\,\bigg[ 1 
    - \frac{\alpha_s^{(n_f)}(\hat\mu_h)}{6\pi}\,L + {\cal O}(\alpha_s^2) \bigg] \,.
\end{equation}
Performing this scheme transformation we find that relation (\ref{Marco}) is replaced by
\begin{equation}
\begin{aligned}
   C_F \left[ \frac{C_F}{90}\,L^6 + \dots \right]
   &= 0.01975\spac L^6 - 0.31111\spac L^5 - 8.74342\spac L^4 - 68.6182\spac L^3
    + \dots \\[-1mm]
   &\quad\mbox{}+ \big( 0.02963\spac L^5 + 0.79012\spac L^4 + 3.57918\spac L^3 
    + \dots \big)\,n_f + \big( 0 + \dots \big)\,n_b \,,
\end{aligned}
\end{equation}
where $n_f=n_l+1=5$. Our analytic result in (\ref{amplOS}) is in perfect agreement with this expression. We emphasize that the coefficient of the $L^3$ term is sensitive to the two-loop anomalous dimensions of the jet and soft functions. The observed agreement thus presents a highly non-trivial cross check of our conjecture for the two-loop anomalous dimension of the soft function made in \cite{Liu:2020eqe}. 

As a side remark, we stress that the above results indicate that the leading double logarithms generally do not provide the dominant contributions to the decay amplitude. Using $m_b=4.8$\,GeV for the $b$-quark pole mass and $M_h=125.1$\,GeV for the mass of the Higgs boson, and indicating powers $L^n$ with the help of a subscript $n$, we find numerically
\begin{equation}
\begin{aligned}
   {\cal M}_b
   &\approx \frac{N_c\spac\alpha_b}{\pi}\,\frac{m_b^2}{v}\,
    \varepsilon_\perp^*(k_1)\cdot\varepsilon_\perp^*(k_2)\,\bigg\{ 
    \Big[ (16.3 - 20.5\spac i)_2 - 2_0 \Big] \\[-1mm]
   &\quad\mbox{}+ \alpha_s\,\Big[ (5.4 + 23.7\spac i)_4 + (-8.9 + 39.2\spac i)_3
    + (-22.8 + 28.6\spac i)_2 + (26.2 - 12.6\spac i)_1 - 3.7_0 \Big] \\[-1mm]
   &\quad\mbox{}+ \alpha_s^2\,\Big[ (-16.2 - 7.8\spac i)_6 + (12.8 + 16.0\spac i)_5
    + (18.6 + 81.2\spac i)_4 + (-27.1 + 118.8\spac i)_3 + \dots \Big] \bigg\} \spac .
\end{aligned}
\end{equation}
Here $\alpha_s\equiv\alpha_s(-M_h^2-i0)\approx 0.107+0.024\spac i$ is itself a complex number. It follows that in order to obtain reliable results it is essential to perform the resummation of logarithmic terms beyond the leading double-logarithmic approximation using RG-improved perturbation theory, such that all large logarithms are exponentiated.

The above comparison provides a highly non-trivial cross check of our factorization theorem (\ref{renfact}) derived in SCET. Previous authors have analyzed the leading and subleading logarithms in the $h\to\gamma\gamma$ decay amplitude using more traditional resummation techniques \cite{Kotsky:1997rq,Akhoury:2001mz}. They worked in a different renormalization scheme, in which the factor $m_b(\mu)$ in the prefactor ${\cal M}_0(\mu)$ is eliminated in favor of the $b$-quark pole mass, whereas the Yukawa coupling $y_b(\mu)$ and the running coupling $\alpha_s(\mu)$ are evaluated at $\mu=M_h$. In this scheme OS$\spac '$, our result for the decay amplitude assumes the form
\begin{equation}\label{amplOSpr}
\begin{aligned}
   {\cal M}_b
   &= \frac{N_c\spac\alpha_b}{\pi}\,\frac{y_b(M_h)}{\sqrt2}\,m_b\,
    \varepsilon_\perp^*(k_1)\cdot\varepsilon_\perp^*(k_2) \\
   &\times \Bigg\{ \frac{L^2}{2} - 2  
    + \frac{C_F\spac\alpha_s(M_h)}{4\pi}\,\bigg[ - \frac{L^4}{12} + \frac{L^3}{2} 
    + \left( 2 - \frac{2\pi^2}{3} + \frac{3i\pi}{2} \right) L^2 \\
   &\hspace{4.9cm}\mbox{}+ \left( 6 + \frac{2\pi^2}{3} + 16\zeta_3 \right) L 
    - 28 + 4\zeta_3 - \frac{\pi^4}{5} - 6i\pi \bigg] \\
   &\hspace{0.8cm}\mbox{}+ C_F \left(\! \frac{\alpha_s(M_h)}{4\pi} \!\right)^2
    \left[ \frac{C_F}{90}\,L^6 + \left( - \frac{3 C_F}{20} - \frac{\beta_0}{30} \right) L^5
    + d_4^{\spac\rm OS'}\spac L^4 + d_3^{\spac\rm OS'}\spac L^3 + \dots \right] \!\Bigg\} \,,
\end{aligned}
\end{equation}
where
\begin{equation}
\begin{aligned}
   d_4^{\spac\rm OS'} &= \left( \frac{5}{12} + \frac{\pi^2}{18} - \frac{i\pi}{4} \right) C_F 
    + \left( - \frac{67}{108} + \frac{\pi^2}{36} - \frac{11 i\pi}{36} \right) C_A
    + \left( \frac{5}{27} + \frac{i\pi}{9} \right) T_F\spac n_f \,, \\
   d_3^{\spac\rm OS'} &= \left( \frac94 - \frac{11\pi^2}{9} + \frac{20}{3}\,\zeta_3 
    + \frac{3i\pi}{2} \right) C_F
    + \left( \frac{157}{36} - \frac{22\pi^2}{27} - 4\zeta_3 + \frac{11 i\pi}{6} \right) C_A \\
   &\quad\mbox{}+ \left( - \frac{17}{9} + \frac{8\pi^2}{27}  - \frac{2 i\pi}{3} \right) 
    T_F\spac n_f \,.
\end{aligned}
\end{equation}
The leading double logarithms of ${\cal O}(\alpha_b\spac\alpha_s^n\spac L^{2n+2})$ have been correctly obtained in \cite{Kotsky:1997rq,Akhoury:2001mz}. In \cite{Akhoury:2001mz} also the next-to-leading logarithms (NLL) of ${\cal O}(\alpha_b\spac\alpha_s^n\spac L^{2n+1})$ were analyzed and an all-order formula for them was proposed. The result reads
\begin{equation}\label{Akhoury}
\begin{aligned}
   {\cal M}_b^{\rm NLL}
   &= \frac{N_c\spac\alpha_b}{\pi}\,\frac{y_b(M_h)}{\sqrt2}\,m_b\,
    \varepsilon_\perp^*(k_1)\cdot\varepsilon_\perp^*(k_2)\,\frac{L^2}{2}\,
    \Bigg\{ \sum_{n=0}^\infty\,\frac{2\Gamma(n+1)}{\Gamma(2n+3)} 
    \left( - \frac{C_F\spac\alpha_s(M_h)}{2\pi}\,L^2 \right)^n \\
   &\quad\mbox{}- \frac{1}{L}\,\sum_{n=1}^\infty\,\frac{\Gamma(n+1)}{\Gamma(2n+2)} 
    \left( - \frac{C_F\spac\alpha_s(M_h)}{2\pi}\,L^2 \right)^n
    \left[ 3 - \beta_0\,\frac{\alpha_s(M_h)}{2\pi}\,L^2\,\frac{n(n+1)}{(2n+2)(2n+3)} \right] 
    \!\Bigg\} \,.
\end{aligned}
\end{equation}
When expanded to ${\cal O}(\alpha_s^2)$ this formula yields
\begin{equation}
\begin{aligned}
   {\cal M}_b^{\rm NLL}
   &= \frac{N_c\spac\alpha_b}{\pi}\,\frac{y_b(M_h)}{\sqrt2}\,m_b\,
    \varepsilon_\perp^*(k_1)\cdot\varepsilon_\perp^*(k_2) \\
   &\times \Bigg\{ \frac{L^2}{2}   
    + \frac{C_F\spac\alpha_s(M_h)}{4\pi} \left( - \frac{L^4}{12} + \frac{L^3}{2} \right) 
    + C_F \left(\! \frac{\alpha_s(M_h)}{4\pi} \!\right)^2
    \left[ \frac{C_F}{90}\,L^6 + \left( - \bm{\frac{C_F}{10}}
    - \frac{\beta_0}{30} \right) L^5 \right] \!\Bigg\} \,.
\end{aligned}
\end{equation}
Interestingly, the coefficient of the $C_F^2\spac\alpha_s^2\spac L^5$ term (marked in bold face) does not agree with our result shown in (\ref{amplOSpr}). It is difficult to trace the origin of this discrepancy, given that in \cite{Akhoury:2001mz} the derivation of the subleading logarithmic contributions is only sketched. The authors start from an analysis of the off-shell Sudakov form factor \cite{Smilga:1979uj}, i.e.\ the quark form factor of a vector current in the limit where $Q^2=|q^2|\gg|p_i^2|$, and then take the limit where the two external legs go on-shell ($p_i^2\to 0$). They also need to account for the kinematic differences between quark scattering off a vector current and photon scattering off a scalar (Higgs) current. In general, a consistent framework based on effective field theory, such as the SCET approach developed here, is certainly helpful to derive consistent results for corrections appearing beyond the leading order in both logarithmic counting and power counting in $\lambda=m_b/M_h$. In our approach, the coefficients of the leading and subleading logarithms in (\ref{amplOSpr}) are determined in terms of one-loop coefficients of anomalous dimensions. We find that
\begin{equation}
   C_F \left[ \frac{C_F}{90}\,L^6 + \left( - \frac{3 C_F}{20} - \frac{\beta_0}{30} \right) L^5
    + \dots \right] 
   = \frac{\Gamma_0^2}{1440}\,L^6
    + \frac{\Gamma_0}{120} \left( 2\gamma_{q,0} + \gamma_0' - \frac{\gamma_{m,0}}{4} 
    - \beta_0 \right) L^5 + \dots \,,
\end{equation}
where $\Gamma_0=4C_F$, $\gamma_{q,0}=-3C_F$, $\gamma_0'=0$ and $\gamma_{m,0}=-6C_F$ are the one-loop coefficients of the cusp anomalous dimensions and the anomalous dimensions of the collinear quark field, the jet function and the running quark mass, respectively, and we have used relation (\ref{gams}) to eliminate the one-loop anomalous dimension $\gamma_{s,0}$ of the soft function. In \cite{Liu:2020tzd} we have extended the prediction of the NLL terms to higher orders of perturbation theory, finding that (\ref{Akhoury}) must be replaced by
\begin{equation}
\begin{aligned}
   {\cal M}_b^{\rm NLL} 
   &= \frac{N_c\spac\alpha_b}{\pi}\,\frac{y_b(M_h)}{\sqrt2}\,m_b\,
    \varepsilon_\perp^*(k_1)\cdot\varepsilon_\perp^*(k_2)\,\frac{L^2}{2}\,
    \sum_{n=0}^\infty (-\rho)^n\,\frac{2\Gamma(n+1)}{\Gamma(2n+3)} \\
   &\quad\times \left[ 1 + \frac{3\rho}{2L}\,\frac{2n+1}{2n+3}
    - \frac{\beta_0}{C_F}\,\frac{\rho^2}{4L}\,\frac{(n+1)^2}{(2n+3)(2n+5)} \right] ,
\end{aligned}
\end{equation}
where $\rho=\frac{C_F\spac\alpha_s(M_h)}{2\pi}\spac L^2$, and in the prefactor $m_b$ denotes the pole mass. The second term inside the brackets in the second line is not in agreement with (\ref{Akhoury}).

In the recent paper \cite{Anastasiou:2020vkr} the resummation approach of \cite{Akhoury:2001mz} was extended to predict the leading and subleading logarithms in the $gg\to h$ amplitude. The authors showed that in an appropriate ``abelian limit'' their result reproduces the formula for the subleading logarithms shown in (\ref{Akhoury}). We thus believe that also in this work the subleading logarithms at three-loop order and beyond are not correctly accounted for. Matching their results with the calculation of \cite{Czakon:2020vql}, the authors have concluded that the coefficient of the $L^5$ term in equation~(C.1) of this paper should take the form $(C_A-C_F) \big(\frac{11}{9}\,C_A - \frac32\,C_F-2T_F\spac n_b\big)/640\approx 0.0017361111$. Adjusting the color factors in our result (\ref{amplOSpr}) to those relevant for the $gg\to h$ amplitude, we find instead the result
\begin{equation}\label{ourprediction}
   \frac{C_A-C_F}{640} \left( \frac{11}{9}\,C_A - C_F - \frac{10}{3}\,T_F\spac n_b \right)
   \approx 0.0017361111 \,.
\end{equation}
The authors of \cite{Czakon:2020vql} have confirmed to us in a private communication that this is indeed the correct expression for the coefficient of the subleading logarithmic contribution.

\section{Conclusions}
\label{sec:concl}
\renewcommand{\theequation}{7.\arabic{equation}}
\setcounter{equation}{0}

We have derived the first renormalized factorization theorem for an observable described at subleading order in SCET power counting. We have focused on the contribution to the radiative Higgs-boson decay amplitude induced by the Higgs coupling to light bottom quarks; however, the methods we have developed are more general and can be applied to other subleading-power factorization theorems. Endpoint-divergent convolution integrals arising when the factorized decay amplitude is expressed in terms of bare matching coefficients and bare operator matrix elements have been tamed by introducing rapidity regulators on the convolution integrals. We have proved two $D$-dimensional refactorization conditions for the bare matching coefficient $H_2^{(0)}(z)$ and the matrix element $\langle O_2^{(0)}(z)\rangle$ in the endpoint region $z\to 0$, which ensure that the dependence on the rapidity regulator cancels out to all orders of perturbation theory. With the help of these relations the factorization formula can be recast into a form where the endpoint divergences are removed by means of suitably chosen subtraction terms (for $T_2$) and cutoffs on the convolution integrals (for $T_3$). 

The main accomplishment of the present work has been to show that such an endpoint-regularized factorization theorem can be consistently formulated in terms of renormalized matching coefficients and operator matrix elements. This is a highly non-trivial point, because endpoint regularization and renormalization do not commute. We have derived the RG evolution equations satisfied by the renormalized matching coefficients and operator matrix elements and derived most of the anomalous dimensions at two-loop order (and the remaining ones at one-loop order). We have then used our results to predict in analytic form the logarithmically enhanced three-loop contributions to the $b$-quark induced $h\to\gamma\gamma$ decay amplitude of ${\cal O}(\alpha_b\spac\alpha_s^2\spac L^k)$ with $k=6,5,4,3$, finding perfect agreement with a numerical computation of these terms performed by the authors of \cite{Czakon:2020vql}. On the other hand, our findings for the structure of the coefficient of the subleading term (with $k=5$) disagrees with the predictions of previous authors \cite{Akhoury:2001mz,Anastasiou:2020vkr}, who had attempted to study the structure of the subleading logarithmic contribution using conventional tools. This demonstrates the usefulness of having a fully systematic approach based on effective field theory to study factorization beyond the leading power in scale ratios.

We are confident that the results presented in this work are a major step forward in the quest for a consistent formulation of SCET factorization theorems at subleading power. For the particular example considered -- the $b$-quark induced $h\to\gamma\gamma$ decay amplitude -- they form the theoretical basis for a systematic resummation of large double and single logarithms beyond leading order in RG-improved perturbation theory. The technical details of this resummation will be discussed in future work. It will also be important to generalize our analysis to the non-abelian case of the Higgs-boson production in the gluon-gluon fusion channel $gg\to h$, extending the approach of \cite{Liu:2017vkm,Liu:2018czl} to higher logarithmic accuracy.

\subsubsection*{Acknowledgements}

We are grateful to Marco Niggetiedt for providing us with the numerical results shown in (\ref{Marco}) and for the permission to present these results prior to their publication in \cite{Niggetiedt:2020sbf}. One of us (M.N.) thanks Gino Isidori, the particle theory group at Zurich University and the Pauli Center for hospitality during a sabbatical stay. This research has been supported by the Cluster of Excellence PRISMA$^+$\! funded by the German Research Foundation (DFG) within the German Excellence Strategy (Project ID 39083149). The research of Z.L.L.\ is supported by the U.S.\ Department of Energy under Contract No.~DE-AC52-06NA25396, the LANL/LDRD program and within the framework of the TMD Topical Collaboration.

\begin{appendix}

\renewcommand{\theequation}{A.\arabic{equation}}
\setcounter{equation}{0}

\section{Bare matching coefficients and matrix elements}
\label{app:bareresults}

For completeness, we list here the expressions for the $h\to\gamma\gamma$ matrix elements of the bare operators $O_i^{(0)}$ and the corresponding bare matching coefficients $H_i^{(0)}$ as derived in \cite{Liu:2019oav}. These expressions are needed to obtain the corresponding renormalized quantities derived in the present work.

\subsubsection*{Bare matrix elements}

Omitting the photon polarization vectors, the $h\to\gamma\gamma$ matrix element of the bare operator $O_1^{(0)}$ is to all orders of perturbation theory simply given by
\begin{equation}\label{O1bare}
   \langle\gamma\gamma|\,O_1^{(0)}\,|h\rangle
   = m_{b,0}\,g_\perp^{\mu\nu} \,,
\end{equation}
where $m_{b,0}$ is the bare $b$-quark mass. The reason is that $O_1$ does not contain any fields with color charges, and hence there are no QCD corrections to the matrix element. 

The bare matrix elements of the remaining operators are known to first order in $\alpha_s$ only. For the case of $O_2^{(0)}$ one finds (with $0\le z\le 1$)
\begin{equation}\label{O2mel}
   \langle\gamma\gamma|\,O_2^{(0)}(z)\,|h\rangle 
   = \frac{N_c\spac\alpha_{b,0}}{2\pi}\,m_{b,0}\,g_\perp^{\mu\nu}
    \left[ e^{\epsilon\gamma_E}\,\Gamma(\epsilon)\,\big(m_{b,0}^2\big)^{-\epsilon} 
    + \frac{C_F\alpha_{s,0}}{4\pi}\,\big( m_{b,0}^2 \big)^{-2\epsilon} \big[ K(z) + K(1-z) \big] 
    \right] ,
\end{equation}
where
\begin{equation}
\begin{aligned}
   K(z) &= \frac{1}{\epsilon^2} \left( \ln z + \frac{3}{2} \right)
    + \frac{1}{\epsilon} \left( \frac{\ln^2 z}{2} - \ln z\ln (1-z) - \frac{1}{4} 
    - \frac{\pi^2}{6} \right) \\
   &\quad\mbox{}+ 6\,\text{Li}_3(z) + \left( 1 - 2z - 2\ln z \right) \text{Li}_2(z) 
    + \frac{\ln^3 z}{6} + \big[ z + \ln(1-z) \big] \ln^2 z \\
   &\quad\mbox{}+ \left( 2\,\text{Li}_2(1-z) - \frac{1}{2} \ln(1-z) - \frac{1+3z}{2} 
    - \frac{\pi^2}{6} \right) \ln z    
    + \frac{3}{2} + \frac{\pi^2}{6} - 4\zeta_3 + {\cal O}(\epsilon) \,.
\end{aligned}
\end{equation}
In (\ref{O2mel}) $\alpha_{s,0}$ and $\alpha_{b,0}$ denote the bare QCD and electromagnetic couplings, respectively. Starting at first order in $\alpha_s$ the matrix element contains terms that are singular for $z\to 0$ or $z\to 1$. The former terms are contained in $K(z)$, while the latter ones are contained in $K(1-z)$. In order to compute the matrix element $\braces{\langle\gamma\gamma|\,O_2^{(0)}(z)\,|h\rangle}$ one needs to take the limit $z\to 0$ in the above expressions. In \cite{Liu:2019oav} this limit has been obtained in closed form in the dimensional regulator $\epsilon$. One finds
\begin{equation}
\begin{aligned}
   \braces{K(z) + K(1-z)}
   &= \frac{e^{2\epsilon\gamma_E}}{1-2\epsilon}\,\bigg[\,
    2(2-3\epsilon+2\epsilon^2)\,\Gamma^2(\epsilon)
    + 2(1-\epsilon)\,\Gamma(\epsilon)\,\Gamma(2\epsilon)\,\Gamma(-\epsilon) \\
   &\hspace{1.95cm}\mbox{}+ z^\epsilon\,(2-4\epsilon-\epsilon^2)\,
    \frac{\Gamma(2\epsilon)\,\Gamma^2(-\epsilon)}{\Gamma(1-2\epsilon)} \bigg] \,.
\end{aligned}
\end{equation}

In order to compute the matrix element of the bare operator $O_3^{(0)}$, as given in the third line of (\ref{fact4}), one needs the expressions for the bare jet and soft functions at NLO in $\alpha_s$. For the bare jet function one obtains
\begin{equation}
   J^{(0)}(p^2) = 1 + \frac{C_F\alpha_{s,0}}{4\pi} \left( -p^2 -i0 \right)^{-\epsilon}
    e^{\epsilon\gamma_E}\,\frac{\Gamma(1+\epsilon)\,\Gamma^2(-\epsilon)}{\Gamma(2-2\epsilon)}\,
   (2-4\epsilon-\epsilon^2) \,.
\end{equation}
This function has a cut along the positive $p^2$ axis starting at $p^2=0$ and extending to infinity. The bare soft function, which is defined in terms of the discontinuity of a soft quark propagator dressed with Wilson lines, can be written in the form
\begin{equation}\label{Ssplit}
   S^{(0)}(w) 
   = - \frac{N_c\spac\alpha_{b,0}}{\pi}\,m_{b,0}\,\Big[ S_a^{(0)}(w)\,\theta(w-m_{b,0}^2)
    + S_b^{(0)}(w)\,\theta(m_{b,0}^2-w) \Big] \,.
\end{equation}
At first order in $\alpha_s$ one finds
\begin{equation}\label{SaSb}
\begin{aligned}
   S_a^{(0)}(w) 
   &= \frac{e^{\epsilon\gamma_E}}{\Gamma(1-\epsilon)}\,\big(w-m_{b,0}^2\big)^{-\epsilon} \\
   &\mbox{}+ \frac{C_F\alpha_{s,0}}{4\pi} \bigg[ 
    \Big[ C_1(\epsilon) + \frac{2}{\epsilon}\,\ln(1-r) \Big] \big(w-m_{b,0}^2\big)^{-2\epsilon} 
    + C_2(\epsilon) \big( m_{b,0}^2 \big)^{1-\epsilon} \big(w-m_{b,0}^2 \big)^{-1-\epsilon} \\
   &\hspace{2.0cm}\mbox{}- 2\,\text{Li}_2(r) + 2 \ln r \ln(1-r) - 3 \ln^2(1-r) + 2 \ln(1-r) 
    + \dots \bigg] \,, \\
   S_b^{(0)}(w) &= \frac{C_F\alpha_{s,0}}{4\pi}\,\big( m_{b,0}^2 \big)^{-2\epsilon}\,\bigg[ 
    - \frac{4}{\epsilon}\,\ln(1-\hat w) + 6 \ln^2(1-\hat w) + \dots \bigg] \,,
\end{aligned}
\end{equation}
where
\begin{equation}\label{C1C2eps}
\begin{aligned}
   C_1(\epsilon) &= \frac{2 e^{2\epsilon\gamma_E}}{\Gamma(1-2\epsilon)}
    \left[\frac{(1+\epsilon)\,\Gamma(-\epsilon)^2}{\Gamma(2-2\epsilon)}
    + 2\Gamma(\epsilon)\,\Gamma(-\epsilon) \right] , \\
   C_2(\epsilon) &= - 2 e^{2\epsilon\gamma_E}\,\frac{3-2\epsilon}{1-2\epsilon}\,
    \frac{\Gamma(\epsilon)}{\Gamma(-\epsilon)} \,,
\end{aligned}
\end{equation}
and we have defined the dimensionless ratios $r=m_{b,0}^2/w$ and $\hat w=w/m_{b,0}^2$, both of which live on the interval $[0,1]$. In both expressions the dots refer to terms of ${\cal O}(\epsilon)$ and higher, which vanish for $r\to 0$ or $\hat w\to 0$, respectively.

\subsubsection*{Bare matching coefficients}

To first order in $\alpha_s$, the expressions for the bare matching coefficients obtained in \cite{Liu:2019oav} read
\begin{align}\label{Hires}
   H_1^{(0)} 
   &= \frac{y_{b,0}}{\sqrt2}\,\frac{N_c\spac\alpha_{b,0}}{\pi} \left( -M_h^2 -i0 \right)^{-\epsilon}
    e^{\epsilon\gamma_E}\,(1-3\epsilon)\,
    \frac{2\Gamma(1+\epsilon)\,\Gamma^2(-\epsilon)}{\Gamma(3-2\epsilon)} \notag\\
   &\quad\times \bigg\{ 1 - \frac{C_F\alpha_{s,0}}{4\pi} \left( -M_h^2 -i0 \right)^{-\epsilon}
    e^{\epsilon\gamma_E}\,\frac{\Gamma(1+2\epsilon)\,\Gamma^2(-2\epsilon)}{\Gamma(2-3\epsilon)} \notag\\
   &\hspace{1.45cm}\times \bigg[ 
    \frac{2(1-\epsilon)(3-12\epsilon+9\epsilon^2-2\epsilon^3)}{1-3\epsilon} 
    + \frac{8}{1-2\epsilon}\,\frac{\Gamma(1+\epsilon)\,\Gamma^2(2-\epsilon)\,\Gamma(2-3\epsilon)}%
                                  {\Gamma(1+2\epsilon)\,\Gamma^3(1-2\epsilon)} \notag\\
   &\hspace{2.1cm}\mbox{}- \frac{4(3-18\epsilon+28\epsilon^2-10\epsilon^3-4\epsilon^4)}{1-3\epsilon}\,
     \frac{\Gamma(2-\epsilon)}{\Gamma(1+\epsilon)\,\Gamma(2-2\epsilon)} \bigg] \bigg\} \,, \notag\\
   H_2^{(0)}(z) &= \frac{y_{b,0}}{\sqrt2}\,\bigg\{ \frac{1}{z} 
    + \frac{C_F\alpha_{s,0}}{4\pi} \left( -M_h^2 -i0\right)^{-\epsilon}
    e^{\epsilon\gamma_E}\,\frac{\Gamma(1+\epsilon)\,\Gamma^2(-\epsilon)}{\Gamma(2-2\epsilon)} \notag\\
   &\hspace{1.9cm}\times
    \left[ \frac{2-4\epsilon-\epsilon^2}{z^{1+\epsilon}} - \frac{2(1-\epsilon)^2}{z}
     -2(1-2\epsilon-\epsilon^2)\,\frac{1-z^{-\epsilon}}{1-z} \right] \bigg\} + (z\to 1-z) \,, \notag\\
   H_3^{(0)} &= \frac{y_{b,0}}{\sqrt 2} \left[ - 1 
    + \frac{C_F\alpha_{s,0}}{4\pi} \left( -M_h^2 -i0\right)^{-\epsilon}
    e^{\epsilon\gamma_E}\,2(1-\epsilon)^2\,
    \frac{\Gamma(1+\epsilon)\,\Gamma^2(-\epsilon)}{\Gamma(2-2\epsilon)} \right] ,
\end{align}
where $y_{b,0}$ is the bare $b$-quark Yukawa coupling. These expressions are exact to all orders in $\epsilon$. From the second relation one obtains
\begin{equation}\label{H2bar}
   \braces{\bar H_2^{(0)}(z)} 
   = \frac{y_{b,0}}{\sqrt2}\,\bigg\{ 1 + \frac{C_F\alpha_{s,0}}{4\pi} 
    \left( - M_h^2 \right)^{-\epsilon}
    e^{\epsilon\gamma_E}\,\frac{\Gamma(1+\epsilon)\,\Gamma^2(-\epsilon)}{\Gamma(2-2\epsilon)}\, 
    \Big[ (2-4\epsilon-\epsilon^2)\,z^{-\epsilon} - 2(1-\epsilon)^2\Big] \bigg\} 
\end{equation}
for the $z\to 0$ limit of the function $\bar H_2^{(0)}(z)$ introduced in (\ref{eq09}). Finally, in the rearranged factorization formula (\ref{fact4}) one needs the infinity-bin subtraction term $\Delta H_1^{(0)}$, which is given by 
\begin{equation}\label{DeltaH1}
\begin{aligned}
   \Delta H_1^{(0)} 
   &= - \frac{y_{b,0}}{\sqrt2}\,\frac{N_c\spac\alpha_{b,0}}{\pi} \left( -M_h^2 -i0 \right)^{-\epsilon}
    \frac{e^{\epsilon\gamma_E}}{\epsilon^2\,\Gamma(1-\epsilon)}\,
    \bigg\{ 1 + \frac{C_F\alpha_{s,0}}{4\pi} \left( -M_h^2-i0 \right)^{-\epsilon} \\
   &\qquad\times 
    \frac{e^{\epsilon\gamma_E}\,\Gamma(-\epsilon)\,\Gamma(1-\epsilon)}{\Gamma(2-2\epsilon)} 
    \left[ (1-2\epsilon+3\epsilon^2)\,\Gamma(\epsilon) 
    + \frac{1+\epsilon}{2}\,\frac{\Gamma(-\epsilon)}{\Gamma(1-2\epsilon)} \right] \bigg\} \,.
\end{aligned}
\end{equation}

\renewcommand{\theequation}{B.\arabic{equation}}
\setcounter{equation}{0}

\section{\boldmath Definitions of the jet and soft functions}
\label{app:jetsoft}

In the analysis of the refactorization conditions in Section~\ref{sec:refact} we have made use of the definitions of the (bare) radiative jet function $J^{(0)}(p^2)$ and the (bare) soft-quark soft function $S^{(0)}(\ell_+\ell_-)$ introduced in \cite{Liu:2019oav}. The two jet functions needed in (\ref{resujet1}) and (\ref{eq19}) are defined via
\begin{equation}
\begin{aligned}
   &\langle\gamma(k_2)|\,T\,\X_{n_2}^{\beta k}(r) 
    \big[\bar\X_{n_2}(y)\,\big(\Asl_{n_2}^\perp(y)+\Gsl_{n_2}^\perp(y)\big)\big]^{\gamma l}\,
    |0\rangle \\
   &= e_b\,\delta^{kl}\,\Big[\,\frac{\nsl_2}{2}\,\rlap/\varepsilon_\perp^*(k_2) \Big]^{\beta\gamma}
    \int\frac{d^Dp}{(2\pi)^D}\,\frac{i\bar n_2\cdot p}{p^2+i0}\,J^{(0)}\big(p^2,(p+k_2)^2\big)\,
    e^{-ip\cdot(r-y)+i k_2\cdot y} 
\end{aligned}
\end{equation}
and
\begin{equation}
\begin{aligned}
   &\langle\gamma(k_1)|\,T\,\big[\big(\Asl_{n_1}^\perp(x)+\Gsl_{n_1}^\perp(x)\big)\,
    \X_{n_1}(x)\big]^{\alpha i}\,\bar\X_{n_1}^{\beta j}(r)\,|0\rangle \\
   &= e_b\,\delta^{ij}\,\Big[\rlap/\varepsilon_\perp^*(k_1)\,\frac{\nsl_1}{2} \Big]^{\alpha\beta} 
    \int\frac{d^Dp}{(2\pi)^D}\,\frac{i\bar n_1\cdot p}{p^2+i0}\,J^{(0)}\big(p^2,(p-k_1)^2\big)\,
    e^{-ip\cdot(x-r)+i k_1\cdot x} \,.
\end{aligned}
\end{equation}
We have written out color indices (roman) and spinor indices (greek) explicitly. In both cases the second argument of the jet function vanishes, because it is equal to the square of the light-like momentum carried by the soft quark (after the multipole expansion). In the main text we have simply dropped this second argument.

The soft-quark soft function needed in (\ref{eq19}) is defined in terms of the soft matrix element
\begin{equation}
\begin{aligned}
   &\frac{e_b^2}{\pi}\,\langle 0|\,T\,\mbox{Tr}\,S_{n_2}(0,y_+)\,q_s^\gamma(y_+)\,
    \bar q_s^\alpha(x_-)\,S_{n_1}(x_-,0)\,|0\rangle \\
   &= i\int\frac{d^D\ell}{(2\pi)^D}\,e^{-i\ell\cdot(y_+ - x_-)}\,
    \bigg[ {\cal S}_1(\ell) + \rlap/\ell\,{\cal S}_2(\ell)
    + \frac{\nsl_1}{n_1\cdot\ell}\,{\cal S}_3(\ell) + \frac{\nsl_2}{n_2\cdot\ell}\,{\cal S}_4(\ell) \\
   &\hspace{4.64cm}\mbox{}+ \frac{\rlap/\ell\nsl_1}{n_1\cdot\ell}\,{\cal S}_5(\ell)
    + \frac{\nsl_2\rlap/\ell}{n_2\cdot\ell}\,{\cal S}_6(\ell) 
    + \frac{\nsl_2\nsl_1}{4}\,{\cal S}_7(\ell) 
    + \frac{\nsl_2\rlap/\ell\nsl_1}{2}\,{\cal S}_8(\ell) \bigg]^{\gamma\alpha} \,,
\end{aligned}
\end{equation}
where the trace in the first line is over color indices, and we have introduced the finite-length soft Wilson lines
\begin{equation}\label{eq25}
\begin{aligned}
   S_{n_1}(x_-,0) \equiv S_{n_1}(x_-)\,S_{n_1}^\dagger(0)
   &= P\exp\left[ ig_s \int_0^{\bar n_1\cdot x/2}\!dt\,n_1\cdot G_s(t n_1) \right] , \\
   S_{n_2}(0,y_+) \equiv S_{n_2}(0)\,S_{n_2}^\dagger(y_+)
   &= P\exp\left[ ig_s \int_{\bar n_2\cdot y/2}^0\!dt\,n_2\cdot G_s(t n_2) \right] , 
\end{aligned}
\end{equation}
Only the first structure function ${\cal S}_1(\ell)$ contributes in (\ref{eq19}) due to the presence of $\nsl_1$ and $\nsl_2$ in the trace over Dirac matrices.

\renewcommand{\theequation}{C.\arabic{equation}}
\setcounter{equation}{0}

\section{\boldmath Details on the derivation of the quantity $\delta H_1^{(0),\,{\rm tot}}$}
\label{app:mismatch}

Here we provide some technical details relevant for the derivation described in Section~\ref{sec:mismatch}. Our starting point is relation (\ref{eq41}). Note that the integrals over the products of $Z_J$ factors in the second and third lines would evaluate to $\delta$-functions if it was not for the upper cutoffs on the integrals over $\ell_\pm$. Using this fact, we can rewrite the result in the form
\begin{equation}
\begin{aligned}
   \delta H_1^{(0)}\spac m_{b,0}
   &= H_3^{(0)} \int_0^\infty\!\frac{d\rho_-}{\rho_-} \int_0^\infty\!\frac{d\rho_+}{\rho_+}\,
    S^{(0)}(\rho_+\rho_-) \int_0^\infty\!d\omega_- \int_0^\infty\!d\omega_+\,
    J^{(0)}(M_h\omega_-)\,J^{(0)}(-M_h\omega_+) \\
   &\quad\times \int_0^{M_h}\!d\ell_-\,Z_J^{-1}(M_h\rho_-,M_h\ell_-)\,Z_J(M_h\ell_-,M_h\omega_-) \\
   &\quad\times \int_0^{\sigma M_h}\!d\ell_+\,Z_J^{-1}(-M_h\rho_+,-M_h\ell_+)\,
    Z_J(-M_h\ell_+,-M_h\omega_+) \\
   &\quad\mbox{}- H_3^{(0)}\! \int_0^{M_h}\!\frac{d\rho_-}{\rho_-}
    \int_0^{\sigma M_h}\!\frac{d\rho_+}{\rho_+}\,S^{(0)}(\rho_+\rho_-)
    \int_0^\infty\!d\omega_- \int_0^\infty\!d\omega_+\,J^{(0)}(M_h\omega_-)\,
    J^{(0)}(-M_h\omega_+) \\
   &\quad\times \int_0^\infty\!d\ell_-\,Z_J^{-1}(M_h\rho_-,M_h\ell_-)\,Z_J(M_h\ell_-,M_h\omega_-) \\
   &\quad\times \int_0^\infty\!d\ell_+\,Z_J^{-1}(-M_h\rho_+,-M_h\ell_+)\,
    Z_J(-M_h\ell_+,-M_h\omega_+) \,.
\end{aligned}
\end{equation}
In this appendix we do not write out the limit $\sigma\to-1$ explicitly, but it is understood in all expressions where $\sigma$ occurs. We now rearrange the limits of the integrals in the following way (in obvious notation):
\begin{align}\label{intsT3}
   & \int_0^\infty\!d\rho_- \int_0^\infty\!d\rho_+ 
    \int_0^{M_h}\!d\ell_- \int_0^{\sigma M_h}\!d\ell_+
    - \int_0^{M_h}\!d\rho_- \int_0^{\sigma M_h}\!d\rho_+
    \int_0^\infty\!d\ell_- \int_0^\infty\!d\ell_+ \notag\\
   \to & \int_0^\infty\!d\rho_- \int_0^\infty\!d\rho_+ 
    \int_{M_h}^\infty\!d\ell_- \int_{\sigma M_h}^\infty\!d\ell_+
    - \int_{M_h}^\infty\!d\rho_- \int_{\sigma M_h}^\infty\!d\rho_+
    \int_0^\infty\!d\ell_- \int_0^\infty\!d\ell_+ \notag\\
   - & \int_0^\infty\!d\rho_- \int_0^\infty\!d\rho_+ 
    \int_{M_h}^\infty\!d\ell_- \int_0^\infty\!d\ell_+
    + \int_{M_h}^\infty\!d\rho_- \int_0^\infty\!d\rho_+
    \int_0^\infty\!d\ell_- \int_0^\infty\!d\ell_+ \notag\\
   - & \int_0^\infty\!d\rho_- \int_0^\infty\!d\rho_+ 
    \int_0^\infty\!d\ell_- \int_{\sigma M_h}^\infty\!d\ell_+
    + \int_0^\infty\!d\rho_- \int_{\sigma M_h}^\infty\!d\rho_+
    \int_0^\infty\!d\ell_- \int_0^\infty\!d\ell_+ \,.
\end{align}
In the next step we consider the contribution to the quantity $\delta' H_1^{(0)}\spac m_{b,0}$ involving the matrix element $\braces{\langle O_2^{(0)}\rangle}$, for which we found the expression  (\ref{deltaprime}). Manipulating this result in a similar way as above, we find
\begin{align}
   &\delta'\hspace{-0.5mm} H_1^{(0)}\spac m_{b,0} \Big|_{\,\braces{\langle O_2^{(0)}\rangle}} 
    \notag\\
   &= -2 H_3^{(0)} \int_0^\infty\!\frac{d\rho_-}{\rho_-} \int_0^\infty\!\frac{d\rho_+}{\rho_+}\,
    S^{(0)}(\rho_+\rho_-) \int_0^\infty\!d\omega_- \int_0^\infty\!d\omega_+\,
    J^{(0)}(M_h\omega_-)\,J^{(0)}(-M_h\omega_+) \notag\\
   &\quad\times \int_0^{M_h}\!d\ell_-\,Z_J^{-1}(M_h\rho_-,M_h\ell_-)\,Z_J(M_h\ell_-,M_h\omega_-) 
    \notag\\
   &\quad\times \int_0^\infty\!d\ell_+\,Z_J^{-1}(-M_h\rho_+,-M_h\ell_+)\,
    Z_J(-M_h\ell_+,-M_h\omega_+) \\
   &\quad\mbox{}+ 2 H_3^{(0)} \int_0^{M_h}\!\frac{d\rho_-}{\rho_-}
    \int_0^\infty\!\frac{d\rho_+}{\rho_+}\,S^{(0)}(\rho_+\rho_-)
    \int_0^\infty\!d\omega_- \int_0^\infty\!d\omega_+\,J^{(0)}(M_h\omega_-)\,
    J^{(0)}(-M_h\omega_+) \notag\\
   &\quad\times \int_0^\infty\!d\ell_-\,Z_J^{-1}(M_h\rho_-,M_h\ell_-)\,Z_J(M_h\ell_-,M_h\omega_-) 
    \notag\\
   &\quad\times \int_0^\infty\!d\ell_+\,Z_J^{-1}(-M_h\rho_+,-M_h\ell_+)\,
    Z_J(-M_h\ell_+,-M_h\omega_+) \,. \notag
\end{align}
In analogy with (\ref{intsT3}) we can rearrange the limits of the integrals as follows:
\begin{equation}\label{intsT2}
\begin{aligned}
   & -2 \int_0^\infty\!d\rho_- \int_0^\infty\!d\rho_+ 
    \int_0^{M_h}\!d\ell_- \int_0^\infty\!d\ell_+
    +2 \int_0^{M_h}\!d\rho_- \int_0^\infty\!d\rho_+
    \int_0^\infty\!d\ell_- \int_0^\infty\!d\ell_+ \\
   \to & +2 \int_0^\infty\!d\rho_- \int_0^\infty\!d\rho_+ 
    \int_{M_h}^\infty\!d\ell_- \int_0^\infty\!d\ell_+
    - 2 \int_{M_h}^\infty\!d\rho_- \int_0^\infty\!d\rho_+
    \int_0^\infty\!d\ell_- \int_0^\infty\!d\ell_+ \,.
\end{aligned}
\end{equation}
Considering now the sum of the results (\ref{intsT3}) and (\ref{intsT2}), we get
\begin{equation}
\begin{aligned}
   & \int_0^\infty\!d\rho_- \int_0^\infty\!d\rho_+ 
    \int_{M_h}^\infty\!d\ell_- \int_{\sigma M_h}^\infty\!d\ell_+
    - \int_{M_h}^\infty\!d\rho_- \int_{\sigma M_h}^\infty\!d\rho_+
    \int_0^\infty\!d\ell_- \int_0^\infty\!d\ell_+ \\
   + & \int_0^\infty\!d\rho_- \int_0^\infty\!d\rho_+ 
    \int_{M_h}^\infty\!d\ell_- \int_0^\infty\!d\ell_+
    - \int_{M_h}^\infty\!d\rho_- \int_0^\infty\!d\rho_+
    \int_0^\infty\!d\ell_- \int_0^\infty\!d\ell_+ \\
   - & \int_0^\infty\!d\rho_- \int_0^\infty\!d\rho_+ 
    \int_0^\infty\!d\ell_- \int_{\sigma M_h}^\infty\!d\ell_+
    + \int_0^\infty\!d\rho_- \int_{\sigma M_h}^\infty\!d\rho_+
    \int_0^\infty\!d\ell_- \int_0^\infty\!d\ell_+ \,.
\end{aligned}
\end{equation}
The terms shown in the last two lines of this expression are related to each other by the substitutions $\rho_\pm\to\rho_\mp$, $\ell_\pm\to\ell_\mp$ and $M_h\to\sigma M_h$, under which the integrand is invariant if we also replace $\omega_\pm\to\omega_\mp$. It follows that these two terms cancel each other, and hence we end up with
\begin{equation}
   \int_0^\infty\!d\rho_- \int_0^\infty\!d\rho_+ 
    \int_{M_h}^\infty\!d\ell_- \int_{\sigma M_h}^\infty\!d\ell_+
    - \int_{M_h}^\infty\!d\rho_- \int_{\sigma M_h}^\infty\!d\rho_+
    \int_0^\infty\!d\ell_- \int_0^\infty\!d\ell_+ \,.
\end{equation}
This proves relation (\ref{difficult}), in which the contribution in the last line is the same as in (\ref{deltaH1def}).

We still need to show that the various terms on the right-hand side of (\ref{difficult}) define hard contributions, which can be associated with the matrix element $\langle O_1^{(0)}\rangle$. This is obvious for the last term, which contains no reference to the $b$-quark mass. It is less obvious for the first two terms, which are given by
\begin{equation}\label{eq51}
\begin{aligned}
   &H_3^{(0)} \int_0^\infty\!\frac{d\rho_-}{\rho_-} \int_0^\infty\!\frac{d\rho_+}{\rho_+}\,
    S^{(0)}(\rho_+\rho_-) \int_0^\infty\!d\omega_- \int_0^\infty\!d\omega_+\,
    J^{(0)}(M_h\omega_-)\,J^{(0)}(-M_h\omega_+) \\
   &\quad\times \int_{M_h}^\infty\!d\ell_-\,Z_J^{-1}(M_h\rho_-,M_h\ell_-)\,
    Z_J(M_h\ell_-,M_h\omega_-) \\
   &\quad\times \int_{\sigma M_h}^\infty\!d\ell_+\,Z_J^{-1}(-M_h\rho_+,-M_h\ell_+)\,
    Z_J(-M_h\ell_+,-M_h\omega_+)\,\Big|_{\rm leading\;power} \\
   &\mbox{}- H_3^{(0)} \int_{M_h}^\infty\!\frac{d\rho_-}{\rho_-}
    \int_{\sigma M_h}^\infty\!\frac{d\rho_+}{\rho_+}\,
    S^{(0)}(\rho_+\rho_-)\,J^{(0)}(M_h\rho_-)\,J^{(0)}(-M_h\rho_+)\,\Big|_{\rm leading\;power} \,.
\end{aligned}
\end{equation}
Via the soft function, these terms are in principle sensitive to the soft scale $m_{b,0}$. However, it is important to remember that we only need the leading-power terms in this expression. In the second integral the variables $\rho_\pm$ are both in the hard region, and hence the arguments of the soft and jet functions are all of ${\cal O}(M_h^2)$. For the first integral, we focus first on the integral over the $Z_J$ factors. Using the explicit expression in (\ref{ZJexpr}) we find
\begin{equation}
\begin{aligned}
   &\int_{M_h}^\infty\!d\ell_-\,Z_J^{-1}(M_h\rho_-,M_h\ell_-)\,Z_J(M_h\ell_-,M_h\omega_-)
    = \delta(\rho_- -\omega_-)\,\theta(\rho_- -M_h) \\
   &\hspace{0.5cm}\mbox{}+ \frac{C_F\alpha_s}{2\pi\epsilon}\,\Big[ \theta(\rho_- > M_h > \omega_-) 
    - \theta(\omega_- > M_h > \rho_-) \Big]\,
    \rho_-\,\Gamma(\rho_-,\omega_-) + {\cal O}(\alpha_s^2) \,.
\end{aligned}
\end{equation}
The first term restricts both $\rho_-$ and $\omega_-$ to be in the hard region. For the second term this is not obvious but still true, because the factor $\rho_-$ in front of the plus distribution $\Gamma(\rho_-,\omega_-)$ removes the factor $1/\rho_-$ in the measure of the integral. We thus get (focussing only on the integrals over ``minus'' momenta)
\begin{equation}\label{prettygood}
\begin{aligned}
   &\quad \int_0^\infty\!\frac{d\rho_-}{\rho_-} S^{(0)}(\rho_+\rho_-) \int_0^\infty\!d\omega_- 
    J^{(0)}(M_h\omega_-) \int_{M_h}^\infty\!d\ell_-\,Z_J^{-1}(M_h\rho_-,M_h\ell_-)\,
    Z_J(M_h\ell_-,M_h\omega_-) \\
   &= \int_{M_h}^\infty\!\frac{d\rho_-}{\rho_-} S^{(0)}(\rho_+\rho_-)\,J^{(0)}(M_h\rho_-) \\
   &\quad\mbox{}+ \frac{C_F\alpha_s}{2\pi\epsilon}
    \int_{M_h}^\infty\!d\rho_-\,S^{(0)}(\rho_+\rho_-) \int_0^{M_h}\!d\omega_- 
    J^{(0)}(M_h\omega_-) \left[ \frac{1}{\rho_-(\rho_- -\omega_-)} \right]_+ \\
   &\quad\mbox{}- \frac{C_F\alpha_s}{2\pi\epsilon}
    \int_0^{M_h}\!d\rho_-\,S^{(0)}(\rho_+\rho_-) \int_{M_h}^\infty\!d\omega_- 
    J^{(0)}(M_h\omega_-) \left[ \frac{1}{\omega_-(\omega_- -\rho_-)} \right]_+ 
    + {\cal O}(\alpha_s^2) \,.
\end{aligned}
\end{equation}
In these expressions we can drop the plus prescription, because the integrand of the integral over $\omega_-$ contains a term $\theta(M_h-\omega_-)$ if we write this integral as $\int_0^\infty\!d\omega_-\cdots$. The plus prescription then gives a subtraction term involving $\theta(M_h-\rho_-)$, which vanishes since the integral over $\rho_-$ runs from $M_h$ to infinity. The first integral on the right-hand side of (\ref{prettygood}) is clearly in the hard region. For the second integral $\omega_-$ must be treated as a hard variable of ${\cal O}(M_h)$, because the jet function does not contain any reference to the mass of the $b$ quark and $\rho_-$ is in the hard region. In other words, the region where $\omega_-={\cal O}(m_b)$ gives rise to a power-suppressed contribution and hence must be dropped. (Recall that we must only keep the leading-power terms in the result.) Finally, for the third integral $\omega_-$ is in the hard region, and the region where $\rho_-={\cal O}(m_b)$ or smaller gives rise to a power-suppressed contribution. For this to be true, it is important that the measure is $d\rho_-$ and not $d\rho_-/\rho_-$. An analogous argument holds for the second integral over $Z_J$ factors in (\ref{eq51}).

With all integration variables restricted to the hard region, we can replace the bare soft function $S^{(0)}(w)$ by its asymptotic form $S_\infty^{(0)}(w)$ defined in (\ref{Sinfty}). When the dust settles, we obtain from (\ref{eq51}) 
\begin{equation}
\begin{aligned}
   \delta H_1^{(0),\,{\rm tot}}
   &= H_3^{(0)}\,\Bigg\{ \frac{C_F\alpha_s}{2\pi\epsilon} 
    \int_{\sigma M_h}^\infty\!\frac{d\rho_+}{\rho_+}\,
    \Bigg[ \int_{M_h}^\infty\!d\rho_-\,\frac{S_\infty^{(0)}(\rho_+\rho_-)}{m_{b,0}}
    \int_0^{M_h}\!d\omega_- \frac{1}{\rho_-(\rho_- -\omega_-)}\\
   &\hspace{4.9cm}\mbox{}- \int_0^{M_h}\!d\rho_-\,\frac{S_\infty^{(0)}(\rho_+\rho_-)}{m_{b,0}}
    \int_{M_h}^\infty\!d\omega_-\,\frac{1}{\omega_-(\omega_- -\rho_-)} \Bigg] \\
   &\hspace{1.4cm}\mbox{}+ \frac{C_F\alpha_s}{2\pi\epsilon} 
    \int_{M_h}^\infty\!\frac{d\rho_-}{\rho_-}\,
    \Bigg[ \int_{\sigma M_h}^\infty\!d\rho_+\,\frac{S_\infty^{(0)}(\rho_+\rho_-)}{m_{b,0}}
    \int_0^{\sigma M_h}\!d\omega_+ \frac{1}{\rho_+(\rho_+ -\omega_+)}\\
   &\hspace{4.9cm}\mbox{}- \int_0^{\sigma M_h}\!d\rho_+\,\frac{S_\infty^{(0)}(\rho_+\rho_-)}{m_{b,0}} 
    \int_{\sigma M_h}^\infty\!d\omega_+\,\frac{1}{\omega_+(\omega_+ -\rho_+)} \Bigg] \\
   &\hspace{1.4cm}\mbox{}+ {\cal O}(\alpha_s^2) \Bigg\} \,,
\end{aligned}
\end{equation}
where at this order we can use the lowest-order expressions for $S_\infty^{(0)}(w)$ given in (\ref{Sinfty}). Performing the integrals then leads to (\ref{eq55}).

\renewcommand{\theequation}{D.\arabic{equation}}
\setcounter{equation}{0}

\section{Two-loop anomalous dimensions}
\label{app:anomdims}

We define the expansion coefficients of the cusp anomalous dimensions as
\begin{equation}
   \Gamma_{\rm cusp}(\alpha_s) 
   = \Gamma_0\,\frac{\alpha_s}{4\pi} + \Gamma_1 \left( \frac{\alpha_s}{4\pi} \right)^2 + \dots \,, 
\end{equation}
and similarly for all other anomalous dimensions. Below we list relevant expansion coefficients needed in Section~\ref{sec:largelogs} in the $\overline{{\rm MS}}$ renormalization scheme. The expansion coefficients of the cusp anomalous dimension $\Gamma_{\rm cusp}$ are given by \cite{Korchemskaya:1992je}
\begin{equation}
   \Gamma_0 = 4 C_F \,, \qquad
   \Gamma_1 = 4 C_F \left[ C_A \left( \frac{67}{9} - \frac{\pi^2}{3} \right) 
    - \frac{20}{9}\,T_F n_f \right] .
\end{equation}
For the coefficients of the anomalous dimension of the running quark mass, which is related to the anomalous dimension of the operator $O_1(\mu)$ by $\gamma_{11}=-\gamma_m$, one finds \cite{Tarrach:1980up}
\begin{equation}
   \gamma_{m,0} = - 6 C_F \,, \qquad
   \gamma_{m,1} = -3 C_F^2 - \frac{97}{3}\,C_F C_A + \frac{20}{3}\,C_F T_F n_f \,.
\end{equation}
The anomalous dimension $\gamma_q$ of the collinear quark field entering in (\ref{gamma33}) has coefficients \cite{Moch:2005id,Becher:2009qa}
\begin{equation}
\begin{aligned}
   \gamma_{q,0} &= -3 C_F \,, \\
   \gamma_{q,1} &= C_F^2 \left( -\frac{3}{2} + 2\pi^2 - 24\zeta_3 \right)
    + C_F C_A \left( - \frac{961}{54} - \frac{11\pi^2}{6} + 26\zeta_3 \right)
    + C_F T_F n_f \left( \frac{130}{27} + \frac{2\pi^2}{3} \right) ,
\end{aligned}
\end{equation}
while the coefficients of the anomalous dimension $\gamma'$ entering in (\ref{gammaJ}) read \cite{Liu:2019oav}
\begin{equation}
   \gamma'_0 = 0 \,, \qquad
   \gamma'_1 = C_F \left[ C_A \bigg( \frac{808}{27} - \frac{11\pi^2}{9} - 28\zeta_3 \bigg) 
    - T_F\,n_f \bigg( \frac{224}{27} - \frac{4\pi^2}{9} \bigg) \right] .
\end{equation}
The anomalous dimension $\gamma_s$ then follows from (\ref{gams}). One finds the expansion coefficients \cite{Liu:2020eqe}
\begin{equation}
\begin{aligned}
   \gamma_{s,0} &= - 6 C_F \,, \\
   \gamma_{s,1} &= C_F^2\,\big( - 3 + 4\pi^2 - 48\zeta_3 \big) 
    + C_F C_A \left( \frac{655}{27} - \frac{55\pi^2}{9} - 4\zeta_3 \right) 
    + C_F T_F\,n_f \left( - \frac{188}{27} + \frac{20\pi^2}{9} \right) .
\end{aligned}
\end{equation}

\end{appendix}

\end{document}